\documentclass[
amsmath, amssymb, aps,prd,nofootinbib,11pt]{revtex4-2}

\usepackage[T1]{fontenc}
\usepackage{csquotes}
\usepackage{graphicx,color}
\usepackage[dvipsnames]{xcolor}
\usepackage{xcolor}
\usepackage{rotating}
\usepackage{amsfonts}
\usepackage{amsmath}
\usepackage{xspace}
\usepackage{mathtools} 
\usepackage{verbatim}
\usepackage{gensymb}
\usepackage{soul}
\usepackage{amsthm}
\usepackage{placeins}

\newtheorem{theorem}{Theorem}
\newtheorem{definition}[theorem]{Definition}
\DeclareMathOperator{\link}{\prec\hspace{-0.4em}\raisebox{-0.31em}{*}\hspace{0.2em}}

\usepackage[colorlinks=true, 
linktoc=page,
linkcolor=MidnightBlue, 
citecolor=OliveGreen!80!Blue, 
urlcolor=MidnightBlue
]{hyperref}

\begin{document}

\title{Towards black-hole horizons and geodesic focusing in causal sets
}

\author{Astrid Eichhorn}
\email[]{eichhorn@thphys.uni-heidelberg.de}
\affiliation{Institute for Theoretical Physics, Heidelberg University, Philosophenweg 12 and 16, 69120 Heidelberg, Germany}

\author{Pedro Gamito}
\email[]{gamito@thphys.uni-heidelberg.de}
\affiliation{Institute for Theoretical Physics, Heidelberg University, Philosophenweg 12 and 16, 69120 Heidelberg, Germany}

\author{Nawder Stokes}
\email[]{stokes@thphys.uni-heidelberg.de}
\affiliation{Institute for Theoretical Physics, Heidelberg University, Philosophenweg 12 and 16, 69120 Heidelberg, Germany}

\begin{abstract}
The event horizon of a black hole is arguably the most dramatic manifestation of the fact that in General Relativity, causal structure is dynamical and spacetimes can be separated into distinct regions by causal boundaries.
Causal set quantum gravity is an approach to quantum gravity in which causal relations between spacetime points constitute the basic structure on which the theory is based. This raises the question how a discrete horizon can be identified in a causal set. In our paper, we first construct a local diagnostic to approximate a global concept, namely the event horizon, based on discrete timelike curves. 
 We then  turn
 to the concept of an apparent horizon, which is based on  
 local properties of geodesics, rather than global properties of the entire spacetime. We undertake first steps towards detecting apparent horizons in causal sets, using so-called ladders as tracers of null geodesics. We find that a discrete counterpart of the expansion changes sign across the black-hole horizon, as it should. Finally, we introduce the notion of a \emph{fuzzy ladder}, which enables us to track null geodesics for larger intervals of the affine parameter. Thereby, we construct a  portion of a discrete horizon in a toy-model for a black-hole spacetime in 1+1 dimensions.
\end{abstract}

\maketitle
\tableofcontents

\section{Motivation}
\label{sec:intro}
Horizons are key predictions of General Relativity (GR) and particularly interesting for quantum gravity due to several reasons.

First, Penrose's strong cosmic censorship conjecture \cite{Penrose:1969pc,Penrose:1979azm} implies that horizons shield  spacetime regions with large quantum-gravity effects from outside observers, effectively ``hiding'' quantum gravity from us. According to the conjecture, curvature singularities are not visible to asymptotic observers and are instead hidden behind horizons. This is key to ensure a deterministic evolution outside the horizon. Therefore, formulating the conjecture precisely and finding a formulation that can be proven to hold in GR is an active area of research, see, e.g., \cite{Dafermos:2017dbw,Sbierski:2024fko}  as well as the review \cite{VandeMoortel:2025ngd}. From a quantum-gravity perspective, the cosmic censorship conjecture may be rephrased as the assertion that in a black-hole spacetime, spacetime regions in which quantum-gravity effects may dominate over GR, are shielded from asymptotic observers by a horizon.

Second, black-hole horizons also lie at the heart of the information paradox \cite{Hawking:1976ra}. Within a semiclassical setting, information, once it has crossed the horizon, appears irretrievably lost to the outside observer.
In a discrete setting, where only an approximate, fuzzy notion of a horizon can exist, it is an important question whether information can ``leak'' from the interior into the exterior.

Third, it is universally accepted lore that the entropy of black holes is given by the Bekenstein--Hawking formula (up to subleading corrections) and thus the entropy is proportional to the area of the horizon. This is a strong motivation for the holographic principle in quantum gravity. In addition, this scaling of entropy with horizon area lies at the heart of many swampland conjectures in quantum gravity, see \cite{Palti:2019pca,vanBeest:2021lhn} for reviews. In a discrete quantum-gravity setting, one may expect that the entropy is effectively encoded in the discrete degrees of freedom that cross the horizon, see, e.g., \cite{Dou:2003af,Barton:2019okw}.

Fourth, horizons determine the observational properties of black holes: For black-hole images, they imply that a black-hole shadow exists \cite{Broderick:2009ph}. For gravitational waves, boundary conditions at a horizon must be purely ingoing, which determines, e.g., the properties of quasinormal modes \cite{Cardoso:2016rao}. In a discrete quantum-gravity context, the fuzzy nature of a horizon may leave imprints (albeit very likely only very tiny ones) in observational quantities. 

Understanding black-hole horizons, their existence, properties and dynamics, therefore is an important goal within quantum gravity.

Here, we approach this goal from the perspective of causal set quantum gravity \cite{Bombelli:1987aa}. 
Causal set quantum gravity is an intrinsically Lorentzian, discrete approach \cite{Bombelli:2006nm}. It postulates that continuous spacetime manifolds together with a metric are emergent from an underlying discrete set of points, on which only causal structure exists. All notions of continuum geometry, including spacetime curvature, geodesic curves and black-hole horizons, are ultimately emergent and the continuum only constitutes an approximation to the fundamentally discrete setting. This approximate relation entails that all notions of continuum geometry are ``fuzzy'', i.e., the expectation values of the discrete quantities fluctuate around their continuum counterparts, and on any given realization of a discrete ``geometry", the discrete quantity and its continuum approximation deviate from each other. Thus, in particular, the horizon, which is an idealized, infinitesimally thin boundary, corresponds to a ``fuzzy'' discrete structure. Therefore, one may expect that several of the idealized properties of horizons that one is used to in continuum physics undergo changes in this setting, with potential consequences for the various topics listed above. This makes it highly worthwhile to work on the concrete construction of a discrete black-hole horizon in causal sets.

Because this theory is based on causal relations between spacetime points as its basic structure, one might expect that black-hole horizons are straightforward to find and characterize in a causal set. This is, however, not the case. In our paper, we discuss several notions of horizons, including event horizons and apparent horizons, and deploy a set of different tools to identify such horizons.\newline

This paper is structured as follows: In Sec.~\ref{sec:intro2}, we introduce causal set quantum gravity. In Sec.~\ref{sec:eventhorizon} we highlight the difficulty with the concept of an event horizon in a causal set and discuss how a local diagnostic that approximates a global concept can be found. We then focus on the apparent horizon, which is defined through the properties of the geodesic expansion. We therefore resort to the definition of ladders
as a tracer of a null geodesics, first introduced in \cite{Bhattacharya:2023xnj}, in Sec.~\ref{sec:ladders}. We identify such ladders for the first time in a 1+1 dimensional toy model of a Schwarzschild spacetime and then construct a discrete analogue of geodesic expansion applied to ladders. We show that the discrete geodesic expansion can distinguish the black-hole interior from the black-hole exterior. Finally, we pull together concepts from Sec.~\ref{sec:eventhorizon} and Sec.~\ref{sec:ladders} in Sec.~\ref{sec:discretehorizon}. There, we introduce the notion of a fuzzy ladder and show that such fuzzy ladders can approximate the black-hole horizon.
We conclude and point out directions for future work in Sec.~\ref{sec:conclusions}.

\section{Introduction to causal set quantum gravity}\label{sec:intro2}
Causal set quantum gravity is based on the causal structure of spacetimes. According to the Hawking--Malament theorem \cite{Hawking:1976fe, malament1977class}, under physically reasonable assumptions, the causal structure encodes the full information on the metric (modulo diffeomorphisms) up to a (local) conformal factor. In metric gravity, all invariant characterizations of the spacetime geometry (at fixed topology), i.e., all curvature invariants and geodesic curves, can be deduced from the metric. Thus, knowing the causal structure and the conformal factor is sufficient to fully characterize the spacetime geometry.

This theorem holds in the continuum. Causal set theory \cite{Bombelli:1987aa}, reviewed in \cite{Dowker:2005tz,Surya:2019ndm,Surya:2025knk}, is based on the idea that the causal structure continues to play a similarly powerful role in a discrete setting. Thus, causal set theory is based on causal relations between a discrete set of spacetime points. From a continuous line element, only the information on the \emph{sign} is kept and the entire structure of a causal set can be encoded in an adjacency matrix, which contains only $0$'s (when two elements are causally unrelated) and $1$'s (when two elements are causally related).
In addition, spacetime discreteness enables us to recover the conformal factor by \emph{counting}: on average, a spacetime volume $V = n \,V_{\rm Planck}$ with the Planck volume $V_{\rm Planck}$, contains $n$ spacetime points. More generally, the discreteness scale $\ell = V_{\rm Planck}^{1/d}$ (with spacetime dimensionality $d$), need not equal the Planck scale, but the latter constitutes the ``natural'' expectation for the discreteness scale in quantum gravity.

A causal set $C$ is defined as a partially ordered set of spacetime points, in which the partial order is based on the relation $\prec$, with $x \prec y$ meaning that the spacetime point $x$ precedes the spacetime point $y$ in a causal sense, i.e., they are either timelike or null related.\footnote{In practice, the probability for any given relation to be null is zero.}
We accordingly have that
\begin{itemize}
\item[i)] {\bf Transitivity:} If $x \prec y$ and $y\prec z$, then $x\prec z$, $\forall\, x,y,z\, \in C$.
\item[ii)]{\bf No closed timelike curves:} If $x\prec y$ and $y \prec x$, then $x =y,\, \forall \, x,y\, \in C$. For this definition to make sense, the causal relation is defined such that each element is related to itself.
\item[iii)] {\bf Discreteness:} $|[x,y]|<\infty$, with $[x,y] = \{z\in C: \, x\prec z\prec y \}$.
\end{itemize}
The first two requirements are shared between the causal order of spacetime points in a physically viable continuum manifold and a causal set. The last requirement is what makes a causal set discrete, because causal intervals always contain infinitely many points in the continuum.
Explicitly, a causal interval $[x,y]$ can be defined as the intersection of the future of $x$ and past of $y$, i.e.,
\begin{align}
{\rm Fut}(x)&= \{z \in C: x \prec z\},\\
{\rm Past}(x)&=\{z \in C: z \prec x\},\\
[x,y]&= {\rm Fut}(x) \cap {\rm Past}(y).
\end{align}

A crucial building block of causal sets is a \textbf{link}. The link, denoted $\link$, is a nearest-neighbor relation. This means that for two elements of the causal set $x, y \in C$, $x \link y \implies \not\!\exists \, z : x \prec z \prec y$. All other causal relations can be deduced from links; thus, the most minimal representation of a causal set is through the link matrix. Its transitive completion gives rise to the adjacency matrix.

The definition of a causal set encompasses many partial orders that bear little resemblance to discretizations of physically relevant spacetimes. In a sum-over-histories approach to quantization, one may hope that a suitable choice of the action results in destructive interference of non-manifoldlike causal sets \cite{Carlip:2022nsv,Carlip:2023zki}. Alternatively, one may consider restricting the measure in the sum-over-histories to manifold-like causal sets, aka \emph{sprinklings}. This becomes particularly appealing, when one considers a causal set as a regularization of a continuum manifold, rather than as a fundamentally discrete manifold, which is, e.g., motivated by the search for a universal continuum limit in quantum gravity \cite{Surya:2011du,Glaser:2017sbe,Eichhorn:2017bwe,Cunningham:2019rob,Eichhorn:2019xav} in the spirit of a convergence of ideas from different quantum-gravity approaches \cite{deBoer:2022zka}.

A sprinkling is a causal set that is constructed from a continuum spacetime manifold. From a fundamental perspective, in which differentiable manifolds are emergent structures, a sprinkling should be understood as an auxiliary process. Once it is understood how to construct manifoldlike causal sets from a dynamical quantum principle \cite{Rideout:1999ub,Zalel:2023uwy} --- a quantum growth model --- the process of sprinkling becomes obsolete.

To construct a sprinkling in practice, one distributes elements of the causal set into a finite region of the continuum manifold according to a Poisson process, i.e., the probability for a spacetime region with volume $V$ to contain $n$ points is 
\begin{equation}
P(n) = \frac{1}{n!}(\rho V)^n \, \exp(-\rho V),
\end{equation}
where the density is $\rho = \ell^{-d}$.
Using a Poisson process ensures that Lorentz invariance can remain intact in a statistical sense \cite{Bombelli:2006nm}.
Subsequently, the relation $\prec$ of the various elements in the sprinkling is inferred from the continuum line element. Thus $x \prec y$, if there exists a causal curve from $x$ to $y$ in the continuum. Finally, the continuum manifold is removed, such that the information on how the elements of the causal set are embedded into the manifold is not part of the definition of a sprinkling. Thus, a sprinkling is represented by a (non-embedded) graph or the equivalent adjacency matrix. In this paper, we will rely on the embedding information sometimes, and will also show sprinklings within an embedding in order to illustrate agreement between discrete and continuum quantities.

Manifoldlike causal sets are those that have a high probability of having arisen from a Poisson-process. Checking manifoldlikeness is a key challenge in causal set quantum gravity. To this end, several geometric and topological observables have been developed, see, e.g., \cite{Myrheim:1978ce,Brightwell:1990ha,Reid:2002sj,Major:2006hv,Sorkin:2007qi,Rideout:2008rk,Major:2009cw,Benincasa:2010ac,Aslanbeigi:2014zva,Eichhorn:2018doy,deBrito:2023axj,Surya:2025mvt}. In addition, quantum field theory for a scalar field can be formulated on a causal set \cite{Johnston:2009fr,Sorkin:2011pn,Aslanbeigi:2014zva}, resulting in additional observables \cite{Albertini:2024srq}.
 A special role in causal sets is played by those observables related to black-hole entropy. This is because
a main motivation to consider spacetime discreteness is the Bekenstein--Hawking entropy formula which is interpreted as providing the leading term to the entropy of a black hole (rather than just the entanglement entropy of quantum fields living \emph{on top} of a black-hole spacetime \emph{without backreaction}). Thus, horizon-molecules have been introduced in causal-set theory as candidates for the discrete degrees of freedom accounting for the entropy \cite{Dou:2003af,Barton:2019okw,Dou:2023puz}. 

A complete list of observables which would enable a distinction of manifoldlike to non-manifoldlike as well as a mapping to a particular continuum manifold is still missing.\footnote{Such a mapping, as well as the distinction of manifoldlike from non-manifoldlike always has to be subject to a (suitably defined) coarse-graining procedure.} 
For an overview of the current status of these developments, see the general reviews \cite{Surya:2019ndm,Surya:2025knk}, as well as more specialized reviews \cite{Yazdi:2022hhv,Dou:2023puz,Jubb:2023mlv,Glaser:2023pcl,X:2023ewv,Zalel:2023uwy,Ashmead:2024pmh,Ahmed:2024gzc,Mathur:2024gmk}.

\subsection{Sprinkling into a black-hole spacetime}

To sprinkle into a black-hole spacetime, we require knowledge of the causal relations of spacetime points, see \cite{He:2008ku}. For the Schwarzschild spacetime, because of the high degree of symmetry, the expressions for null geodesics are known analytically. We use this to construct a sprinkling into a toy model of $(1+1)$-dimensional Schwarzschild spacetime. 

The reasons for only considering a toy two-dimensional spacetime, instead of the full $(3+1)$-dimensional Schwarzschild spacetime, are several. First and foremost, the discrete approximation of null geodesics introduced in \cite{Bhattacharya:2023xnj} that we will use and further develop, is defined for ($1+1$)-dimensional settings.
Second, working in a toy $(1+1)$-dimensional example allows us to be numerically more efficient and to more easily illustrate the discrete spacetime structures we consider.
Finally, we note that the salient features of the causal structure of Schwarzschild spacetime are found in the $(t, r)$-slices  as a consequence of spherical symmetry. For details on how the full $(3+1)$-dimensional Schwarzschild spacetime may be sprinkled numerically, see \cite{Homsak:2024tce}. 

Hence, we construct a sprinkling of the (1+1)-dimensional spacetime with line-element corresponding to the induced metric on a submanifold at constant angular coordinates, given in 
Schwarzschild coordinates
by 
\begin{align} \label{eq:schwarzschild-lineleement}
\mathrm ds^2 = - \left(1- \frac{r_S}{r} \right) \mathrm dt^2 + \left(1- \frac{r_S}{r} \right)^{-1} \mathrm dr^2 \,,
\end{align}
with $r_S$ denoting the radial location of the event horizon. 
Sprinkling into such a two-dimensional spacetime is distinct from taking a two-dimensional slice through a four-dimensional sprinkling. In a sprinkling, a submanifold of lower dimension than the original manifold corresponds to a set of points of measure zero. Thus, we first restrict to the two-dimensional hypersurface, calculate the induced metric and then sprinkle according to it.
The induced metric on the (1+1)-dimensional submanifold spanned by $(t,r)$ has a constant determinant, implying a constant sprinkling density. This remains true, also if we use $(t_{\ast}, r)$ coordinates, as we do below.

\begin{figure}[!t]
    \includegraphics[width=\linewidth]{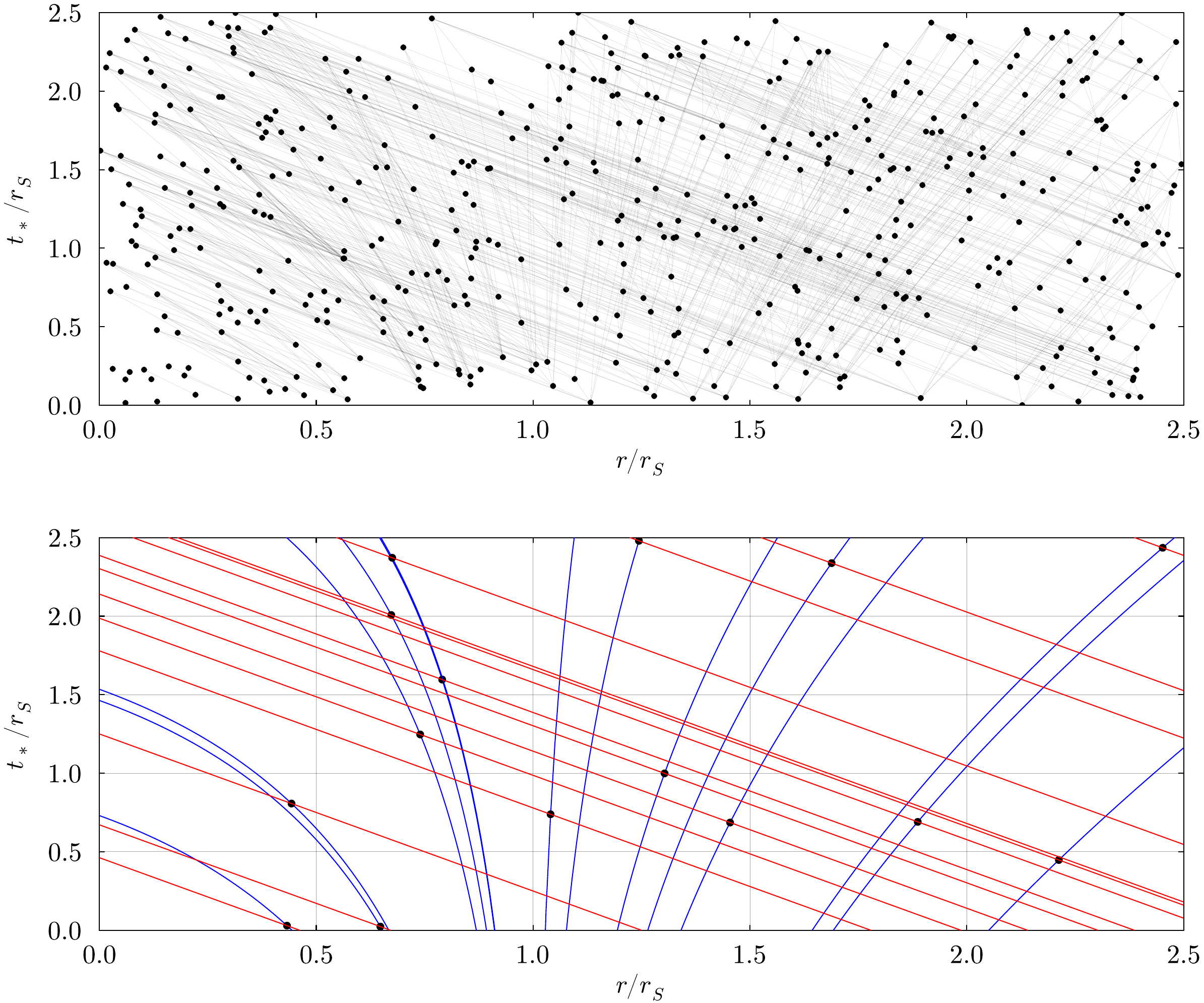}
    \caption{\label{fig:geodesics-and-links-schwarzschild} 
    We show a sprinkling of $500$ points, together with their links, into a patch of $(1+1)$-dimensional Schwarzschild spacetime (upper panel). The corresponding continuum null geodesics (outgoing in blue and ingoing in red) are used to infer the links (bottom panel). The horizon lies at $r/r_S=1$.
    }
\end{figure}

The null geodesics of this spacetime can be expressed analytically in terms of $(t_*, r)$ with
\begin{align} \label{eq:edd-fink-tstar}
t_* 
= t + 2M \ln \left(\left|\frac{r - 2M}{2M}\right|
\right)\,.
\end{align}
In these coordinates, the ingoing and outgoing null geodesics are, respectively
\begin{align} \label{eq:schwarzschild_geods}
t_{*\text{in}}(r) = c_1 - r \,, \qquad t_{*\text{out}}(r) = r + 4M \ln \left(\left|\frac{r - 2M}{2M}\right|\right) + c_2 \,,
\end{align}
where $c_1$ and $c_2$ are constants of integration that characterize the individual geodesics. For a given sprinkled point at $(T_{\ast}, R)$, we determine $c_1$ and $c_2$ such that the ingoing and outgoing geodesics pass through this point. Then, any point to the future of both geodesics is causally connected to the point at $(T_{\ast}, R)$. Finally, to determine the links, we use a transitive reduction and remove all relations that are implied by other relations.
In Fig.~\ref{fig:geodesics-and-links-schwarzschild} we show an example for a resulting sprinkling, together with the continuum expressions for the null geodesics used to infer the causal relations. We can already see an imprint of the continuum lightcones, and their ``tilting over'' at the horizon, in the structure of the links in the causal set. 

As a final point, note how the only information we required of the spacetime is the expression for its null geodesics. This indicates that this method may be straightforwardly adapted to other spacetimes of the same form 
as the one used here, i.e.,
\begin{align}
\mathrm ds^2 = - f(r) \mathrm dt^2 + f(r)^{-1} \mathrm dr^2 \,,
\end{align}
where $f(r)$ is some function of the radial coordinate $r$. Such spacetimes, with the requirement $f(r)\sim r^2+\mathcal{O}(r^3)$, constitute regular black holes. These have received a lot of attention in the literature, see, e.g., \cite{Olmo:2022cui,Eichhorn:2022bgu,Ashtekar:2023cod,Carballo-Rubio:2023mvr,Carballo-Rubio:2025fnc} for reviews and \cite{Bueno:2025zaj,Eichhorn:2025pgy,Carballo-Rubio:2025ntd,Borissova:2026wmn} for theories beyond General Relativity with regular, four-dimensional solutions. 
In causal set theory, one may first wonder why sprinkling into a regular black-hole spacetime should be of interest. After all, spacetime discreteness results in the absence of curvature singularities, in the sense that there is a vanishing probability that a point of the sprinkling lands at $r=0$  where the classical singularity is located. However, absence of curvature singularities is not the only desired feature of physical black-hole spacetimes. In addition, these should be geodesically complete. Geodesic completeness is not implied by the absence of curvature singularities. In fact, for regular black holes, geodesics cannot automatically be continued across $r=0$, because continuous differentiability of geodesics is often lost at this point. \\
\begin{figure}[t!]
\includegraphics[width=\linewidth]{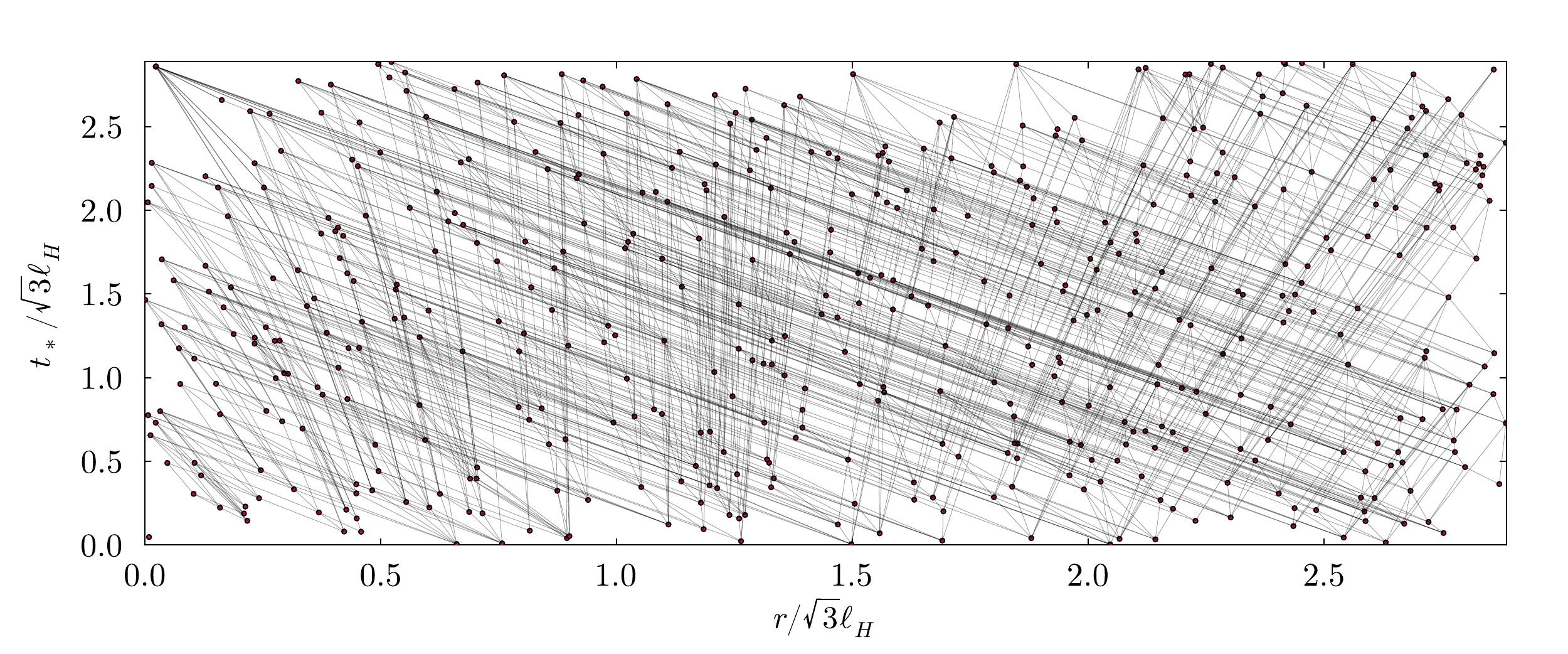}
\caption{\label{fig:hayward-causalset} We show a sprinkling into (1+1)-dimensional Hayward spacetime. Because the sprinkling is not into a (3+1)-dimensional setting, our conjecture that an antipodal continuation of geodesics is possible across $r=0$, cannot be directly tested. We do, however, observe an imprint of regularity in that light cones keep larger opening angles than in the corresponding Schwarzschild sprinkling in Fig.~\ref{fig:geodesics-and-links-schwarzschild} close to $r=0$.}
\end{figure}
As a specific example, we pick the Hayward spacetime \cite{Hayward:2005gi}, with $f(r) = 1- \frac{2Mr^2}{r^3+2M\ell_H^2}$, where $\ell_H$ is a new length scale that can be associated with the regularizing effect. A sprinkling into such a spacetime is shown in Fig.~\ref{fig:hayward-causalset}. In this case, continuum null geodesics are not infinitely many times differentiable, instead, they are only $C^{(5)}$ functions, if one wants to perform an antipodal continuation  across $r=0$ \cite{Zhou:2022yio,Torres:2022twv}. However, in a causal set, there are no smooth geodesics. Thus, there is no meaningful difference between $C^{(5)}$ and $C^{(\infty)}$. We thus conjecture that a sprinkling into such a regular black-hole spacetime is actually a physically viable system in the sense that it does not exhibit any (discrete counterparts of) geodesics that end in finite proper time. We leave a more detailed study of such questions to future work and here limit ourselves to pointing out that constructing such sprinklings is actually possible.

\section{The event horizon: approximating a global concept through local diagnostics}\label{sec:eventhorizon}

An event horizon is defined as the boundary of the causal past of future null infinity $\mathcal{J}^+$. To identify an event horizon, one therefore needs to find the full past of future null infinity, i.e., the full past of any  non-spacelike
curve that ends in future null infinity. 
The complement of the union of these pasts is the black-hole interior and the boundary of the union of the pasts is the event horizon. Thus, to define an event horizon in a causal set, an infinite sprinkling is required.\footnote{As an alternative, one may consider a sprinkling into a Penrose diagram of a black-hole spacetime. The underlying metric is related to the original black-hole metric by a conformal transformation and thus leaves causal relations intact. To the best of our knowledge, this idea has not yet been explored.}

\begin{figure}[!t]
\begin{center}
    \includegraphics[width=0.9\linewidth]{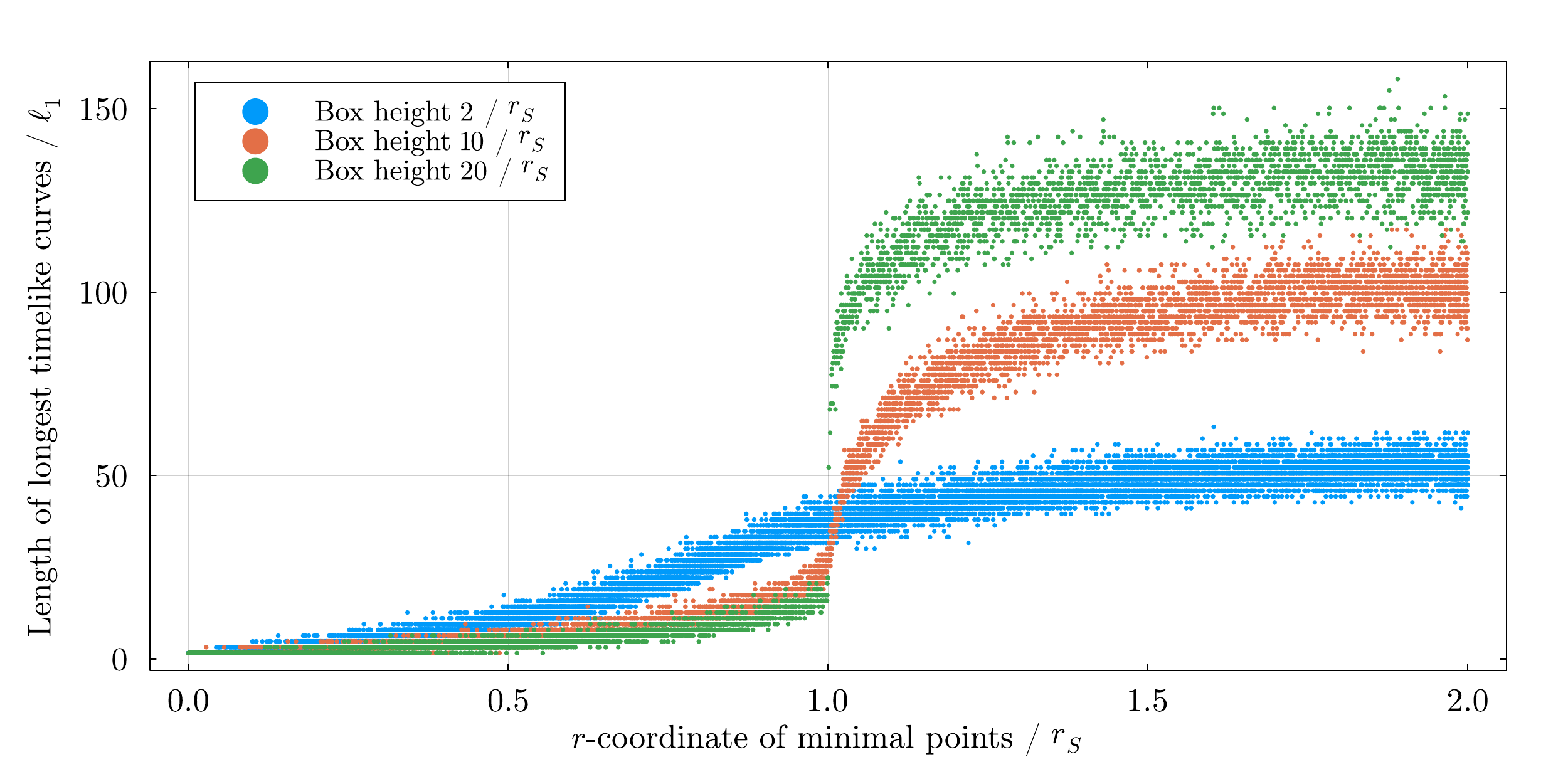}\\
    \includegraphics[width=0.9\linewidth]{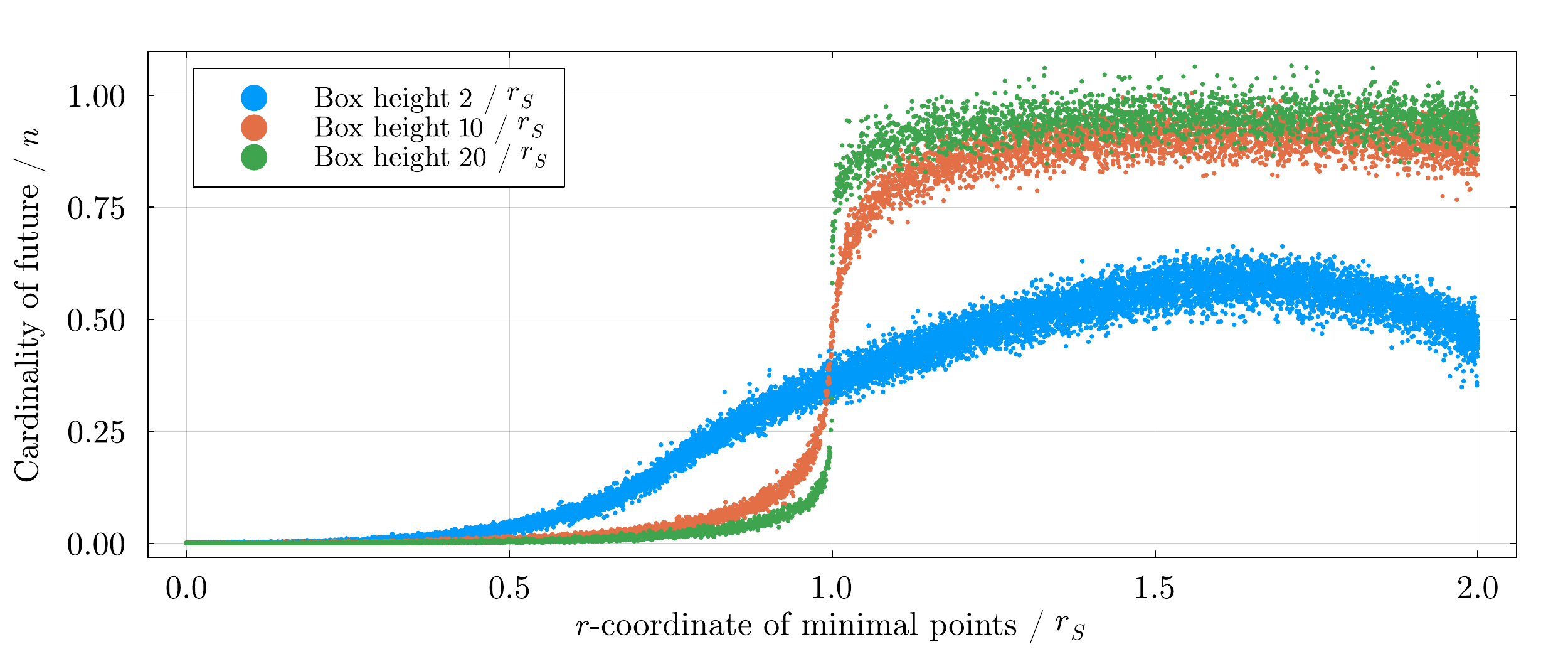}
    \end{center}
\caption{\label{fig:averagechainlength} Upper panel: The length of the longest timelike curves for three sprinklings into patches of $(1+1)d$ Schwarzschild spacetime of different size organized by the $r$-coordinate of the start of the chain, at minimal points (points with no past in the sprinkling). The lengths are given in terms of the discreteness scale $\ell_1 = \sqrt{n/V_1}$ of the causal set sprinkled into the smallest patch $V_1$, whereas the $r$-coordinates are in terms of the Schwarzschild radius. Lower panel: The cardinality of the future of minimal points 
organized by their $r$-coordinate.  The cardinalities are given in terms of the average sprinkling size $n$, whereas the $r$-coordinates are given in terms of the Schwarzschild radius. The data points for both plots were collected in 400 sprinklings of average size $n = 10^3$.
}
\end{figure}

Assuming that we had at our disposal an infinite causal set that is a sprinkling into, e.g., Schwarzschild spacetime, there is another difficulty, namely that one can identify all points which have no future, but one cannot straightforwardly distinguish those (futureless) points at future null infinity $\mathcal{J}^+$ from those futureless points which lie close to the singularity. Therefore, finding a partition of the causal set into the black-hole interior and the black-hole exterior is not straightforward, even given an infinite sprinkling. 
In~\cite{Barton:2019okw}, this point was addressed by defining a \textit{causal horizon} $\mathcal{H}_c$ as the boundary of the causal past of a future-inextendible curve $\gamma_0$ of infinite proper length
\begin{align}
\mathcal{H}_c = \partial \operatorname{Past}(\gamma_0) \,.
\end{align}
Selecting curves of infinite proper length is exactly what differentiates those curves that end at $\mathcal{J}^+$ from those that end at $r=0$.

We expand upon this idea for a causal set of finite size, because this is all that is available to us in practice. In a given sprinkling into Schwarzschild spacetime, we identify all minimal elements, i.e., those elements without a past. We calculate the length of the longest chain that starts at each of those minimal elements. For minimal elements inside the horizon, the length of the longest chain is limited, because each timelike curve inside the horizon must reach $r=0$ within a finite amount of proper time. In contrast, the longest chains starting at minimal elements outside the horizon are in practice only limited by the overall size of the causal set.
In the upper panel of Fig.~\ref{fig:averagechainlength}, we observe a sharp transition between the length of the longest chain exactly at the location of the horizon. The jump becomes even more pronounced, when we increase the timelike extent of the sprinkling, i.e., the timelike extent of the box that we sprinkle into. While in Fig.~\ref{fig:averagechainlength} we sort the minimal elements by their radial coordinate, the same information could of course be extracted without reference to coordinates. Calculating the length of the longest chain starting at all minimal elements sprinkled into a box with large enough timelike extent is expected to produce a bimodal distribution according to which one can sort minimal elements into those inside and those outside the horizon.

We note that instead of the length of the longest chain, we can also calculate the cardinality of the future of each minimal element, cf.~lower panel in Fig.~\ref{fig:averagechainlength}. Again, the distribution is bimodal. We expect, however, that this distribution may be affected more strongly by the choice of boundary. For instance, if we sprinkle into a causal diamond, the cardinality of the future of the minimal elements varies between $n$ and $\sqrt{n}$ already for Minkowski spacetime.

However, such a definition would no longer work for regular black holes, because in their case, we expect that timelike curves can also be continued for arbitrarily long proper time inside the horizon. Thus, a regular black-hole spacetime, e.g., of the Hayward type, when considered in $3+1$ dimensions, likely does not allow a partition of the corresponding causal set through these diagnostics.

\section{Apparent horizon and geodesic focusing}\label{sec:ladders}
In practical calculations with finite regions of black-hole spacetimes, e.g., in numerical relativity, event horizons are useless. In addition, in time-dependent settings, they largely lose their physical meaning \cite{Ashtekar:2004cn,Booth:2005qc}. Instead, a concept that we will now focus on as well is that of an apparent horizon. We expect that it will also continue to make sense in sprinklings into regular black-hole spacetimes, where the previous method of partitioning a causal set  into interior and exterior elements is likely to fail.

The apparent horizon, contrary to the event horizon, is defined locally and is, technically speaking, a marginally outer trapped surface \cite{Poisson:2009pwt}. 
On a trapped surface, the expansions of
both congruences of future-directed null geodesics (ingoing and outgoing ones) are negative everywhere. 
The geodesic expansion is
\begin{align} \label{eq:expansion1}
\Theta = \nabla_\alpha k^\alpha \,,
\end{align}
where $k^\alpha$ is the tangent vector to the affinely parameterized null geodesic. If we 
consider all past and future trapped surfaces, we obtain a trapped region. The boundary of this four-dimensional region is called trapping horizon. 

\begin{figure}[!t]
  \centering
\includegraphics[width=0.7\linewidth]{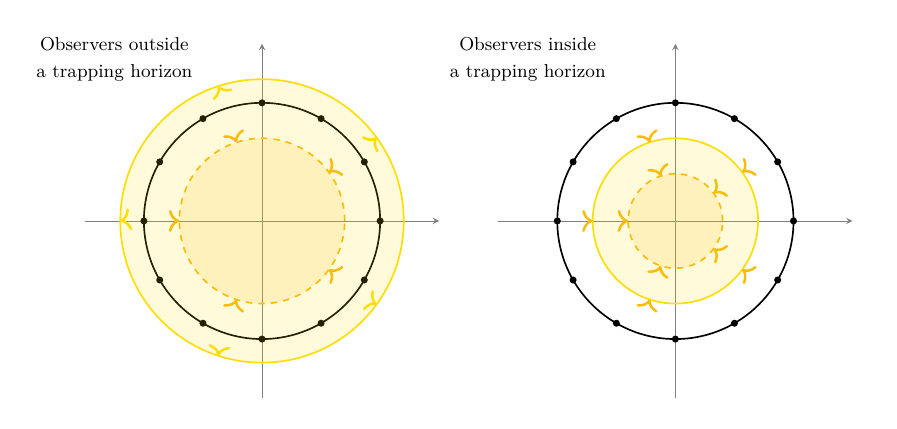}
  \caption{Illustration of a trapped surface. In the left image, there is a family of observers along a spacelike closed curve (black line) not inside an apparent horizon. They send out outgoing and ingoing light rays. The outgoing light rays form a closed spacelike curve of larger circumference after some time. In contrast, if the observers sit inside an apparent horizon, both ingoing and outgoing light rays trace out closed spacelike curves of smaller circumference after some time, as in the right image.}
  \label{fig:apparent-horizon}
\end{figure}

An illustration of a trapped surface is given in Fig.~\ref{fig:apparent-horizon} in terms of ingoing and outgoing light rays emitted from points lying on a closed spacelike curve.

This is the motivation for the alternative definition of geodesic expansion 
\begin{align} \label{eq:expansion2}
    \Theta = \frac{1}{A} \frac{\mathrm d A}{\mathrm d\lambda} \,,
\end{align}
as the logarithmic change in the cross-sectional area $A$ of the congruence of null geodesics along the geodesic parameterized by $\lambda$ \cite{Poisson:2009pwt}.

One can explicitly compute the null geodesic expansion of both ingoing and outgoing congruences in Schwarzschild spacetime. The result is
\begin{align} \label{eq:diff-expansion-schwarzschild}
\Theta_\text{in}(r) = - \frac{2}{r} \,, \qquad \qquad  \, \Theta_\text{out}(r) = \frac{1}{r} \left(1 - \frac{2M}{r} \right) \,,
\end{align}
Thus, we find that the expansion of ingoing radial null geodesics is negative everywhere, whereas the expansion of outgoing radial null geodesics is positive in the region $r > 2M$ and negative in the region $r < 2M$. Harking back to our definition of an apparent horizon as a marginally outer trapped surface, we find that for Schwarzschild spacetime, the apparent horizon is characterized by the radial coordinate $r = 2M$, for which $\Theta_\text{out}(r) = 0$. For the Schwarzschild spacetime, this is identical to the location of the event (and Killing) horizon.

In the present setting, we do not have spatial two-surfaces available, because we are considering (1+1)-dimensional sprinklings. Therefore, we cannot compute the expansions, but instead we consider the one-dimensional spatial distance between neighboring geodesics along a sequence of spatial hypersurfaces at increasing values of the time coordinate. In Minkowski spacetime, we expect this distance to be constant in time if we use pairs of only ingoing or only outgoing null geodesics. In contrast, when there is geodesic focusing, we expect this distance to decrease with time. To calculate spatial distances, we make use of the construction from \cite{Eichhorn:2018doy}, in which spatial distances are calculated from causal information in a causal set. As a first step towards this, we must find the causal-set approximations of ingoing and outgoing null geodesics.

\subsection{Null geodesics via ladders}

A proposal to approximate continuum null geodesics in a causal set which is a sprinkling into (1+1)-dimensional spacetime was put forward in \cite{Bhattacharya:2023xnj}. It is based on so-called \emph{ladders}.
We review the definition of ladders and then introduce a notion of geodesic expansion tailored to causal ladders. Later, we will modify this definition and introduce the concept of a fuzzy ladder, which on average leads to longer ladders than the rigid-ladder definition we provide now. 

Ladders can be built iteratively, starting from a first \emph{rung}, which is given by a link.
Given such a link $p_1 \link q_1$, we search for another link $(p_2, q_2)$ (the second rung) that satisfies the properties 
\begin{align}
p_1 \link p_2 \,, \qquad q_1 \link q_2 \,, \qquad |[p_1, q_2]| = 2\,.
\end{align}
The motivation for these choices is to \enquote{stack} a second link on top of $(p_1, q_1)$ in such a way that there are no other points in the causal interval between $p_1$ and $q_2$, i.e., the intersection of the future of $p_1$ with the past of $q_2$ is empty. An immediate consequence of the third property is $p_1 \prec q_2$, i.e., that $p_1$ and $q_2$ are causally related (but not linked). 

Due to its shape, we will refer to this construction forthwith as a \textbf{causal ladder}. This comes with related ladder jargon, such as \textbf{rung} to refer to the tuple $(p_i, q_i)$ and \textbf{side} to refer to the families $\{p_i\}_{i = 1, \ldots, k}$ and $\{q_j\}_{j = 1, \ldots, k} $. The causal ladder is formally defined as follows:

\begin{definition}[Causal ladder $L_k$] \label{def:causal-ladders}
Let $C$ be a causal set and $I \subsetneq \mathbb{N}$ an index set of cardinality $|I| = k$. A causal ladder, denoted $L_k$, of length $k$ is a set of tuples $\{(p_1, q_1), \ldots, (p_k, q_k)\}$ with $p_i, q_i \in C$ and $i \in I$, which satisfies the conditions
\begin{enumerate}
    \item $p_{i-1} \link p_i$, for all $i \in I$,
    \item $q_{i-1} \link q_i$, for all $i \in I$,
    \item $|[p_i, q_j]| = 2(j-i)$, for all $j \leq i \in I$
    \item $|[p_i, p_j]| = j-i-1$, for all $i \leq j \in I$,
    \item $|[q_i, q_j]| = j-i-1$, for all $i \leq j \in I$.
\end{enumerate}
\end{definition}

\begin{figure}[!t]
    \centering
\includegraphics[width=0.7\linewidth,clip=true, trim=1cm 16cm 33cm 0cm]{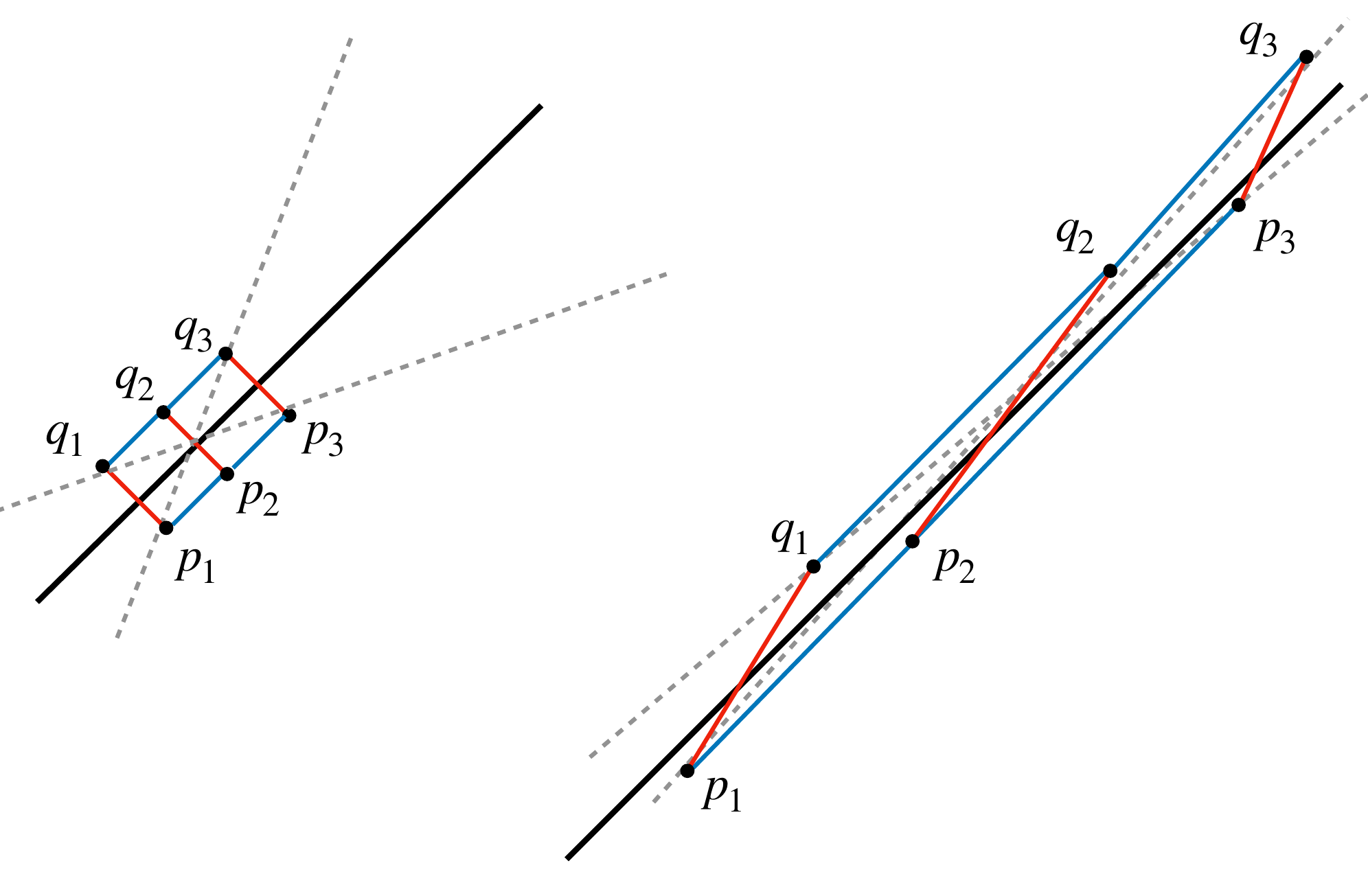}
    \caption{
    We compare an illustration of a ladder in which the rungs are lightlike (left side) to a ladder in which the rungs are clearly timelike (right side). The first is a ladder as one might imagine it based on the definition. However, because links have vanishing probability to be lightlike, such ladders do not typically occur in sprinklings. Instead, a typical ladder looks like the one on the right-hand-side. We note in passing that due to this property, fewer rungs are sufficient to strongly constrain the opening angle of all straight lines that intersect all rungs.}
    \label{fig:typical-ladder}
\end{figure}

The ladder approximates a null geodesic as follows, in the simplest example of Minkowski spacetime: Given the first rung, the opening angle of all possible straight lines that intersect the rung is large. One of these lines is a null geodesic, the others are timelike and spacelike lines. As successive rungs are added, the range of opening angles decreases, until, at sufficiently many rungs, the opening angle is constrained enough so that the ladder provides a good approximation to a null geodesic.

\begin{figure}[!t]
    \centering
    \includegraphics[width=0.9\linewidth]{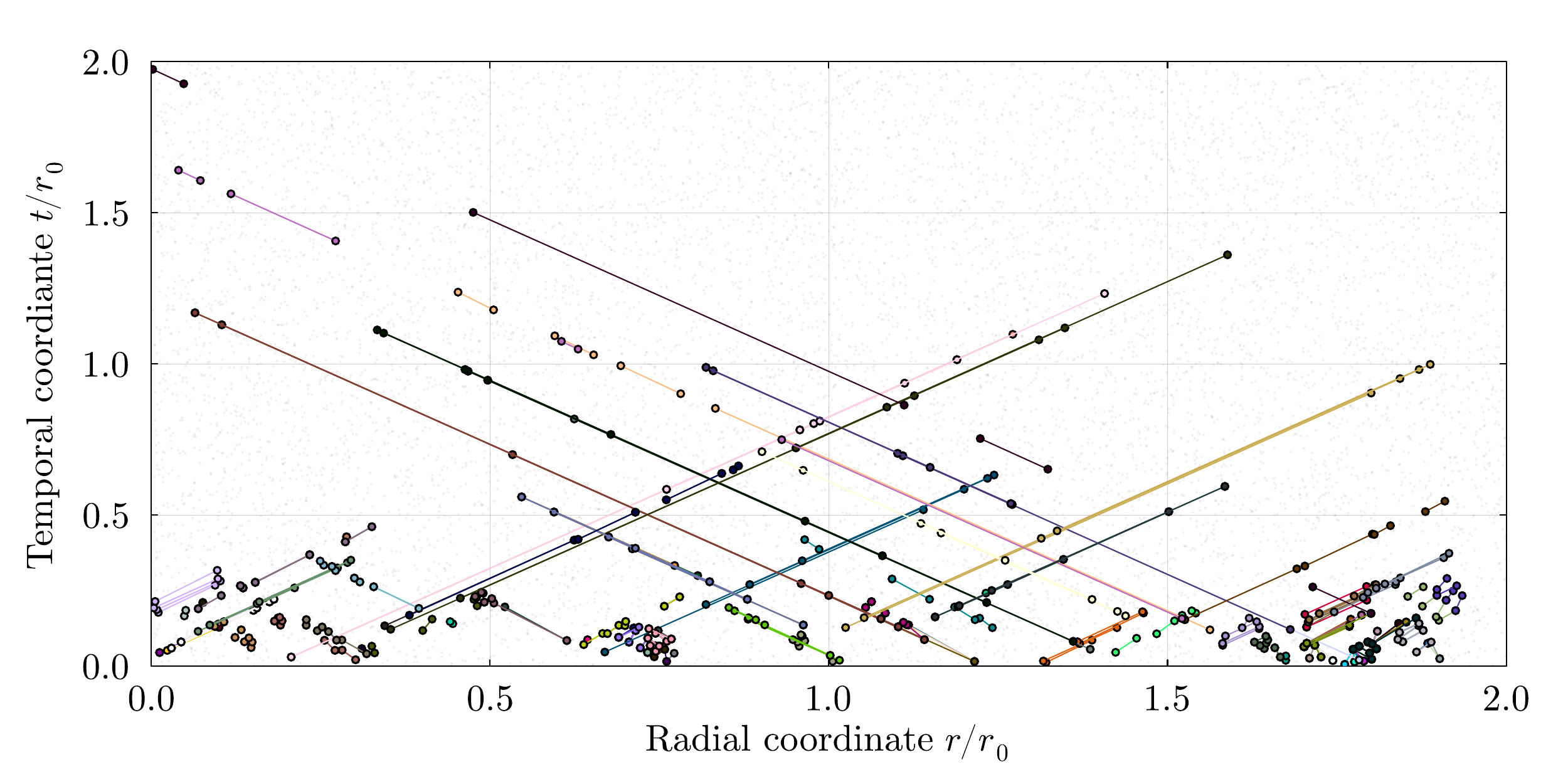}

    \centering
    \includegraphics[width=0.9\linewidth]{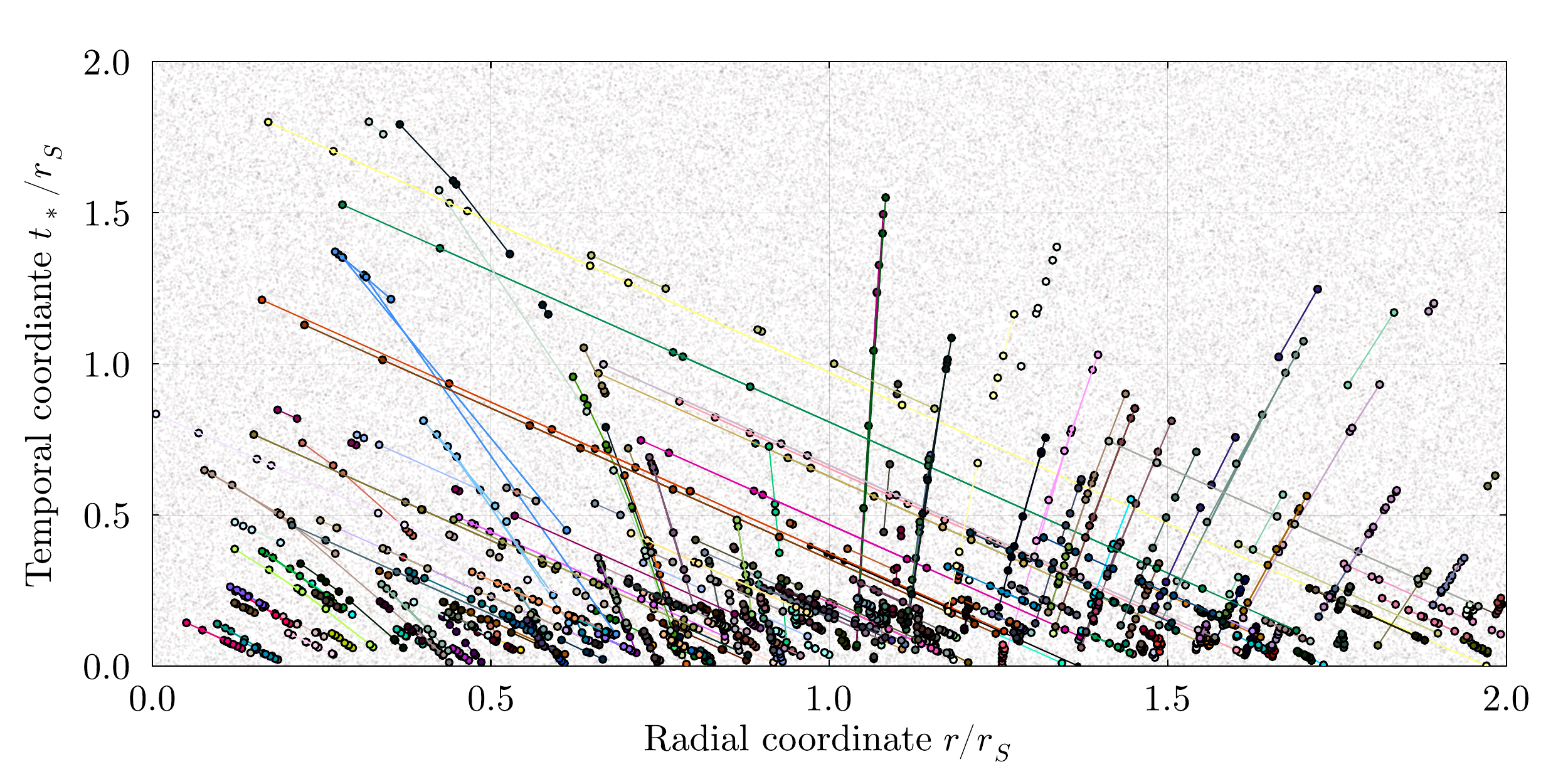}
\caption{\label{fig:rigid_ladders} We show ladders in a sprinkling of size $10^4$ into Minkowski spacetime (upper panel), where $r_0$ is some characteristic length scale, as well as ladders in a sprinkling of size $10^5$ into Schwarzschild spacetime (lower panel), where $r_S$ is the Schwarzschild radius. We show the rungs of the ladders, but do not include their sides to avoid making the figure too cluttered. Different ladders are distinguished by different colors.}
\end{figure}

From the definition of a ladder, and the above discussion, one may expect that the rungs of a ladder are generically oriented at approximately $90 \degree$ with respect to the null geodesic that is being approximated. Instead, we find, as already seen in \cite{Bhattacharya:2023xnj}, that rungs typically run nearly parallel to the null geodesic they are approximating, as illustrated in Fig.~\ref{fig:typical-ladder}.

In \cite{Bhattacharya:2023xnj}, the ladder definition was tested numerically in sprinklings into Minkowski spacetime. Here, we test it for the first time in a spacetime with nonvanishing curvature, namely the (1+1)-dimensional toy model of Schwarzschild spacetime. Numerically, we only search for the first rung in the bottom region of the causal set (in the range $t_{\ast}/r_S \in [0, 0.2]$ on average, such that the points were chosen among the bottom 1/10 of the causal set points), because we are interested in ladders that extend for as many rungs as possible. We are successful in identifying ladders, see Fig.~\ref{fig:rigid_ladders}, which clearly follow the null geodesics in the spacetime shown in Fig.~\ref{fig:geodesics-and-links-schwarzschild}.

\subsection{Towards geodesic focusing and geodesic expansion with ladders}
As a step towards detecting geodesic focusing and measuring geodesic expansion, we use pairs of causal ladders and compute how the spatial distance between them changes as a function of affine parameter, which we approximate by the label of the rung. 
Within the pairs of ladders we use, both ladders are either ingoing or outgoing. In flat spacetime, the spatial distance between such ladders stays constant as a function of time.

Within a sprinkling, we use the predistance function defined in \cite{Eichhorn:2018doy} to approximate the spatial distance between pairs of rungs on pairs of ladders.\\
We label ladders with capital Latin indices $I, J,...$ and rungs by small Latin indices $a, b,...$. Specifically, given the 
$a$th rung of ladder $I$ and the $b$th rung of ladder $J$, we denote $D_{IJ}^{ab}$ the mean of all four possible spatial distances that can be computed from the two rungs.\footnote{The case in which these distances are not spacelike is addressed below; here we assume that they are spacelike.}

\begin{figure}
    \centering
    \includegraphics[width=0.7\linewidth]{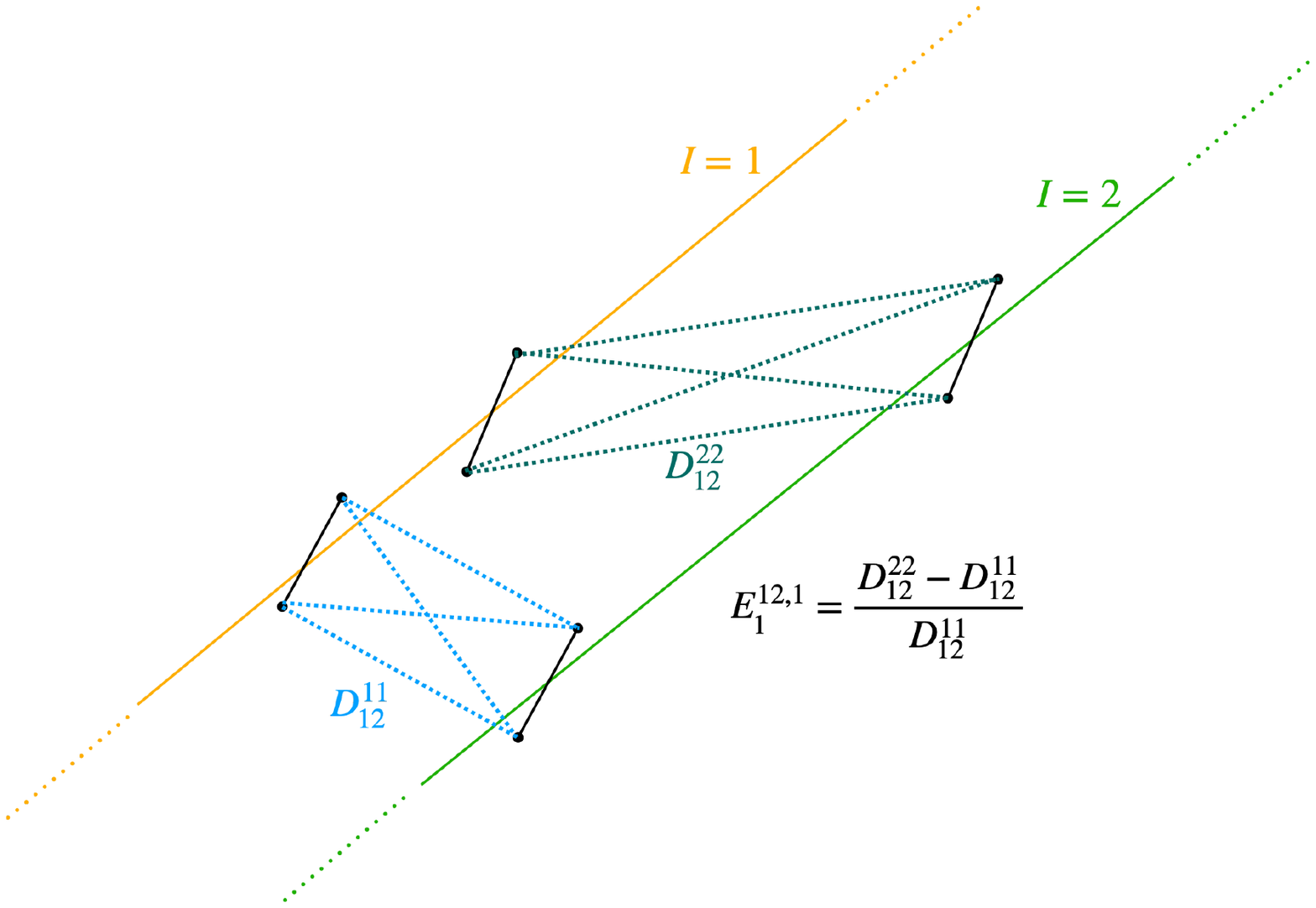}
    \caption{Illustration of our discrete expansion  given
    in Eq.~\ref{eq:discrete-expansion} for the example of a ladder with a single nearest neighbor. The colored solid lines represent the continuum null geodesic that the ladders (rungs in black) are approximating. The dotted lines illustrate the four (spatial) distances needed to compute $D_{IJ}^{ab}$. If any of these distances is timelike, as can happen due to the nature of the ladders, we discard the full $D_{IJ}^{ab}$ associated with that rung pair.}
    \label{fig:illustration-E}
\end{figure}

In a spacetime with nontrivial curvature, we expect that pairs of ingoing (or outgoing) null geodesics approach each other, if there is geodesic focusing, or are driven apart, if there is defocusing. To measure this, we define, analogously to the continuum expansion defined in Eq.~\ref{eq:expansion2}, the quantity
\begin{equation} \label{eq:discrete-expansion}
E_{I}^{ab,n} = \frac{1}{N_J} \sum_{J} \frac{D_{IJ}^{(a+n)(b+n)} - D_{IJ}^{ab}}{D^{ab}_{IJ}},
\end{equation}
where the sum is taken over a set of ladders which each start within a given spatial radius of the first rung of a fixed reference ladder, of which we are computing the expansion at rung $I$, and $N_J = \sum_J$, the total number of ladders we are comparing the reference ladder to. The computation of $E_{I}^{ab,n}$ is illustrated with an example in Fig.~\ref{fig:illustration-E}. To obtain a measurement that is most similar with the continuum expansion $\Theta$ in Eq.~\ref{eq:expansion2}, we additionally disregard any $D_{IJ}^{ab}$ between timelike separated rungs. This can happen in our context due to the rungs not being lined up consistently, see e.g., Fig.~\ref{fig:rigid_ladders}. 

In the following, we will compute the average discrete geodesic expansion over all ladders $I$, over all rung pairs $a, b$ and over all separations $n$. We will denote this by $\operatorname{mean}(E)$. We will also average over many sprinklings to form the expectation value of this quantity over sprinklings. This addresses statistical fluctuations within an individual  sprinkling. The full averaged discrete expansion is denoted $\overline{\operatorname{mean}(E)}$ to emphasize the averaging both over ladder pairs within a sprinkling, as well as over  many sprinklings. Before analyzing these quantities, we first make clear what is the exact continuum analogue to the discrete expansion defined in Eq.~\ref{eq:discrete-expansion}.

\subsubsection{The continuum quantity}

The induced spatial distance on a constant $t_*$ hypersurface for $r>r_S$ in a Schwarzschild spacetime is given by the spatial line element
\begin{equation}
ds^2 = \left(1-\frac{r_S}{r}\right)^{-1} dr^2 = \left(1-\frac{r_S}{r(r_*)}\right) \mathrm dr_*^2.
\end{equation}
The continuum analogue to the above discrete expansion in Eq.~\ref{eq:discrete-expansion} is
\begin{align}
\widetilde{\Theta}_{\text{ext}}(t, \delta) = \frac{f(t + \delta) - f(t)}{f(t)} \qquad \text{with} \qquad f(t) = \int_{r_{*1}(t)}^{r_{*2}(t)} \mathrm dr_* \sqrt{1 - \frac{r_S}{r(r_*)}} \,,
\end{align}
where $\delta$ is introduced to represent the shift between rungs and is the analogue to the parameter $n$ in Eq.\ \ref{eq:discrete-expansion}. The switch to tortoise coordinates is useful because $r_*(t+\delta) = r_*(t) + \delta$, as $r_* = \text{const.} \pm t$. Using this, we can rewrite the continuum analogue
\begin{align}
\widetilde{\Theta}_{\text{ext}}(t, \delta) &= \frac{\int_{r_{*1}(t) + \delta}^{r_{*2}(t) + \delta} \mathrm dr_* \sqrt{1 - \frac{r_S}{r(r_*)}}  - \int_{r_{*1}(t)}^{r_{*2}(t)} \mathrm dr_* \sqrt{1 - \frac{r_S}{r(r_*)}} }{\int_{r_{*1}(t)}^{r_{*2}(t)} \mathrm dr_* \sqrt{1 - \frac{r_S}{r(r_*)}} } \,.
\end{align}
In App.~\ref{app:continuumE}, we show that this function satisfies the limits
\begin{align} \label{eq:cont-expansion-ext}
\lim_{t\to+\infty} \widetilde{\Theta}_{\text{ext}}(t, \delta)  = 0 \,, \qquad \lim_{t\to-\infty} \widetilde{\Theta}_{\text{ext}}(t, \delta) = e^{\delta/2r_S} - 1 \,.
\end{align}
Additionally, one can convince oneself that $f(t+\delta) > f(t)$, due to the positivity of the integrand, and thus, that $\widetilde{\Theta}_{\text{ext}}(t, \delta)$ is a strictly positive function.
\\

We can perform an analogous computation for the interior region. The induced spatial distance on a spatial hypersurface, i.e., a constant-$r$ hypersurface for $r<r_S$ in a Schwarzschild spacetime is given by the spatial line element
\begin{equation}
ds^2 = -\left(1-\frac{r_S}{r}\right)dt^2,
\end{equation}
where we remember that for $r<r_S$, the time-coordinate and spatial coordinate flip their role, and surfaces of constant $r$ become spacelike. The analogue to the discrete expansion in this region is 
\begin{align}
\widetilde{\Theta}_{\text{int}}(r, \delta) = \frac{g(r + \delta) - g(r)}{g(r)} \qquad \text{with} \qquad g(r) = \int_{t_1(r)}^{t_2(r)} \mathrm dt \, \sqrt{\frac{r_S}{r} - 1}  \,,
\end{align}
where $t(r)$ is the expression for the outgoing null geodesics given in Eq.~\ref{eq:schwarzschild_geods}. We find 
\begin{align}
\widetilde{\Theta}_{\text{int}}(r, \delta) &= \frac{\int_{t_1(r)}^{t_2(r)} \mathrm dt \, \sqrt{\frac{r_S}{r+\delta} - 1} - \int_{t_1(r)}^{t_2(r)} \mathrm dt \, \sqrt{\frac{r_S}{r} - 1}}{\int_{t_1(r)}^{t_2(r)} \mathrm dt \, \sqrt{\frac{r_S}{r} - 1}} = \frac{\sqrt{\frac{r_S}{r+\delta} - 1} - \sqrt{\frac{r_S}{r} - 1}}{\sqrt{\frac{r_S}{r} - 1}}\,.
\end{align}
In App.~\ref{app:continuumE}, it is shown that this function satisfies the properties
\begin{align} \label{eq:cont-expansion-int}
\lim_{r\to 0} \widetilde{\Theta}_{\text{int}}(r, \delta) = -1 \,, \qquad \widetilde{\Theta}_{\text{int}}(r, \delta) < 0 \,.
\end{align}
In particular, it is a strictly negative function. \\
Finally, note that a strictly positive $\widetilde{\Theta}_{\text{ext}}(t, \delta)$ and a strictly negative $\widetilde{\Theta}_{\text{int}}(r, \delta)$ is expected from what we know of null geodesic expansion in the interior and exterior regions of the Schwarzschild spacetime.\\
In the continuum, we can of course calculate $\widetilde{\Theta}_{\rm int/ext}$ as a function of $r$ (or $t$) and $\delta$ precisely. In causal sets, we do not expect that this dependence is  exactly
reproduced. There are several reasons for this: first, to calculate the discrete analogue of $\widetilde{\Theta}_{\rm int/ext}$, i.e., $\overline{{\rm mean}(E)}$ at different locations $r$ or $t$ we would need a huge number of sprinklings, in order for the average over sprinklings to converge at each location. Thus, we will average over different locations within as well as outside of the horizon.
Second, we expect significant systematic uncertainties, because, as is evident from the illustration in Fig.~\ref{fig:illustration-E}, pairs of rungs on neighboring ladders are not located at equidistant locations along the ladders, such that the spatial distance between ladder pairs can increase strongly even for sprinklings in Minkowski spacetime. Therefore, we expect to at best be able to reproduce the \emph{signs} of $\widetilde{\Theta}_{\rm int}$ and $\widetilde{\Theta}_{\rm ext}$. This is what we attempt to achieve below.

\subsubsection{Discrete expansion in Minkowski spacetime}
We first calculate $\operatorname{mean}(E)$ in sprinklings into Minkowski spacetime. In the continuum, we expect $E=0$, i.e., pairs of outgoing (or ingoing) null geodesics remain parallel to each other in Minkowski spacetime. In a sprinkling, our illustration in Fig.~\ref{fig:illustration-E} already highlights that, because rung pairs do not occur equidistantly along neighboring ladders, the spatial distance between rung pairs can vary, and in particular can increase significantly from one rung pair to the next. We expect that this creates a systematic bias of $\overline{\operatorname{mean}(E)}$, which is ultimately a discreteness effect.

\begin{figure}
    \centering
    \includegraphics[width=\linewidth]{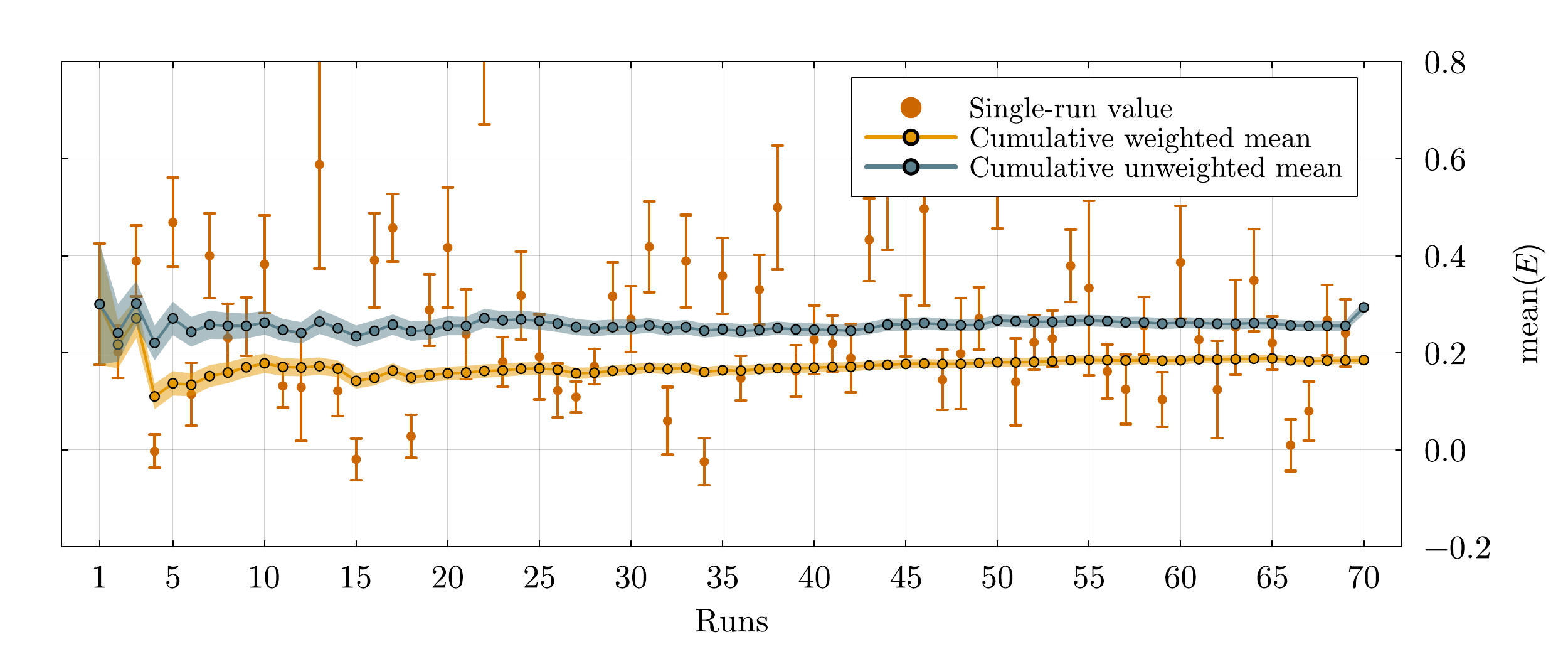}
    \caption{The weighted and unweighted average of $\operatorname{mean}(E)$ among 70 sprinklings into $(1+1)$-dimensional Minkowski spacetime, where the nearest-neighbors were chosen in the interval $[5, 20] \ell$, cf.~App.~\ref{app:robustness}. Additionally, the individual $\operatorname{mean}(E)$ were plotted to illustrate their relatively large fluctuation. The final weighted average is $\overline{\operatorname{mean}(E)}= 0.185 \pm 0.007$.}
    \label{fig:minkowski-expansion-rigid}
\end{figure}

To calculate $\overline{\operatorname{mean}(E)}$, we have to make a number of technical choices. In the spirit of brevity, we will concentrate here on showing the results of our analysis of $E$ in sprinklings into Minkowski spacetime, discussing the details of these technical choices in Appendix~\ref{app:robustness}. 

We find (cf.~Fig.~\ref{fig:minkowski-expansion-rigid}) that the averaged value for $\operatorname{mean}(E)$ over 70 sprinklings is 
\begin{align} \label{eq:expansion-minkowski-rigid}
\overline{\operatorname{mean}(E)}_{\text{Minkowski}}= 0.185 \pm 0.007 \,.
\end{align}
To obtain this average, we weighted each run by 
\begin{align}
w_i = \frac{N_i}{\sigma_i^2}\,,
\end{align}
where, in the $i$th run, $N_i$ is the number of ladders that are being summed over, and $\sigma_i$ the standard deviation of the measurement. The introduction of this weight is tantamount to a higher trust afforded to those measurements with many data points and low standard deviation. To appreciate the effect this weighting has on the final result, the naive ``one bucket'' average is also shown in Fig.~\ref{fig:minkowski-expansion-rigid}. The weighted averages will therefore be used in the following analysis, since it reduces the bias yielding values closer to the expected Minkowski result, namely zero. \\
As we already anticipated, discreteness results in a systematic bias of $\overline{\operatorname{mean}(E)}$ to positive values. When calculating $\overline{\operatorname{mean}(E)}$ for sprinklings into curved spacetimes, we subtract the value obtained in Minkowski spacetime. 

\subsubsection{Discrete expansion in Schwarzschild spacetime}
We expect $\overline{\operatorname{mean}(E)}$ to change sign across the event horizon of a black hole. To test this hypothesis, and to discover whether $\overline{\operatorname{mean}(E)}$ constitutes a discrete observable capable of detecting the (approximate) location of the apparent horizon, we compute $\overline{\operatorname{mean}(E)}$ separately in sprinklings into the interior and sprinklings into the exterior of (1+1)-dimensional Schwarzschild spacetime.

\begin{figure}[!t]
    \centering
    \includegraphics[width=0.95\linewidth]{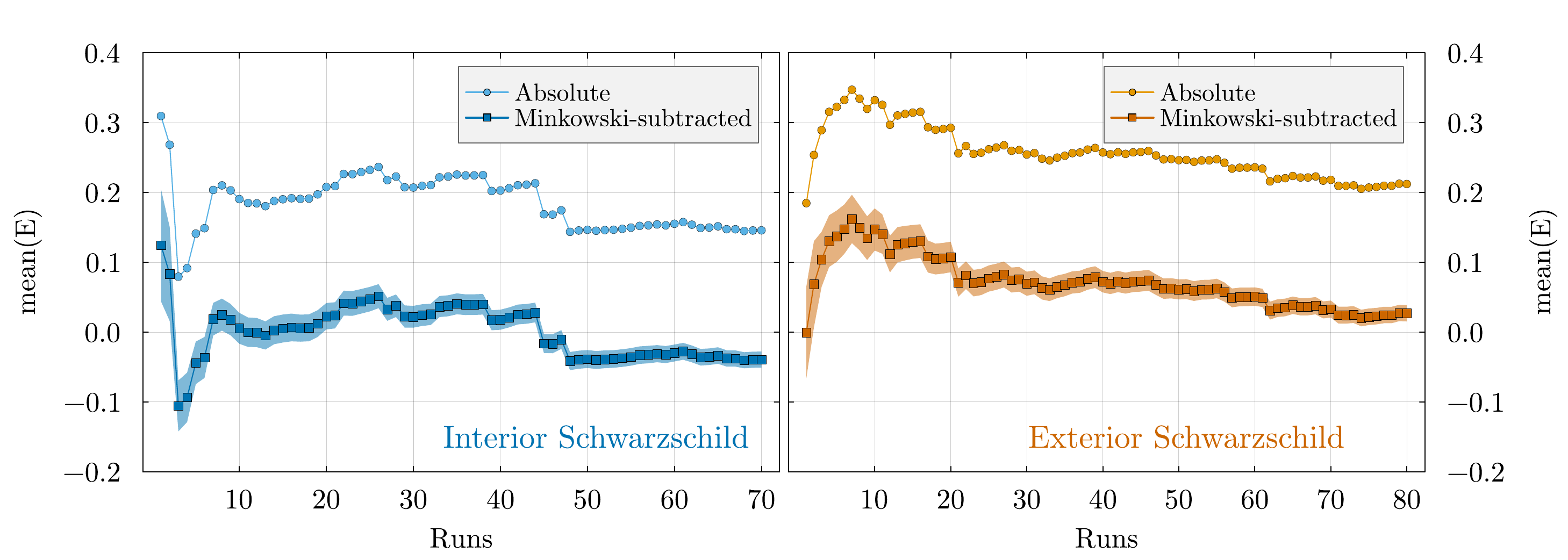}
    \caption{The dependence of $\overline{\operatorname{mean}(E)}$ on the number of sprinklings considered for a sprinkling into interior Schwarzschild spacetime (left) and exterior Schwarzschild spacetime (right). The final values are $\overline{\operatorname{mean}(E)}_{r<r_S} = -0.039 \pm 0.011$ and $\overline{\operatorname{mean}(E)}_{r>r_S} = 0.027 \pm 0.011$.}
    \label{fig:final-expansion}
\end{figure}

One of the main results of our paper is
Fig.~\ref{fig:final-expansion}. There, the absolute
and the Minkowski-subtracted result for $\overline{\operatorname{mean}(E)}$ for the regions interior and exterior to the horizon is shown.  We stress that we consider the latter the physically meaningful quantity, because spacetime discreteness introduces a bias into $\overline{\operatorname{mean}(E)}$ in Minkowski spacetime, and this non-zero baseline must be subtracted before results for sprinklings into other spacetimes can be interpreted.
We find a weighted average
\begin{align}
\overline{\operatorname{mean}(E)}_{r<r_S} = -0.039 \pm 0.011 \,, \qquad \qquad \overline{\operatorname{mean}(E)}_{r>r_S} = 0.027 \pm 0.011 \,,
\end{align}
and thus the expected negative expansion in the interior region, and a positive expansion in the exterior.

The large fluctuations between sprinklings make it necessary to consider a large number of sprinklings to obtain a converged value. From Fig.~\ref{fig:final-expansion}, we see that the mean value does not change much over the last few runs that have been added, and consider this a sign of convergence. More extensive tests of convergence could of course be performed with additional runs, but we do not consider them necessary here, because our target is merely the \emph{sign} of $\overline{\operatorname{mean}(E)}$ in the two regions.

\section{The discrete horizon}\label{sec:discretehorizon}

Finally, to find the subset in a causal set that approximates the horizon, we  combine
the various concepts we have introduced. First, we separate the interior from the exterior spacetime region in two ways: Either, we can use the discrete expansion, or we can use the length of timelike curves. The discrete expansion is applicable, irrespective of whether or not the black-hole spacetime is geodesically incomplete or not. However, the length of timelike curves is a more precise diagnostic for an individual sprinkling, at least in our main case of interest, the toy model of the Schwarzschild spacetime, and with the choice of box boundaries with large ratio of timelike to spacelike extent. Using the length of the timelike curves, we split a minimal antichain into two sets: one that lies in the black-hole interior and one in the exterior, cf.~Fig.~\ref{fig:averagechainlength}. Then, we identify the pair of elements that are closest to each other in spatial distance, but lie on opposite sides of the horizon. If there is a ladder that originates at one of the two elements and approximates an outgoing null geodesic, this null geodesic traces the black-hole horizon. The existence of such a ladder is of course not guaranteed, but occurs in a subset of sprinklings.

\begin{figure}[!t]
  \begin{center}
    \includegraphics[width=0.9\linewidth]{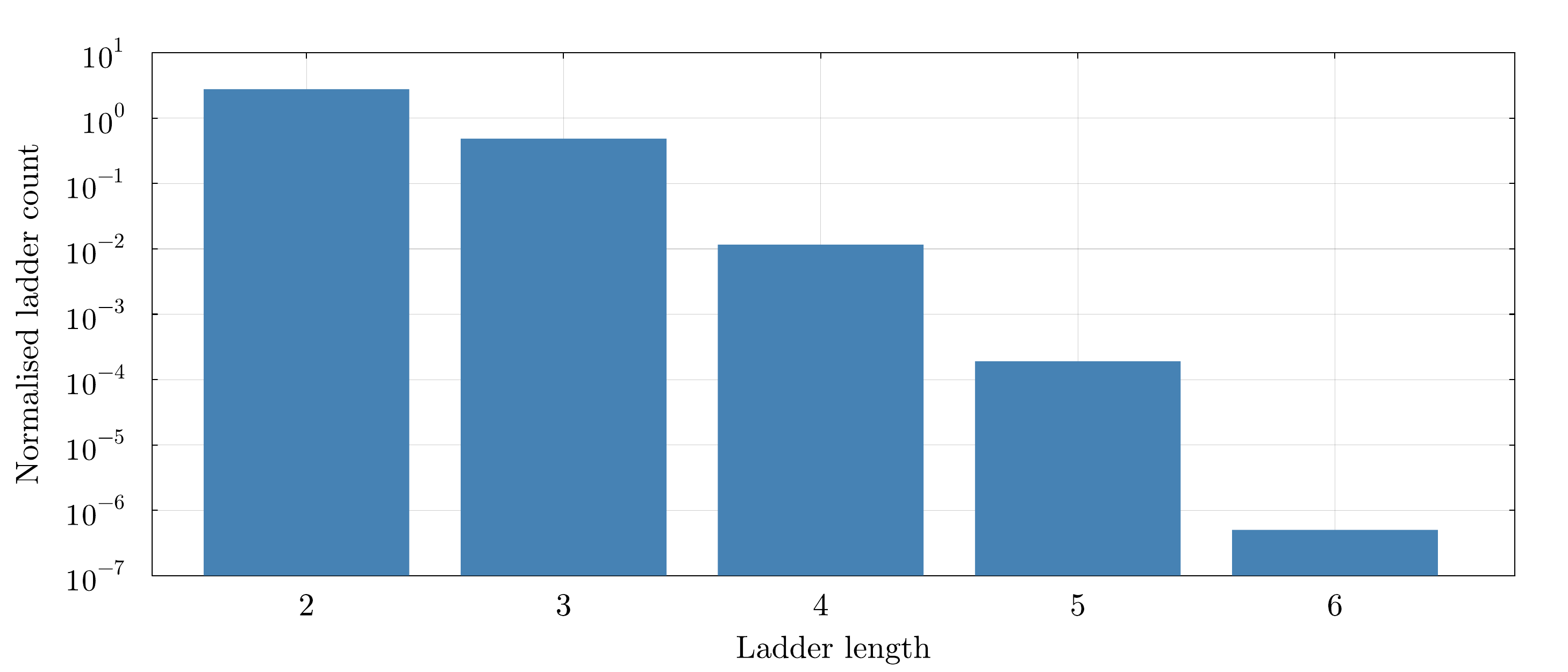}
    \end{center}
    \caption{\label{fig:histogram-mink-original} We show a histogram for the number of ladders of a given length (defined as the number of rungs) for a Minkowski sprinkling with $n=10^4$ points. For this, we have included $200$ sprinklings to ensure that our data is statistically converged, and normalized the ladder count by the total number of points sprinkled, i.e., in this case $2 \cdot 10^6$. The normalization is chosen to illustrate the rarity of longer ladders, as, e.g., only one ladder of length 6 is found for every $2 \cdot 10^6$ sprinkled points on average.}
\end{figure}

When attempting to realize this construction, we come across a main challenge, namely that ladders are typically rather short, i.e., consist only of few rungs. We can see this in Fig.~\ref{fig:rigid_ladders} already, but also highlight this in the histogram in Fig.~\ref{fig:histogram-mink-original}. There, we show the rarity of ladders with more than five rungs, i.e., the length of the ladder counts the number of rungs. Statistically, we only find a single ladder with six rungs in a sprinkling with $2 \cdot 10^6$ points. 

This motivates us to introduce a new definition, namely of a ``fuzzy'' ladder. A fuzzy ladder satisfies fewer rigidity constraints than the original ladder definition. This makes it statistically more likely to find ladders with more rungs. Given their increased lengths, these fuzzy ladders constitute better approximations of the black-hole horizon, because they can trace it for larger ranges of the affine parameter.

\subsection{Definition of fuzzy ladders} 
 The original causal ladder construction, see Def.~\ref{def:causal-ladders}, defines a rigid ladder. The conditions 3.-5.~in Def.~\ref{def:causal-ladders} are too constraining to allow long ladders. Thus, we investigate which conditions should be made less strict in order to increase the probability for a ladder to have significantly more rungs. Our definition below is based on numerical experiments with different modifications of the original ladder definition.

We keep one part of condition 3., namely the constraint that $p_i$ and $q_i$ share a link. We drop the remaining constraints on the size of the causal intervals between points on opposite sides of the ladders. In addition, we loosen the constraints on the causal intervals between points on the same side of the ladder, i.e., conditions 4.~and 5. Our new definition is the following:

\begin{definition}[``Fuzzy causal ladder'' $L_k^{(M)}$]
\label{def:fuzzy-causal-ladders}
Let $C$ be a causal set, 
$I \subsetneq \mathbb{N}$ an index set of cardinality $|I| = k$ and $M \in \mathbb{N}$.
A fuzzy causal ladder, denoted $L_k^{(M)}$, of length $k$, is a set of tuples $(p_1, q_1), \ldots, (p_k, q_k)\}$ with $p_i, q_i \in C$ and $i \in I$, which satisfies the conditions
\begin{enumerate}
    \item $p_{i-1} \link p_i$, for all $i \in I$,

    \item $q_{i-1} \link q_i$, for all $i \in I$,

    \item $p_i \link q_i$, for all $i \in I$,

    \item for all $i \in I$ with $i > M$, one has
    \begin{align}
        M - 1 \le |[p_{i-M}, p_i]| \le 2M - 1, \nonumber
    \end{align}

    \item for all $i \in I$ with $i > M$, one has
    \begin{align}
        M - 1 \le |[q_{i-M}, q_i]| \le 2M - 1. \nonumber
    \end{align}
\end{enumerate}
\end{definition}``Fuzzy ladders'' allow controlled violations of the rigidity conditions. Additional elements can  exist between the two sides of the ladder, because the intervals between $p_i$ and $q_j$ with $j>i+1$ are no longer constrained.
The conditions 4.~and 5.~permit 
up to $M$ additional elements within the causal intervals between elements in a single side of a ladder. In addition, there are no restrictions on the causal intervals formed by the $p_i$ (or $q_i$) that are less than $M$ links apart.

\begin{figure}[!t]
\includegraphics[width=0.48\linewidth,clip=true, trim=7cm 4cm 17cm 0cm]{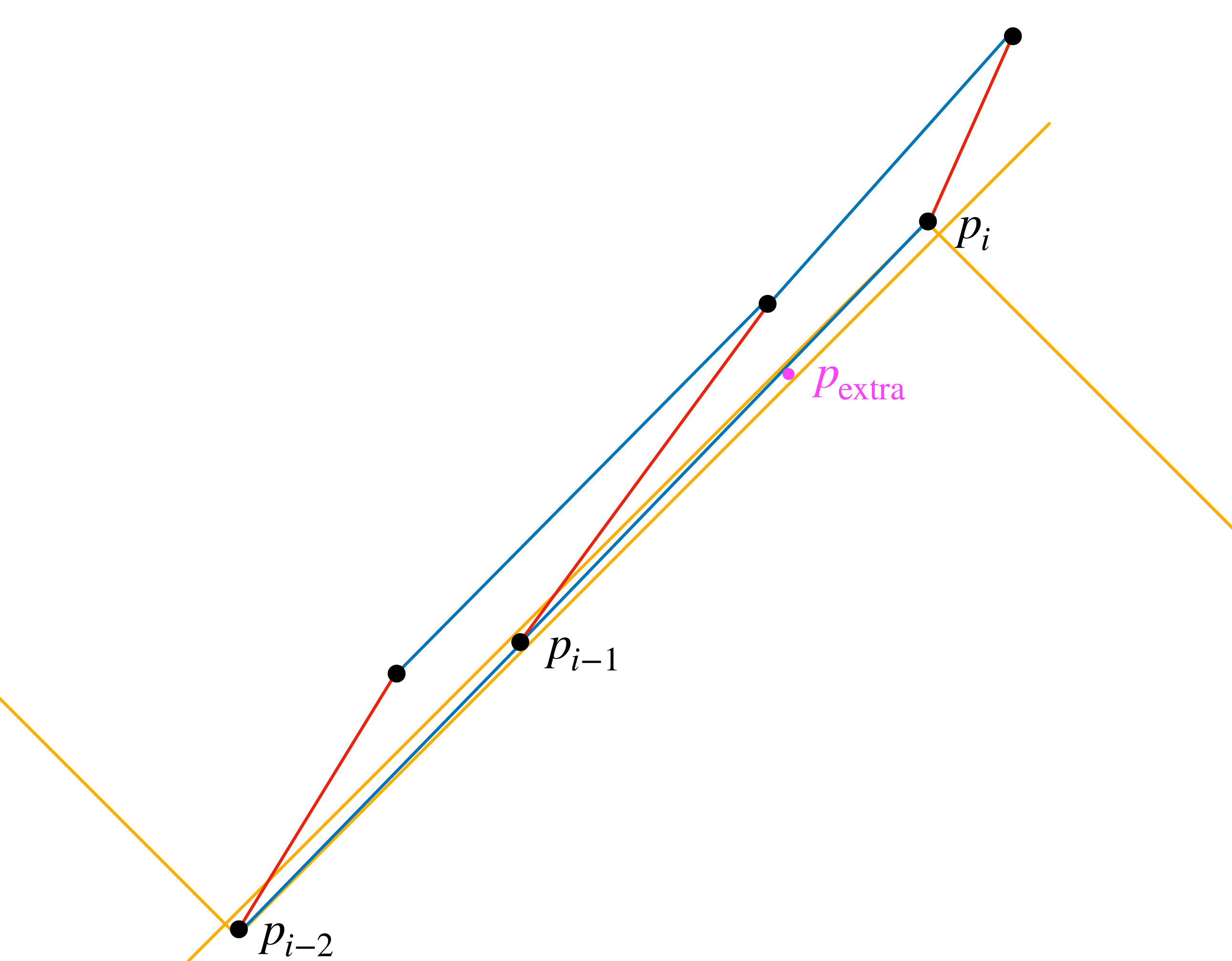}\quad
\includegraphics[width=0.48\linewidth,clip=true, trim=7cm 4cm 17cm 0cm]{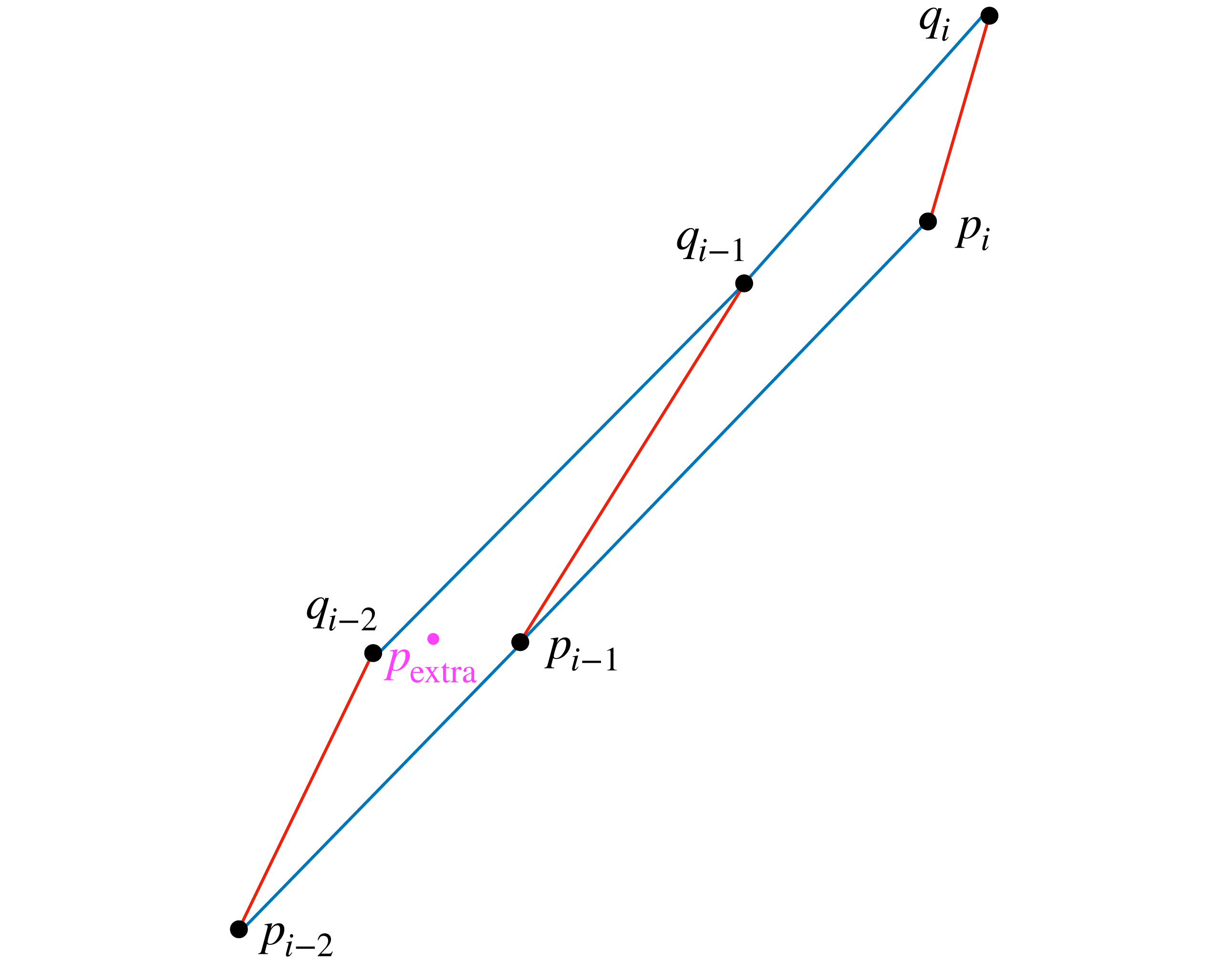}
\caption{\label{fig:fuzzyladderillustration} 
We show two illustrations of fuzzy ladders (with $M=2$), which do not satisfy the original, rigid ladder definition due to the magenta element $p_{\rm extra}$. In the left panel, it lies within the causal interval of $p_i$ and $p_{i-2}$, which is indicated with the help of the past lightcone of $p_i$ and future lightcone of $p_{i-2}$ (indicated by orange lines). In the right panel, it lies within the interval between $q_{i-1}$ and $p_{i-2}$.
}
\end{figure}

To illustrate the consequence of this definition, we show examples of fuzzy ladders in Fig.~\ref{fig:fuzzyladderillustration}, which do not satisfy the original, rigid ladder definition. We discuss further properties of fuzzy ladders in App.~\ref{app:fuzzy}, where we also test the robustness of our results for the discrete expansion with fuzzy ladders.

\begin{figure}
\includegraphics[width=\linewidth]{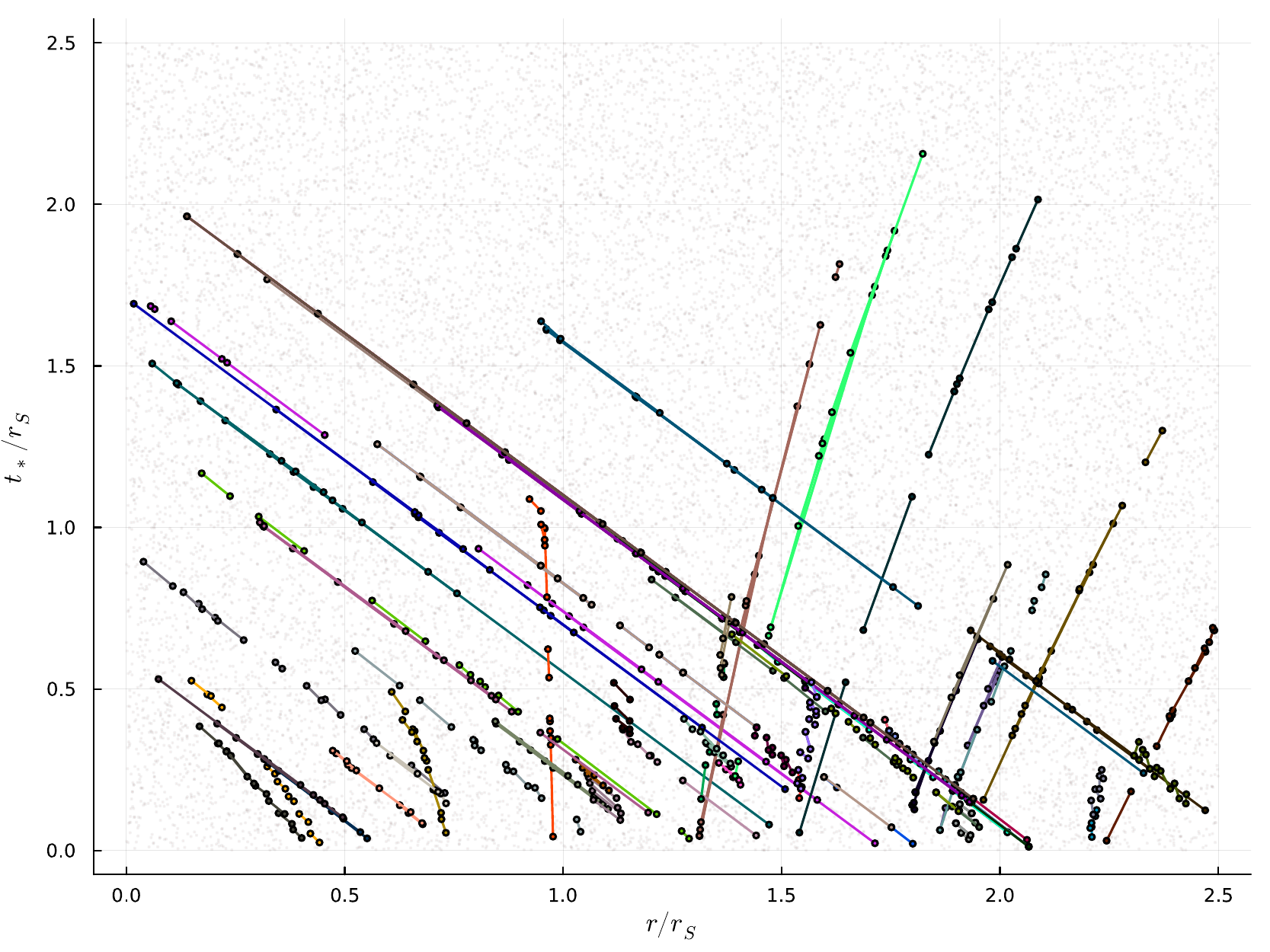}
\caption{\label{fig:fuzzyladdersSchwarzschild} We show an example of a sprinkling into $(1+1)$-dimensional Schwarzschild spacetime. The horizon is located at $r=r_S$. Ladders are highlighted in different colors, with each rung of a ladder shown by a continuous line. We do not show the sides of the ladders. Comparing to the lower panel in Fig.~\ref{fig:geodesics-and-links-schwarzschild}, we  see that ingoing null geodesics correspond to ladders with a slope of negative unit value;  outgoing ladders change from growing towards larger $r/r_S$ for $r/r_S>1$ to growing towards smaller $r/r_S$ for $r/r_S<1$. We even see an outgoing null geodesic (in orange) that is very close to the event horizon at $r=r_S$ and only ``peels off'' at $t_{\ast}/r_S\approx 1$.}
\end{figure}

We find that fuzzy ladders are much more likely to extend over significantly more rungs than rigid ladders. As an example, we show all ladders with length larger or equal than 8 in a sprinkling into Schwarzschild spacetime in Fig.~\ref{fig:fuzzyladdersSchwarzschild}. We highlight that with the previous, rigid ladder definition, already the probability of ladders of length 6 was essentially zero. Fig.~\ref{fig:fuzzyladdersSchwarzschild} also indicates that fuzzy ladders still trace null geodesics well, as one can see by comparing to Fig.~\ref{fig:geodesics-and-links-schwarzschild}. In particular, the property that rungs are often nearly parallel to the null geodesic itself, is preserved. Thus, the new, fuzzy ladders are also expected to very strongly constrain the opening angle of the various lines that can be drawn which intersect all rungs.

\begin{figure}[!t]
\includegraphics[width=\linewidth]{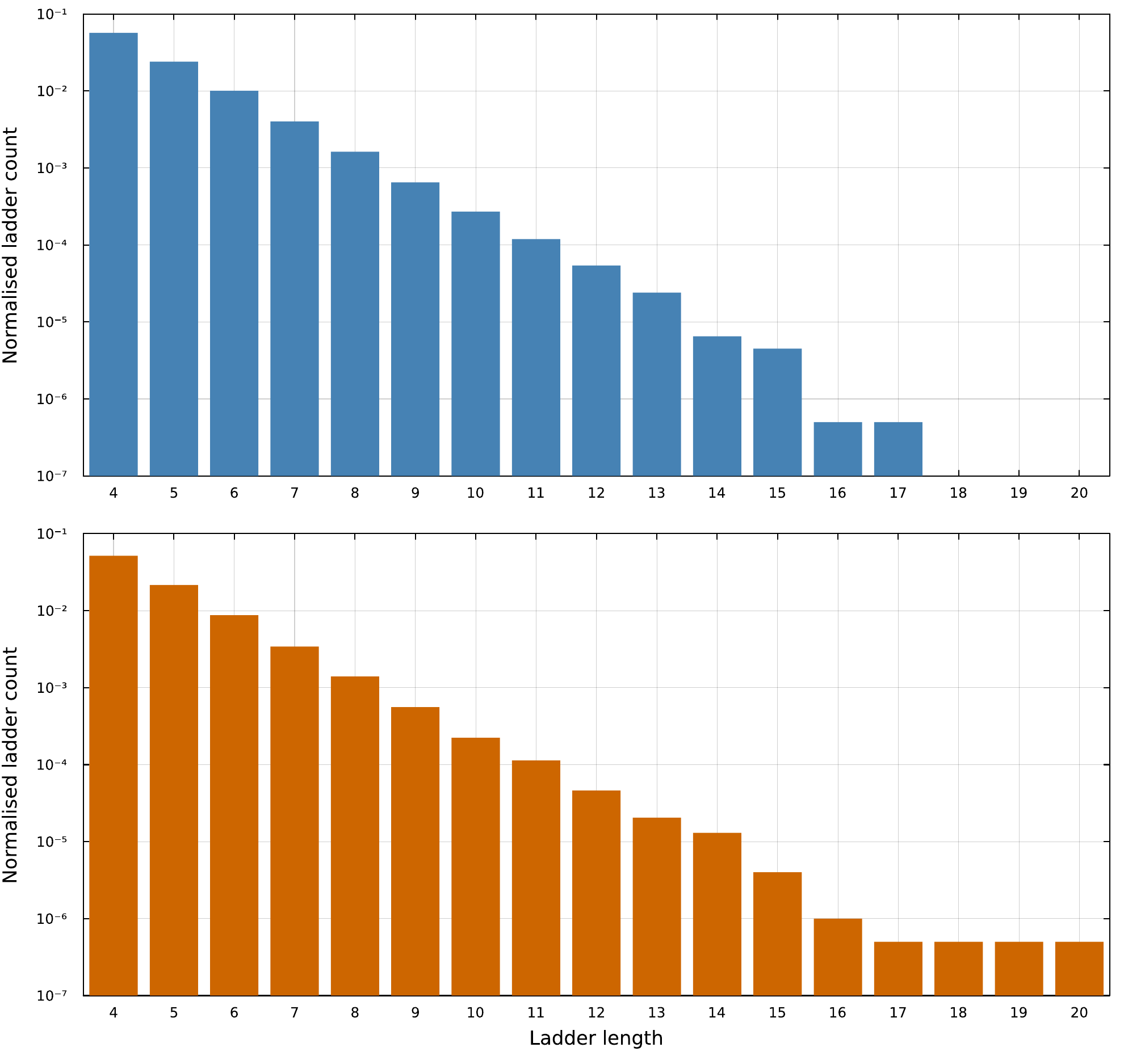}
\caption{\label{fig:fuzzyladderhistogram} 
We show a histogram for the number of fuzzy ladders, with $M=3$, of length $L\geq4$ for Minkowski sprinklings with $n=10^4$ points in the box $t,r\in[0,1]$ (upper panel) and Schwarzschild sprinklings with $n=10^4$ points in the box $t_*,r\in[0,1]$ with $r_S=1$ (lower panel). For each histogram, we have included $200$ sprinklings for each spacetime. The histograms are normalized by the total number of sprinkled points, $200\cdot10^4$. Since we find on average roughly $10^3$ ladders with $L\geq4$ per sprinkling for Minkowski and $9\cdot10^2$ for Minkowski, the length-$8$ bin is still well sampled: there are on average about $16$ length-8 ladders per Minkowski sprinkling and $14$ length-$8$ ladders per Schwarzschild sprinkling.
}
\end{figure}

To highlight that Fig.~\ref{fig:fuzzyladdersSchwarzschild} is a typical example, we provide a histogram showing the typical number of fuzzy ladders of length $L$ in Fig.~\ref{fig:fuzzyladderhistogram}. We conclude from this that our new, fuzzy ladder definition achieves our goal of increasing the absolute count of longer ladders. 

\subsection{Approximating the discrete horizon}
The fuzzy-ladder definition enables us to  approximate the discrete horizon. To that end, we follow the idea outlined above, and distinguish elements along a minimal antichain by whether they are inside or outside the horizon using the length of the longest timelike chain originating in each element. We then identify a fuzzy ladder that originates close to the boundary between the two subsets. 
 To identify this boundary, we could in principle use the causal set distance function, but would have to address the challenge of asymptotic silence \cite{Eichhorn:2018doy} at small spatial distances. Instead, to provide a proof-of-principle that a discrete horizon can be identified, we use the embedding information to identify the origin of an appropriate ladder by simply selecting a ladder that starts very close to $r=r_S$ and that constitutes an outgoing (not ingoing) null geodesics.

\begin{figure}[!t]
    \centering
    \includegraphics[width=0.48\linewidth]{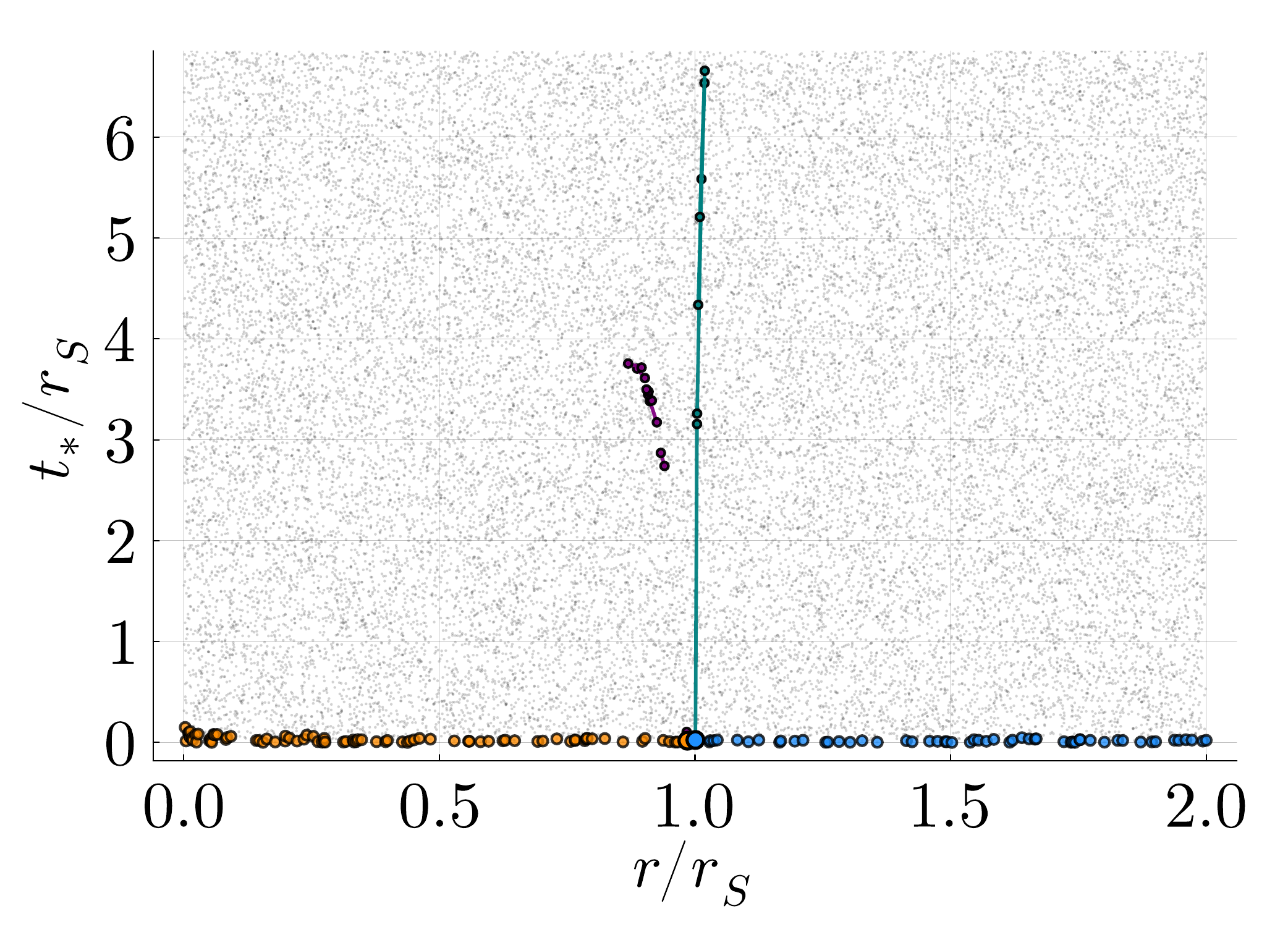}\quad
    \includegraphics[width=0.48\linewidth]{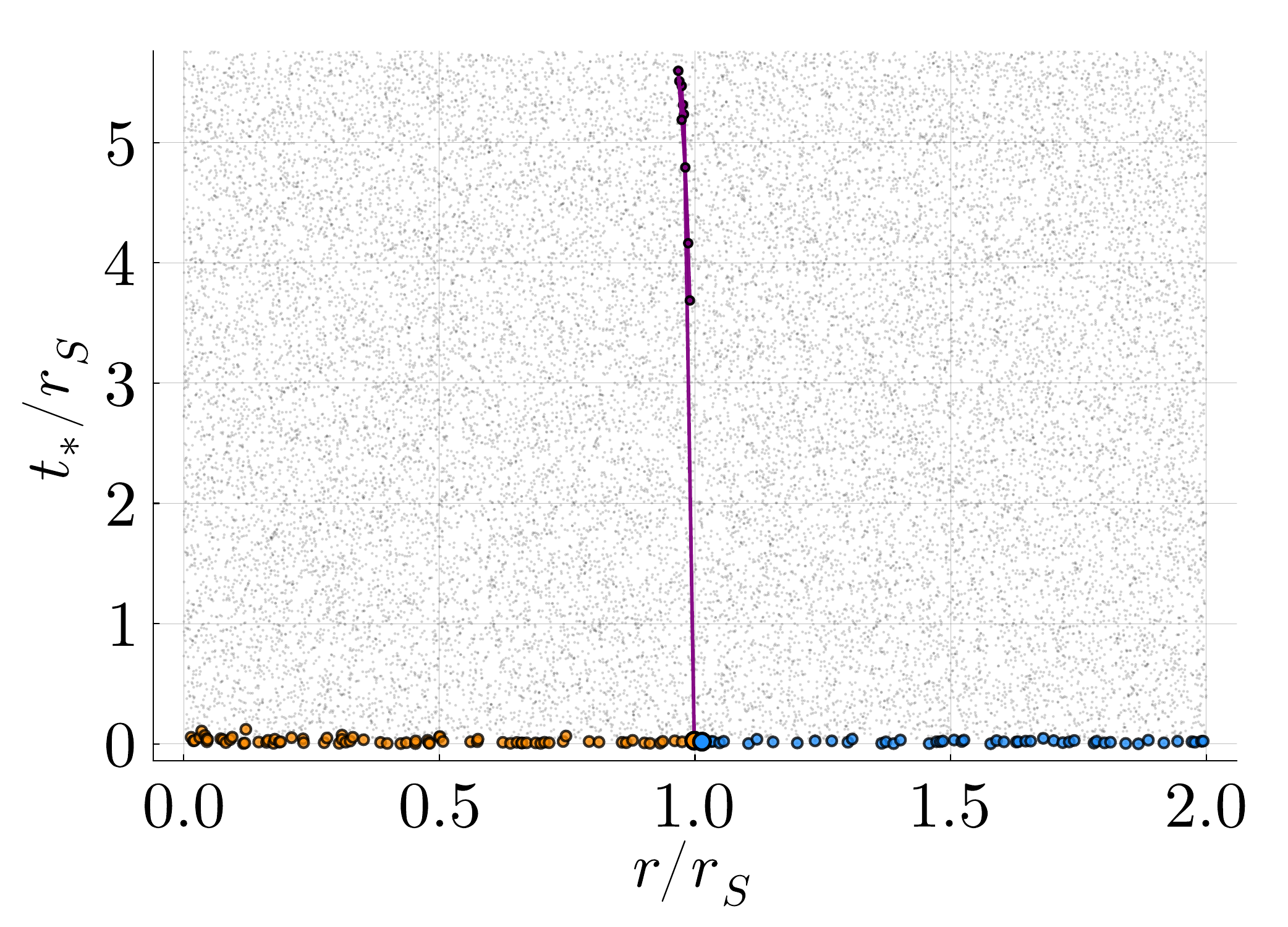}\\
    \includegraphics[width=0.48\linewidth]{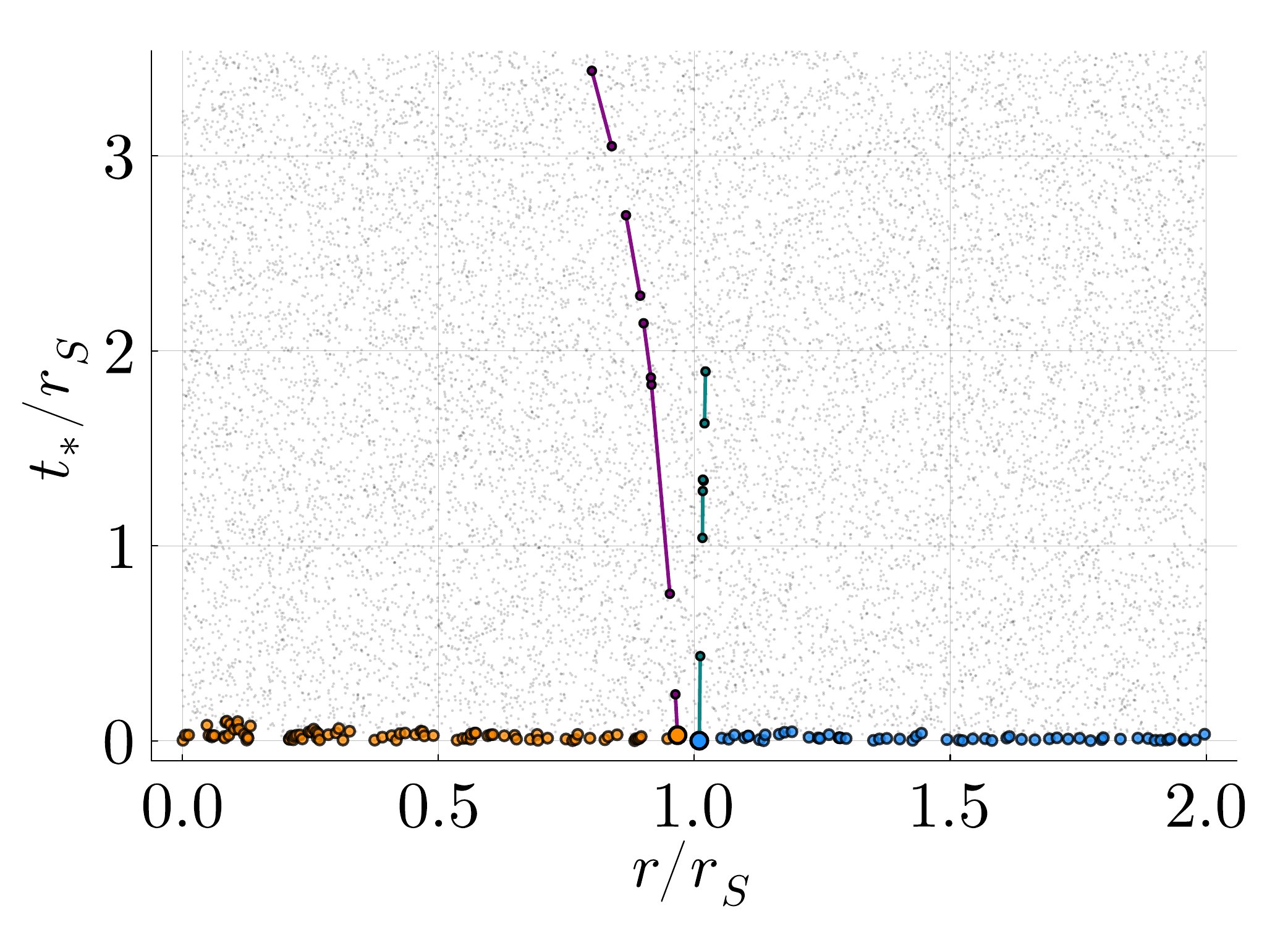}\quad
    \includegraphics[width=0.48\linewidth]{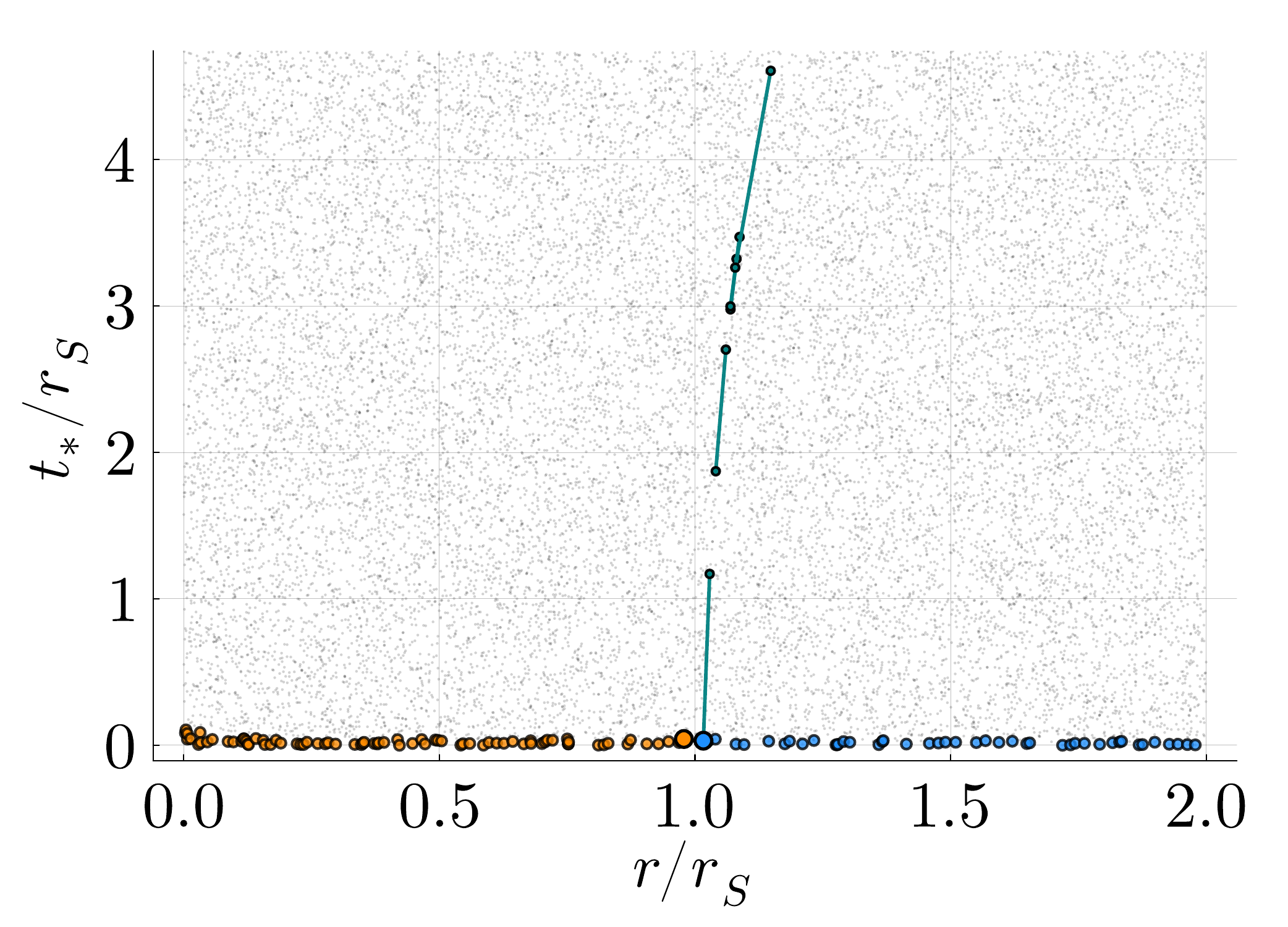}
    \caption{\label{fig:discretehorizon}
    Examples of fuzzy ladders approximating the horizon in sprinklings into $(1+1)$-dimensional Schwarzschild spacetime in a box of $t_*/r_S \in [0,50]$ and $r/r_S \in [0,2]$ with $2\cdot10^5$ points each, ensuring hat the longest-chain diagnostic is working properly  to distinguish horizon-interior from horizon-exterior elements. The minimal antichain is colored according to whether the longest chain starting at a given point is relatively long or short, cf.~Fig.~\ref{fig:averagechainlength}. This separates the minimal antichain into horizon-interior elements (orange) and horizon-exterior elements (blue).
    We only show the lower parts of the sprinkling, where the relevant ladders are located. Purple ladders start on the outermost orange point, teal ladders start on the innermost exterior point. The first two examples are the best-case scenarios found for a ladder approximating the horizon. Bottom left shows an example of both sides having a ladder that approximates at different precisions and extends, and bottom right the most common type of ladders found (excluding sprinklings whose ladders starting on the points of the minimal-antichain were too small or non-existent).}
\end{figure}

As a matter of course, not every sprinkling into a black-hole spacetime contains such a ladder, but we can easily find sprinklings which do. Examples are shown in Fig.~\ref{fig:discretehorizon}. There are two effects visible in these examples: First, even with the fuzzy ladder definition the typical length of ladders is limited. The examples shown are selected from generating $\mathcal{O}(10)$ sprinklings; for a significantly longer ladder approximating the horizon, we would have to generate a significantly larger number of sprinklings. Second, we see that while the ladders originate very close to $r=r_S$, they ``peel off'' from the horizon after a few rungs. This is expected, because the horizon location is an unstable point in the dynamics, i.e., in the continuum, any infinitesimal initial distance $\delta = (r-r_S)$ of an outgoing null geodesic from the horizon grows as a function of affine parameter. In the discrete setting, this results in the ladders ``peeling off'' from $r=r_S$ after a few rungs.  This is in line with our general expectation that spacetime discreteness causes there to be no approximation of the horizon without such effects.\footnote{This is similar to the idea that particles undergo ``swerves'', i.e., a diffusion process in momentum space, when they propagate on a discrete causal set \cite{Philpott:2008vd}. The ladders similarly deviate from following continuum null geodesics perfectly.}
To obtain a discrete horizon that extends over a larger range in $t_{\ast}$, we propose that our construction could be iteratively repeated, by identifying another antichain to the future of the first one and using it as a starting point of new timelike curves as well as another ladder. In this way, a piecewise discrete horizon could be constructed as the union of the ladders identified in this way. We leave such an iterative procedure to future work.

Overall, our results constitute a proof-of-principle that a discrete approximation of a black-hole horizon can be identified in a sprinkling.

\section{Conclusions and outlook}\label{sec:conclusions}

Black-hole horizons are some of the most tantalizing concepts in fundamental physics. They epitomize the extreme consequences that the dynamical nature of causal structure can have in GR. Similarly, in quantum gravity, they are linked to some of the deepest mysteries, such as the interplay between quantum spacetime and information. They also inform conjectures with far-reaching consequences, such as the holographic principle, as well as the no-global-symmetry conjecture. 

Therefore, understanding the nature of black-hole horizons in quantum gravity is an important goal. In this paper, we have approached this goal from the perspective of causal-set quantum gravity. Although built upon the combination of causal structure with spacetime discreteness, the black-hole horizon as a causal boundary is not easy to access in a causal set that is approximated by a black-hole spacetime.\\
Therefore, we have combined three diagnostics to find the discrete counterpart to a black-hole horizon in the continuum. \\
The first diagnostic splits a particular subset of causal-set elements into two subsets that lie within the interior and exterior region. 
This diagnostic is applicable to black-hole spacetimes which are geodesically incomplete. In a sprinkling into such a spacetime, we identify a minimal antichain, i.e., a set of elements with no relations between them, none of which has a past within the sprinkling. For each such element, we compute the longest timelike chain starting at this element. For elements inside the black-hole horizon, the maximum amount of time that can pass before the timelike chain terminates, is relatively short. In contrast, for elements outside the black-hole horizon, timelike chains can be continued up to the boundary of the sprinkled region. This leads to a bimodal distribution for the length of longest chain starting at the elements of the antichain. This allows us to distinguish which elements of the antichain lie inside and which lie outside of the horizon.\\
As a drawback of this diagnostic, it is not applicable to geodesically complete black holes. Ultracompact objects in nature, if they indeed are black holes, are expected to be geodesically complete, because geodesic incompleteness is usually viewed as a sign of the breakdown of GR.\\
The second diagnostic consists in discrete counterparts to null geodesics, so-called ladders, first introduced in \cite{Bhattacharya:2023xnj}. We study them for the first time in sprinklings into black-hole spacetimes and find that they approximate null geodesics well. We therefore use pairs of ladders to calculate a discrete counterpart to the geodesic expansion, namely the spatial distance between neighboring null geodesics as a function of the affine parameter along the null geodesic. We can reproduce the expected signs for this quantity in the regions exterior and interior to the black-hole horizon. This sets us up for a detection of the boundary between the two regions --- the apparent horizon --- in black-hole spacetimes irrespective of whether they are singular or regular.\\
As a drawback of this diagnostic, it requires numerous sprinklings into a given continuum spacetime to converge, because the number of ladders is not sufficiently dense within a given sprinkling.\\
Our third diagnostic is a new definition of ladders, which we call \emph{fuzzy ladders}. In this definition of a ladder, some of the rigidity constraints of the original definition \cite{Bhattacharya:2023xnj} are loosened. The resulting structure has a somewhat larger variability, but still traces null geodesics very well. As a central advantage, the typical length of a fuzzy ladder, i.e., the number of rungs that make it up, is significantly longer than for rigid ladders. 
Because the horizon itself is a null hypersurface (and thus a null geodesic in a 1+1 dimensional setting), it is approximated by one particular (fuzzy) ladder. In combination with our first diagnostic, we can find a fuzzy ladder which originates at the boundary between the interior and exterior region along the minimal antichain. Examples are shown in Fig.~\ref{fig:discretehorizon}. These fuzzy ladders approximate outgoing null geodesics. In practice, we typically find ladders which stay close to the horizon initially, but then ``peel off'' and move into the interior. This is to be expected, because the horizon is a marginally unstable surface --- the angle of an outgoing geodesic has to be fine-tuned exactly for this geodesic to stay on the horizon forever. The probability for a corresponding ladder to exist in the causal set is zero due to spacetime discreteness.

In summary, this collection of diagnostics therefore puts us in a position to identify black-hole horizons in causal sets in geodesically incomplete as well as geodesically complete spacetimes. 
We expect that such a practical identification will enable us to investigate some of the questions surrounding black-hole horizons mentioned in the introductory Sec.~\ref{sec:intro}, including questions linked to (entanglement) entropy, as have already been studied for cosmological horizons in \cite{Surya:2020gjj}.

As a very first step towards black-hole horizons in causal set quantum gravity, there are of course several aspects in which our work can be improved in the future.\\
First, it is an obvious goal to generalize to 3+1 dimensions. A main challenge to solve is to find an appropriate generalization of ladders, which only approximate null geodesics in 1+1 dimensions. A tubular ``binding of ladders'' appears to be an obvious concept to explore, but may be a statistically rare subset of a causal set.\\
Second, when computing the discrete expansion, we have identified pairs of ladders that are both out- (or in-)going by using embedding information, rather than using a causal-set-intrinsic way of removing pairs in which one ladder is in- and the other outgoing. In principle, one can use crossings of ladders, first discussed in \cite{Bhattacharya:2023xnj}, to achieve this.
A crossing of two ladders must necessarily  consist of one ingoing and one outgoing ladder. If there are two subsequent crossings along a single ladder, the two ladders that make up the other ``arms'' of the crossings are both of the same type, i.e., both ingoing or both outgoing. Again, a practical hurdle to using this concept is likely the statistical rarity of such subsequent crossings.\newline

Based on our work, understanding the properties of discrete black-hole horizons may be achievable in the future and deepen our understanding of quantum gravity and its interplay with information.\newline\\

 \textbf{\textit{Acknowledgments.}}
 We gratefully acknowledge helpful discussions with Sumati Surya.
 We acknowledge the European Research Council's (ERC) support under the European Union’s Horizon 2020 research and innovation program Grant agreement No.~101170215 (ProbeQG).
This work is supported by the Deutsche Forschungsgemeinschaft (DFG, German Research Foundation) under Germany’s Excellence Strategy EXC 2181/1 - 390900948 (the Heidelberg STRUCTURES Excellence Cluster).
 
\appendix
\setcounter{equation}{0}
\renewcommand{\theequation}{A\arabic{equation}}

\section{Calculation of the continuum analogue of $E$}
\label{app:continuumE}

We supply further details pertaining to the discrete expansion $E$ introduced in the main text. There, we provided a continuum analogue $\widetilde{\Theta}$ to our construction, and made some statements regarding its sign. We will derive these results now, starting with the analogue defined for the exterior region.

\subsection{Exterior region of the black hole}

\begin{figure}[b!]
    \centering
    \includegraphics[width=\linewidth]{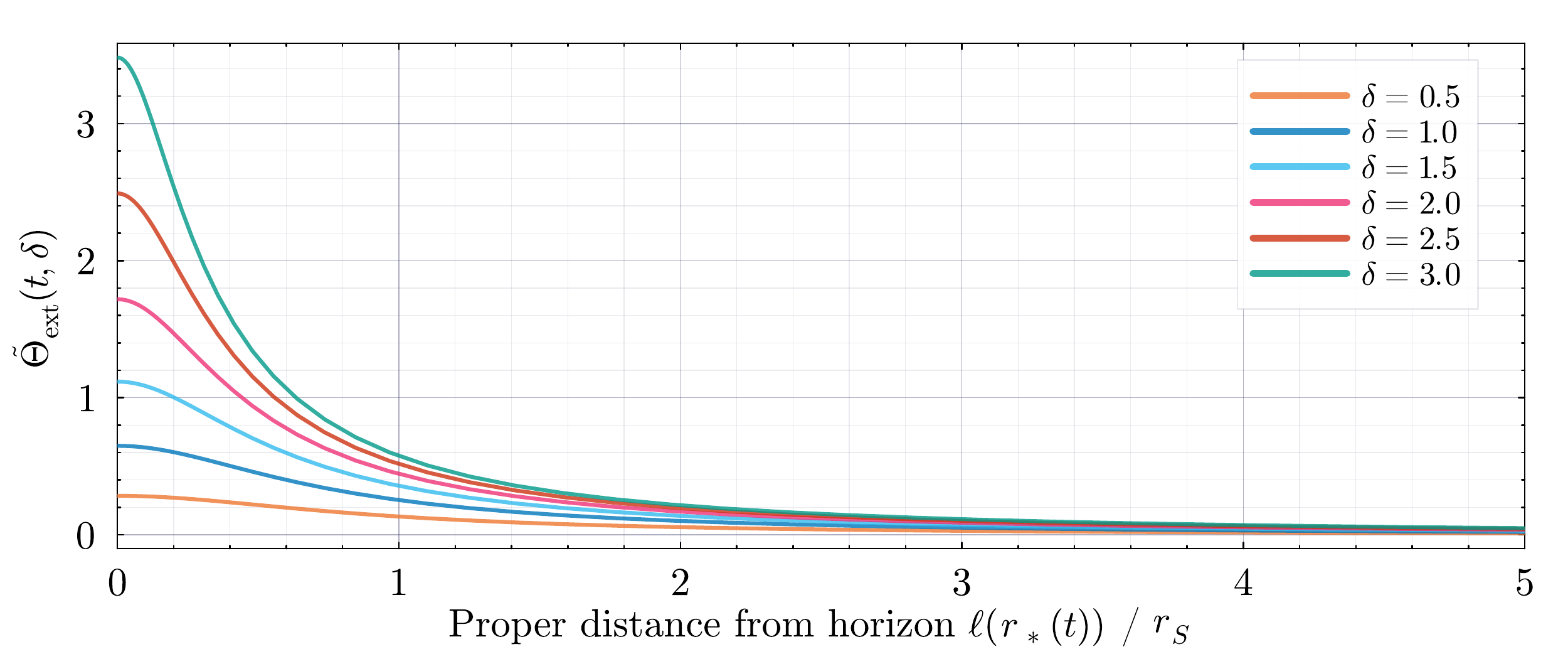}
    \caption{Examples of the exterior analogue of our discrete expansion measurement $\widetilde{\Theta}_{\text{ext}}(t, \delta)$ for various values of the parameter $\delta$, introduced to model the rung separation of the discrete case. The most important
    feature is the positivity of the function.}
    \label{fig:theta-out-illstr}
\end{figure}

In the main text, we defined the continuum analogue of the discrete measurement $E$ in the exterior region of the black hole by
\begin{align}
\widetilde{\Theta}_{\text{ext}}(t, \delta) = \frac{f(t + \delta) - f(t)}{f(t)} \qquad \text{with} \qquad f(t) = \int_{r_{*1}(t)}^{r_{*2}(t)} \mathrm dr_* \sqrt{1 - \frac{r_S}{r(r_*)}} \,.
\end{align}
Inserting the function $f$, we had written
\begin{align}
\widetilde{\Theta}_{\text{ext}}(t, \delta) &= \frac{\int_{r_{*1} + \delta}^{r_{*2} + \delta} \mathrm dr_* \sqrt{1 - \frac{r_S}{r(r_*)}}  - \int_{r_{*1}}^{r_{*2}} \mathrm dr_* \sqrt{1 - \frac{r_S}{r(r_*)}} }{\int_{r_{*1}}^{r_{*2}} \mathrm dr_* \sqrt{1 - \frac{r_S}{r(r_*)}} } \,.
\end{align}
The integral boundaries can be simplified to write
\begin{align}
\widetilde{\Theta}_{\text{ext}}(t, \delta) &= \frac{\int_{r_{*2}}^{r_{*2} + \delta} \mathrm dr_* \sqrt{1 - \frac{r_S}{r(r_*)}}  - \int_{r_{*1}}^{r_{*1}+\delta} \mathrm dr_* \sqrt{1 - \frac{r_S}{r(r_*)}} }{\int_{r_{*1}}^{r_{*2}} \mathrm dr_* \sqrt{1 - \frac{r_S}{r(r_*)}} } \,.
\end{align}
Expressing this function analytically is involved, so we only show the full results in Fig.~\ref{fig:theta-out-illstr} and only discuss the asymptotic behavior explicitly.
We consider the asymptotic behavior in the two limiting cases $t \to \pm \infty$, which correspond to the limits $r \to \pm \infty$. We find
\begin{align}
\lim_{t\to+\infty} \widetilde{\Theta}_{\text{ext}}(t, \delta)  = \lim_{r_* \to +\infty} \widetilde{\Theta}_{\text{ext}}(t(r_*), \delta) = \frac{\delta - \delta}{1} = 0\,,
\end{align}
where we use that $\sqrt{1 - r_S/r} \to 1$ in this limit. For the other limit, we find
\begin{align}
\lim_{t\to-\infty} \widetilde{\Theta}_{\text{ext}}(t, \delta) &= \lim_{r_* \to -\infty} \widetilde{\Theta}_{\text{ext}}(t(r_*), \delta) \\
&= \lim_{r_* \to -\infty} \frac{\int_{r_{*2}}^{r_{*2} + \delta} \mathrm dr_* e^{r_*/2r_S}   - \int_{r_{*1}}^{r_{*1}+\delta} \mathrm dr_* e^{r_*/2r_S}  }{\int_{r_{*1}}^{r_{*2}} \mathrm dr_* e^{r_*/2r_S}} \\
&= \lim_{r_* \to -\infty} \left(e^{\delta/2r_S} - 1 \right) \\
&= e^{\delta/2r_S} - 1 \,.
\end{align}
We used the exact relation
\begin{align}
e^{(r_* - r)/ r_S} = \frac{r}{r_S} - 1 \,,
\end{align}
in the limit $r \to r_S$.

\subsection{Interior region of the black hole}

\begin{figure}[!t]
    \centering
    \includegraphics[width=\linewidth]{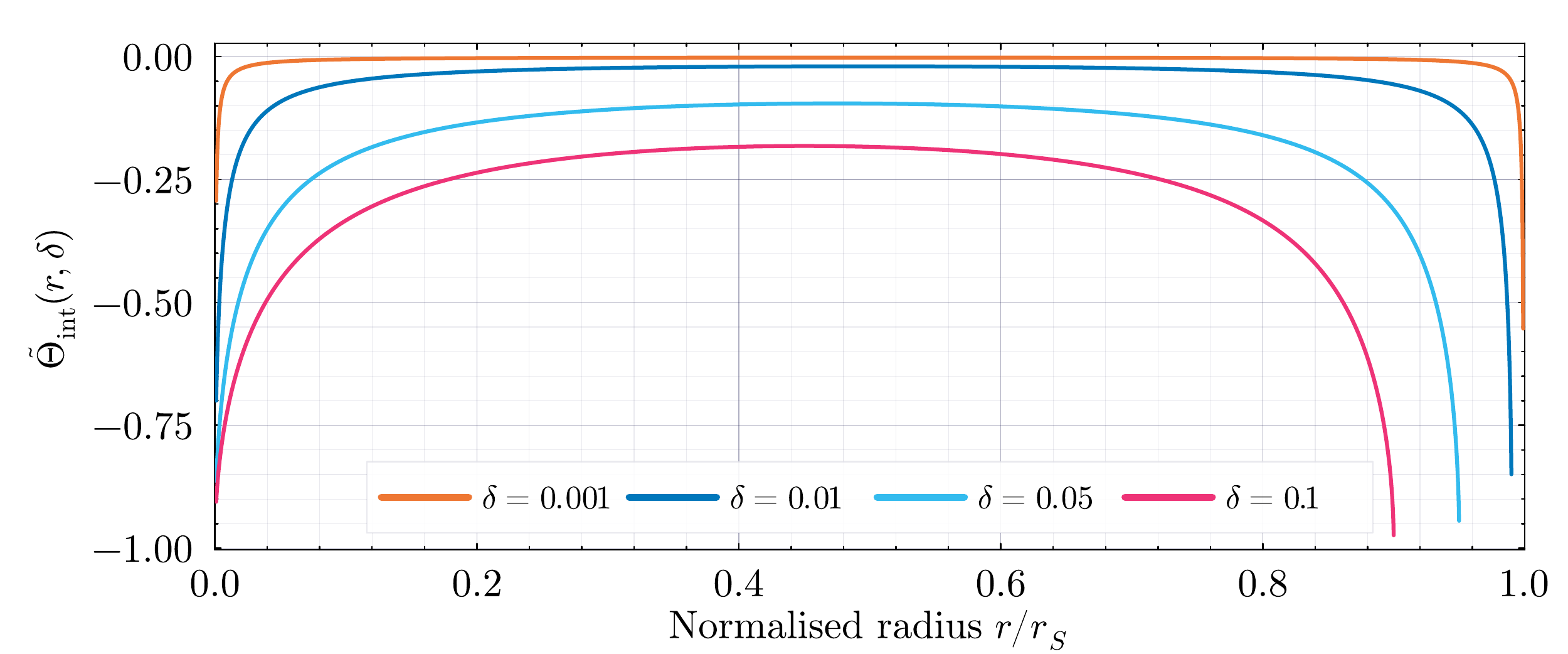}
    \caption{Examples of the interior analogue of our discrete expansion measurement $\widetilde{\Theta}_{\text{int}}(r, \delta)$ for various values of the parameter $\delta$, introduced to model the rung separation of the discrete case. The most important feature is the negativity of the function.}
    \label{fig:theta-in-illstr}
\end{figure}

Analogously, we defined the continuum analogue of the discrete measurement $E$ in the interior region of the black hole by
\begin{align}
\widetilde{\Theta}_{\text{int}}(r, \delta) = \frac{g(r + \delta) - g(r)}{g(r)} \qquad \text{with} \qquad g(r) = \int_{t_1(r)}^{t_2(r)} \mathrm dt \, \sqrt{\frac{r_S}{r} - 1}  \,,
\end{align}
Inserting the function $g$, we found
\begin{align}
\widetilde{\Theta}_{\text{int}}(r, \delta) = \frac{\sqrt{\frac{r_S}{r+\delta} - 1} - \sqrt{\frac{r_S}{r} - 1}}{\sqrt{\frac{r_S}{r} - 1}}\,.
\end{align}
From this expression, we see that $\widetilde{\Theta}_{\text{int}}$ cannot exceed zero and only approaches zero for $\delta \rightarrow 0$, as it should. This is also visible in Fig.~\ref{fig:theta-in-illstr}.

\section{Robustness of results for the discrete expansion}

\label{app:robustness}

\begin{figure}[!b]
    \centering
    \includegraphics[width=0.95\linewidth]{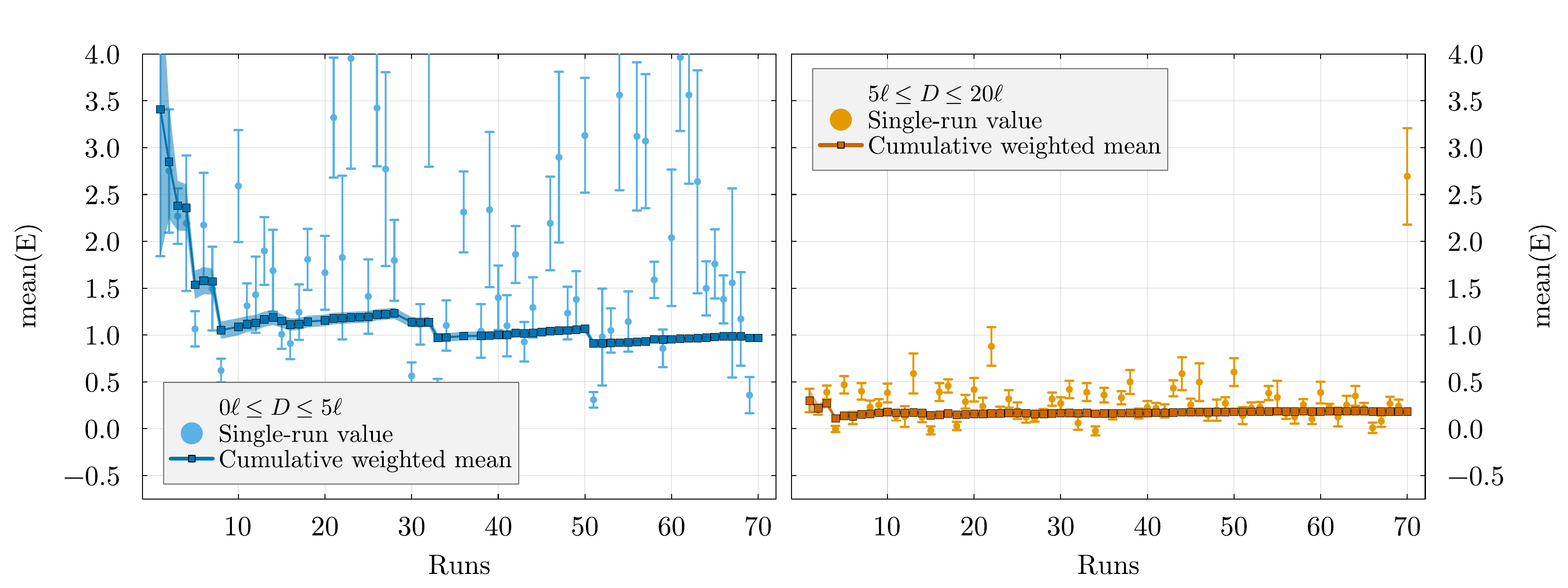}
    \caption{The dependence of the weighted average of $\operatorname{mean}(E)$ on sprinklings into $(1+1)$-dimensional Minkowski spacetime, where the nearest-neighbors were chosen in the interval $[0, 5] \ell$ (right) and  $[5, 20] \ell$ (left). Additionally, the individual $\operatorname{mean}(E)$ for each sprinkling was plotted in blue to illustrate its large fluctuation. The final weighted average is $\overline{\operatorname{mean}(E)}_{\text{left}} = 0.970 \pm 0.033$ and $\overline{\operatorname{mean}(E)}_{\text{right}}= 0.185 \pm 0.007$.}
    \label{fig:DAS-influence}
\end{figure}

We supply here the technical details of implementing the discrete expansion Eq.~\ref{eq:discrete-expansion} in causal sets. We will concentrate on the implementation in sprinklings into Minkowski spacetime, as we have a clear expectation for $E$ in this case, namely $E = 0$. \\
First, we have to decide which pairs of null geodesics we include. We will only consider ladders of a minimum length of four rungs. In the continuum, null geodesics at an arbitrary value of the spatial distance between them produce $E=0$, but in a sprinkling into Minkowski spacetime, we expect a dependence on the spatial distance. The reason lies in a property of the spatial distance function which has been dubbed ``asymptotic silence'' in \cite{Eichhorn:2018doy}: for small continuum values of the spatial distance, the causal-set predistance function significantly overestimates the spatial distance. We therefore only work with pairs of ladders for which the starting points lie within an interval $[x_{\rm min}, x_{\rm max}] \ell$. Numerically, we have found $x_{\rm min}=5$ and $x_{\rm max}=20$ to be the best choice. It produces smaller fluctuations than when choosing $x_{\rm min} = 0$, as can be seen in Fig.~\ref{fig:DAS-influence}. We keep the same choice for sprinklings into Schwarzschild spacetime.
\\
Second, as already stated in the main text, we
disregard pairs of rungs, for which any pair of elements (composed of one element on each of the ladder) are causally related. Recall that $D_{IJ}^{ab}$ is the mean of four distances, corresponding to the four distances that we can compute between the two points of rung $a$ of ladder $I$ and the two points of rung $b$ of ladder $J$ (cf.~illustration in Fig.~\ref{fig:illustration-E}). Thus, if any of these distances is timelike, we remove this pair of
rungs from the calculation of $E$.
 With these two technical choices, there is still the possibility of a systematic offset caused by discreteness effects, pushing the value of $\operatorname{mean}(E)$ away from the value of its continuum analogue $\widetilde{\Theta}$. This is indeed what we find, as the weighted average value of $\operatorname{mean}(E)$ for sprinklings into Minkowski spacetime is a positive value not compatible with zero, as stated in the main text (cf.~Eq.~\ref{eq:expansion-minkowski-rigid}). 

\begin{figure}[!t]
    \centering
    \includegraphics[width=0.95\linewidth]{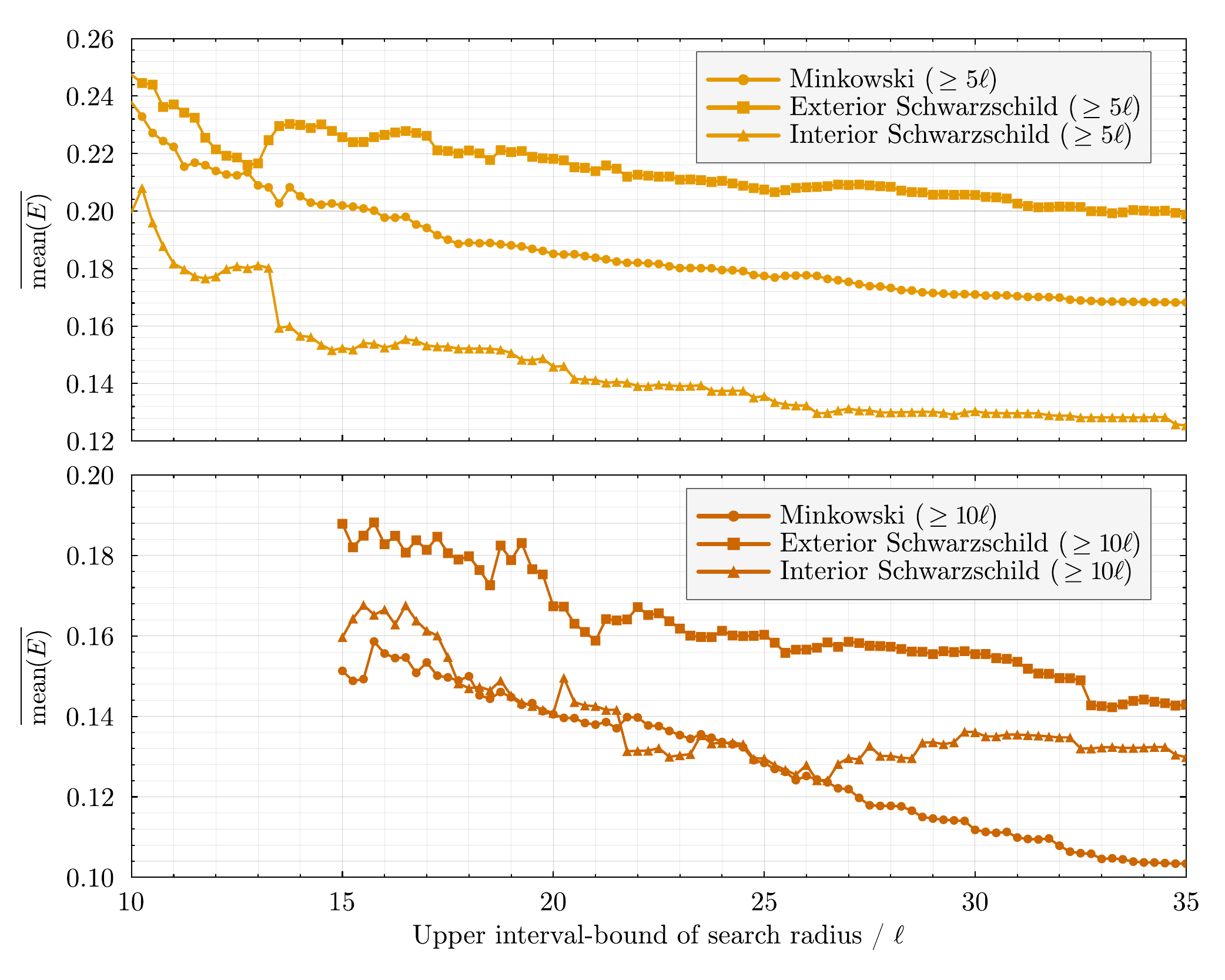}
    \caption{A study of the dependence of the discrete expansion on the choice of $x_{\rm max}$ for the cases $x_{\rm min} = 5$ (upper panel) and $x_{\rm max} = 10$ (lower panel). This region cannot start too close to the reference ladder to avoid an overestimation of the distance due to discrete asymptotic silence. From these two plots, it is clear that our choice of $x_{\rm min} = 5$ is the one with the clearer separation between the various measurements of the discrete expansion, but also the one that satisfies our expectation of Eq.~\ref{eq:E-relation1}.}
    \label{fig:finding-upperlim}
\end{figure}

Our continuum expectation of null geodesic expansion in Schwarzschild spacetime is summarized by the two inequalities 
\begin{align} \label{eq:E-relation1}
\overline{\operatorname{mean}(E)}_{\text{Schwarzschild}, r<r_S} < \overline{\operatorname{mean}(E)}_\text{Minkowski} < \overline{\operatorname{mean}(E)}_{\text{Schwarzschild}, r>r_S} \,,
\end{align}
i.e., the null geodesic expansion in the interior is negative and positive in the exterior. In Fig.~\ref{fig:finding-upperlim} two different classes of sprinklings are shown, the upper panel measurements of $\overline{\operatorname{mean}(E)}$ have $x_{\rm min} = 5$, while the lower panel measurements have $x_{\rm min} = 10$. In Fig.~\ref{fig:finding-upperlim}, we can see the traces of the effect of discrete asymptotic silence in that the measurements in the upper panel have larger absolute values of $\overline{\operatorname{mean}(E)}$ compared to the measurements in the lower panel. \\
We stated above that the interval $[x_{\rm min}, x_{\rm max}] \ell = [5,20]\ell$ was chosen. The reason for choosing $x_{\rm min} = 5$ over $x_{\rm min} = 10$ was because the separation between the different discrete expansions, following Eq.~\ref{eq:E-relation1}, is greatest in that case, as can be clearly seen in Fig.~\ref{fig:finding-upperlim}. The choice of upper interval-bound $ x_{\rm max} \ell$ is more subtle. It needs to be large enough to appreciate the distance between the various discrete expansions and to allow for better statistics (a larger interval compares the reference ladder with more ladders), but small enough to retain the interpretation of the continuum geodesic expansion, as the change in the area of an infinitesimal pencil of geodesics (cf.~Eq.~\ref{eq:expansion2}). 

\section{Fuzzy ladders: properties and discrete expansion}
\label{app:fuzzy}

\subsection{Properties of fuzzy ladders}

For a small subset of ladders, giving up the rigidity conditions of the original ladder definition results in problems.

The first problem is that a single null geodesic may be approximated by a family of ladders, differing from one another only by a small number of points. To remove this redundancy in our results, one ladder was selected as the representative approximation to the geodesic, and any ladder sharing more than half of its points with this representative was discarded.

 The second problem is somewhat more challenging. It consists in the misidentification of two intersecting null geodesics, i.e., an ingoing and an outgoing one, as a single null geodesic. 

\begin{figure}[!t]
    \centering
    \includegraphics[width=\linewidth]{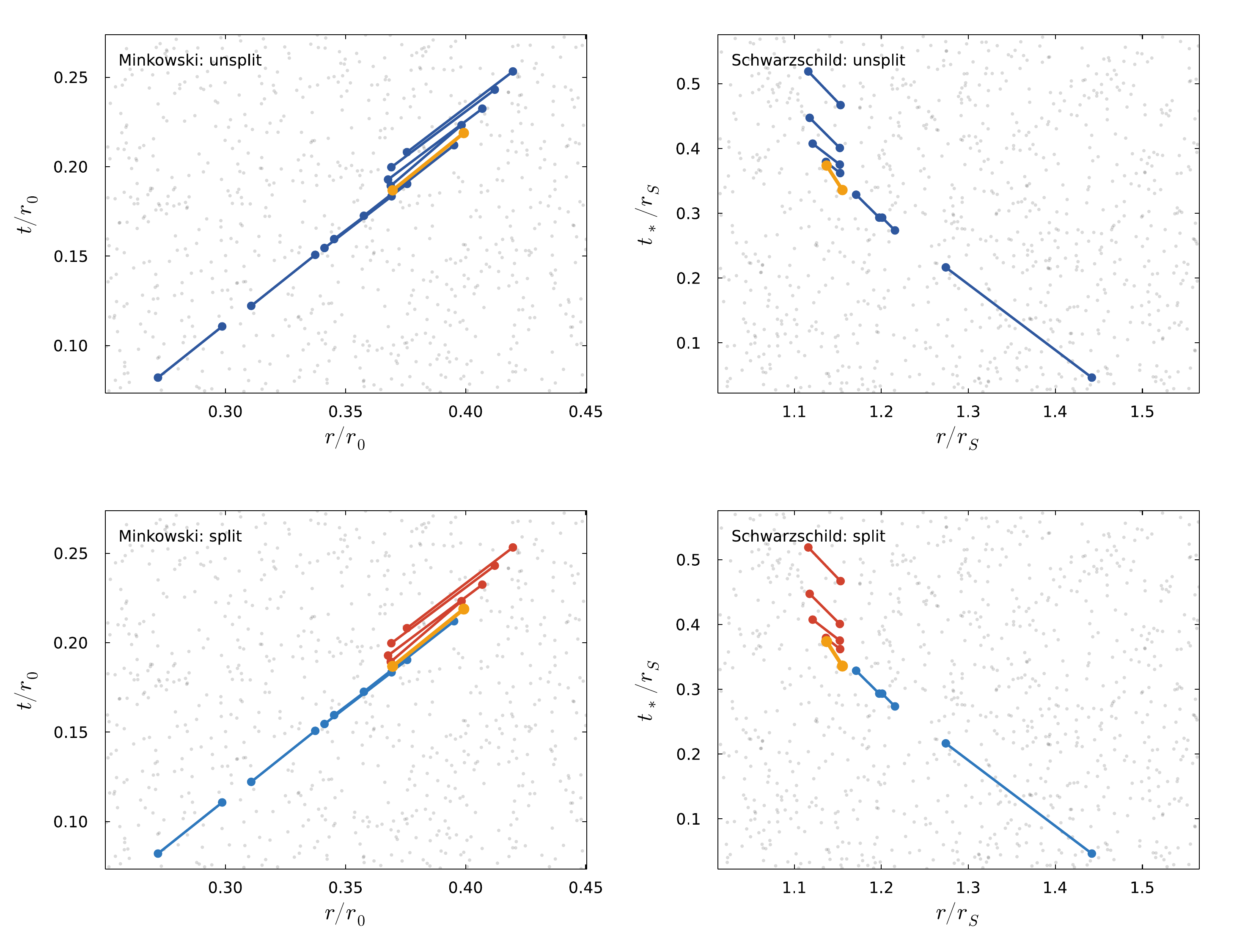}
    \caption{\label{fig:hockey-stick}Examples of a joint configuration of two fuzzy ladders in Schwarzschild and Minkowski. Before separation, these configurations are identified as a single long ladder in dark blue. After splitting at the turning rung, they are two distinct ladder segments, one ingoing and one outgoing. Red and light blue represent the different directed ladders after the splitting (colored by order in $t_\ast$). The yellow rung is the shared one, i.e., where the change in orientation  takes place.}
\end{figure}

 As one can see in Fig.~\ref{fig:hockey-stick}, the ``fuzzy causal ladder'' in Def.~\ref{def:fuzzy-causal-ladders} allows for joint configurations that are identified
 as single long ladders by our algorithm, but which one would  naturally split into two shorter components (one ingoing and one outgoing).
The first of these (counting from the bottom to the top in $t_\ast$) is  nearly parallel to the respective approximate null geodesic, whilst the other is nearly perpendicular, as seen in Fig.~\ref{fig:hockey-stick}. 
This happens because the fuzzy definition enforces mesoscopic rigidity along each of the sides of the ladder through the $M$-scale interval conditions, whilst no longer imposing sufficiently rigid conditions to ensure that the entire object follows a single consistent orientation. 
Consequently, for a given scale $M$, one segment may continue in one direction up to some rung $k$, while a second segment extends away from the same rung in the opposite direction, and this configuration still satisfies all conditions from Def.~\ref{def:fuzzy-causal-ladders}.
As a result, an ingoing segment and an outgoing segment present themselves ``glued'' together at an intermediate rung whilst still satisfying all defining conditions. 

We note in passing that, although problematic for our purposes, such an intersection of two geodesics may be relevant in other settings.  

 In practice, we deal with those configurations by making use of the information on the embedding, where one can  identify whether any given portion of a ladder is ingoing or outgoing and can accordingly split such long ladders. 

\subsection{The discrete expansion with fuzzy ladders}

We now turn our attention towards the discrete expansion with fuzzy ladders. 
For fuzzy ladders, there are two competing effects on the discrete expansion. On the one hand, their greater length compared to rigid ladders is an advantage and should produce a more robust estimate of the discrete expansion already from a single pair of ladders. On the other hand, their reduced rigidity is a disadvantage, because it leads to larger discretization effects in the expansion. 
To see which effect dominates in practice, we follow a similar procedure as for the rigid ladders to extract the discrete expansion.

At the same time, we use fuzzy ladders as another cross-check of our result for the discrete expansion that we obtained in the main text. First, using different ladder definitions (rigid versus fuzzy) tests whether our results are contingent upon using one particular definition, or whether they are qualitatively robust. Second, we test the impact of various technical choices in the calculation.

\begin{figure}[!t]
  \begin{center}
    \includegraphics[width=0.9\linewidth]{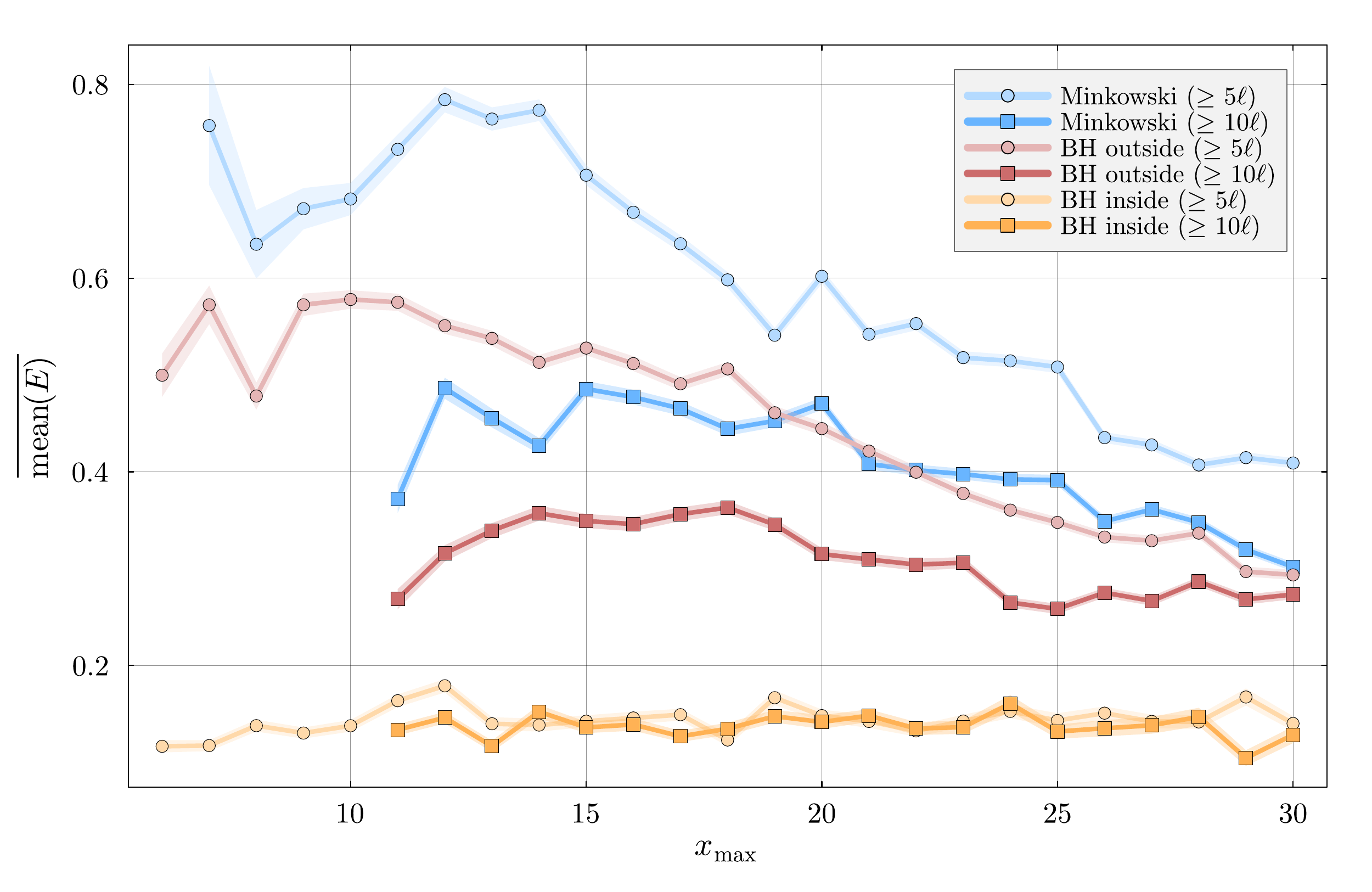}
    \end{center}
\caption{\label{fig:fuzzy_initial_distance}{Dependence of the discrete expansion on the width of the region from which nearest-neighbors are chosen for the fuzzy-ladder definition. The possible starting values were $5\ell$ and $10\ell$ just like for the rigid ladders study.}}
\end{figure}

For fuzzy ladders we take $M=3$, as it yields a sufficiently large number of ladders, including a good number of long ones, without pushing the mesoscopic scale to unnecessarily large values. 
In particular, we use only $8$-rung ladders to compute the discrete expansion.
Moreover, we chose $[x_{\rm min}, x_{\rm max}] \ell = [10,20]\ell$ cf.~Fig.~\ref{fig:fuzzy_initial_distance}. There, we see that choosing $x_{\rm min}=5$ results in a larger value of the mean of the expansion, which we ascribe to the effects of discrete asymptotic silence, just like for rigid ladders.

\begin{figure}[!t]
  \begin{center}
    \includegraphics[width=0.9\linewidth]{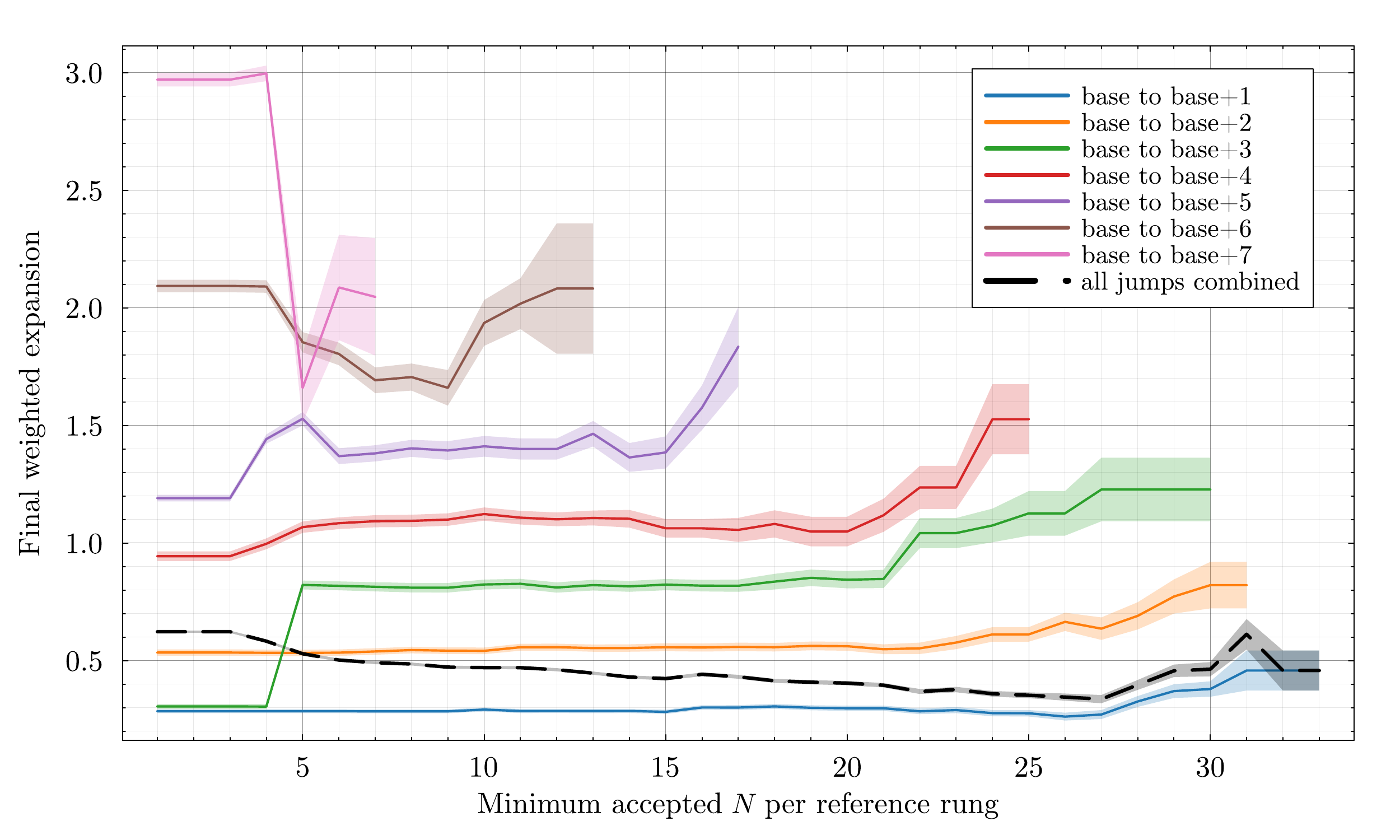}
    \end{center}
    \caption{\label{fig:Mink_jumps_fuzzy} Final weighted expansion for $30$ realizations of a box boundary Minkowski spacetime with $M=3$ as a function of the minimum accepted $N$ for $8$-rungs ladders, with separated curves for each jump size and the combined weighted result shown in black.}
\end{figure}

Another technical choice consists in the introduction of a lower bound on the number of data points from a ladder: we assume that the effects of discreteness and fuzziness are largest, when a given ladder is only compared to one other ladder. Thus, we only accept results for which the expansion associated to a single rung has been computed with at least ten other rungs contributing to this data point. A dependence of the results on this lower bound is shown in Fig.~\ref{fig:Mink_jumps_fuzzy}.

\begin{figure}[!t]
  \begin{center}
    \includegraphics[width=0.9\linewidth]{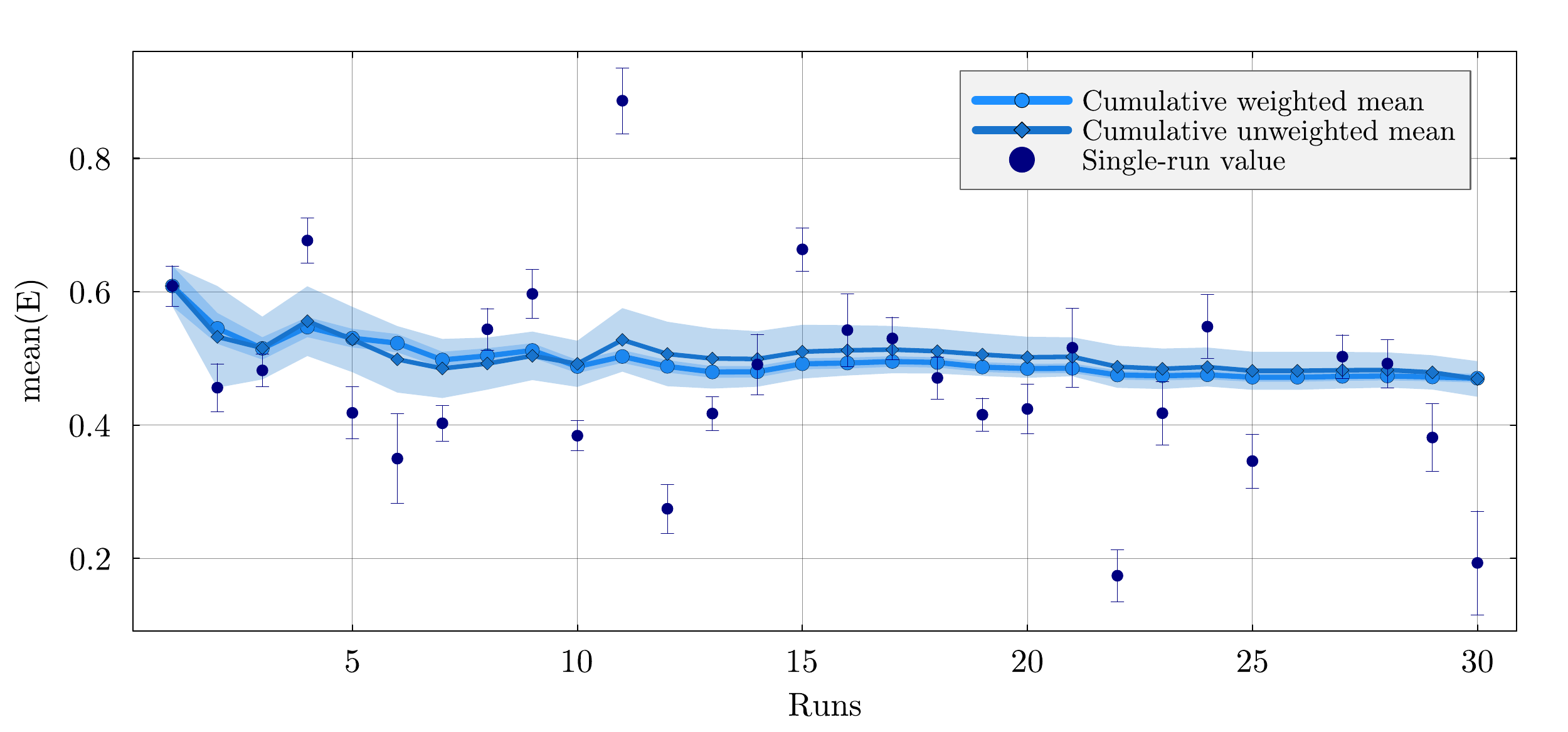}
    \end{center}
    \caption{\label{fig:Minkowski_expansion_fuzzy} Cumulative expansion in Minkowski spacetime as a function of included runs of a squared box with $M=3$, for $8$-rung ladders and $N\geq 10$. The light blue curve and its band show the cumulative weighted mean of the expansion proxy, while the darker blue curve and its band show the corresponding cumulative unweighted mean. The darker blue points show the single-run values of the expansion estimate, $\operatorname{mean}(E)$. }
\end{figure}

The results obtained for the choice $N=10$ are shown in Fig.~\ref{fig:Minkowski_expansion_fuzzy}. We used $30$ realizations and plotted the evolution of the weighted/unweighted average of the discrete expansion as a function of the realizations. From the relative stability of the results from approximately 20 runs onwards, we infer that our results are statistically converged.
We also observe that the weighted and unweighted averages give very similar results. The unweighted average is systematically slightly lower than the weighted one, but the two estimates agree within their error bands.
The individual contributions from each run are also shown as darker blue dots. Notice that there is no dark blue point corresponding to runs $26$ and $27$. This is due to these sprinklings not contributing, since they failed to satisfy the minimum requirement $N\geq10$.

For Minkowski we obtained an expansion value, after the weighted average, of
\begin{equation}
    \overline{\operatorname{mean}(E)}_{\text{Minkowski}} = 0.470 \pm 0.006 \,.
\end{equation}
This value is clearly further away from zero than the one of rigid ladders, cf.~Eq.~\ref{eq:expansion-minkowski-rigid}. This is likely an effect of the ``fuzziness" of ladders. It shows that, while the qualitatively effect that discreteness generates a systematic shift to positive values, the rigid-ladder definition performs better, despite the contributing ladders being shorter.

As was done in the case of the rigid ladders, we will use this value as a baseline. We will subtract it from the result for Schwarzschild spacetime, because we interpret it as a consequence of discreteness, not of spacetime curvature. 

In Fig.~\ref{fig:Mink_jumps_fuzzy}, we also show how the different jump sizes contribute to the final average. 
When comparing different rungs, one may have several possible separations between them, corresponding to different jump sizes. We find that the smallest jumps, namely $2$ and $3$, dominate the average expansion.
This is simply due to their much greater abundance relative to larger jumps, which therefore contribute less significantly to the final result.

 We now turn to the analysis of the Schwarzschild spacetime and focus on outgoing ladders.
Just like for the rigid ladders, we study interior and exterior regions separately, using $40$ sprinklings for each case.
For the interior, the sprinklings were performed in the region $0 \leq r \leq 1$ and $0 \leq t_* \leq 1$ with $r_S=1$. 
For the exterior, also $40$ runs were performed, this time in the region $1 \leq r \leq 2$ and $0 \leq t_* \leq 1$. 
As in the Minkowski case, we used $8$-rungs fuzzy ladders of scale $M=3$.  
Although this may seem directly comparable to the Minkowski case, the amount of available data is smaller. 
This is because in the Minkowski analysis the ingoing and outgoing ladders were treated separately, but both sets of data were used, whereas here only the outgoing sector contributes. Nevertheless, our results are sufficient to achieve convergence.

\begin{figure}[!t]
  \begin{center}
    \includegraphics[width=0.9\linewidth]{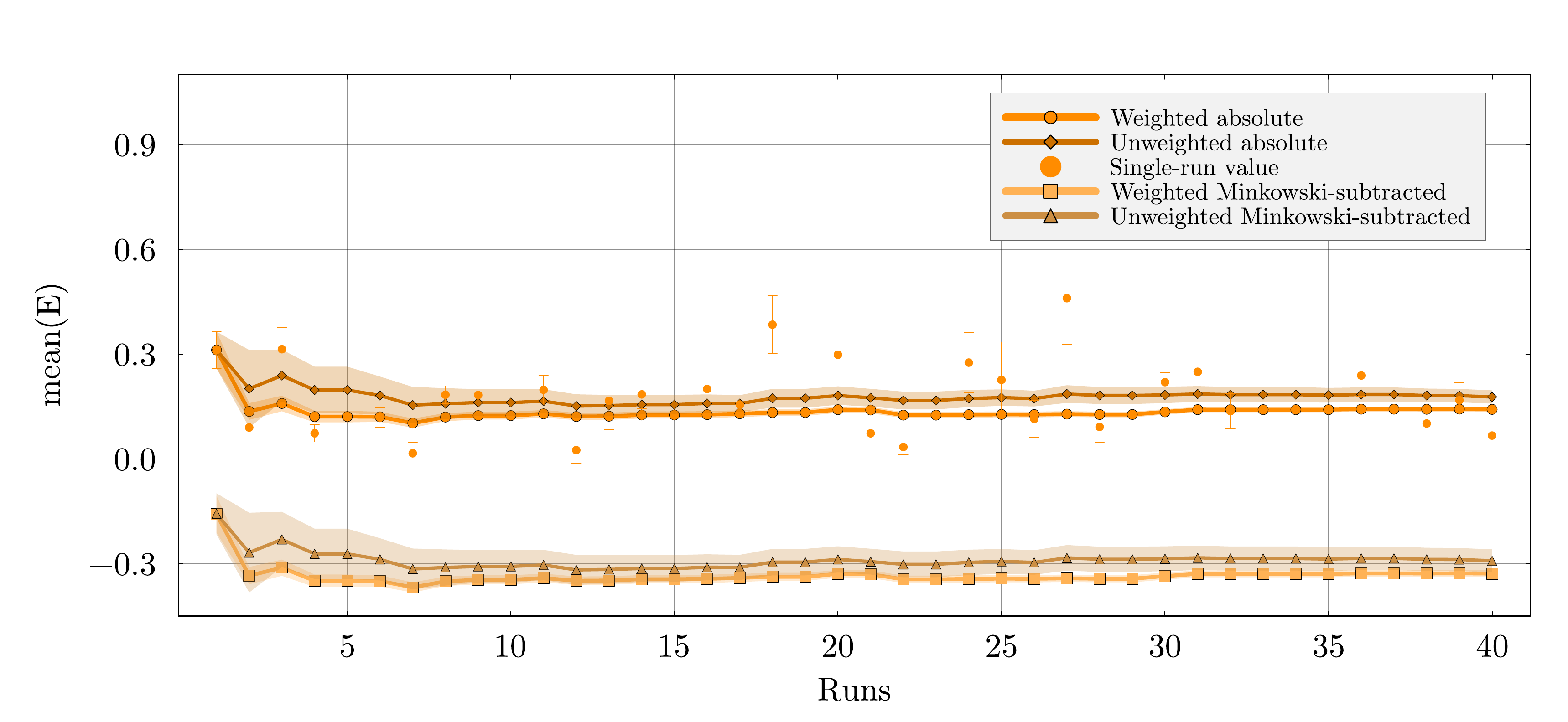}
    \end{center}
    \caption{\label{fig:bhin_jumps_fuzzy} Cumulative expansion in the interior of Schwarzschild spacetime as a function of included runs of a squared box with $M=3$, for $8$-rung ladders and $N\geq 10$. The lighter yellow curve on top (ball points) and its band show the cumulative weighted mean of the expansion proxy, while the darker yellow curve  on top (diamond points) and its band show the corresponding cumulative unweighted mean, as it is labeled. The points show the single-run values of the expansion estimate, $\operatorname{mean}(E)$. The two subtraction curves show the corresponding weighted and unweighted results after subtracting the final Minkowski baseline.}
\end{figure}

\begin{figure}[!t]
  \begin{center}
    \includegraphics[width=0.9\linewidth]{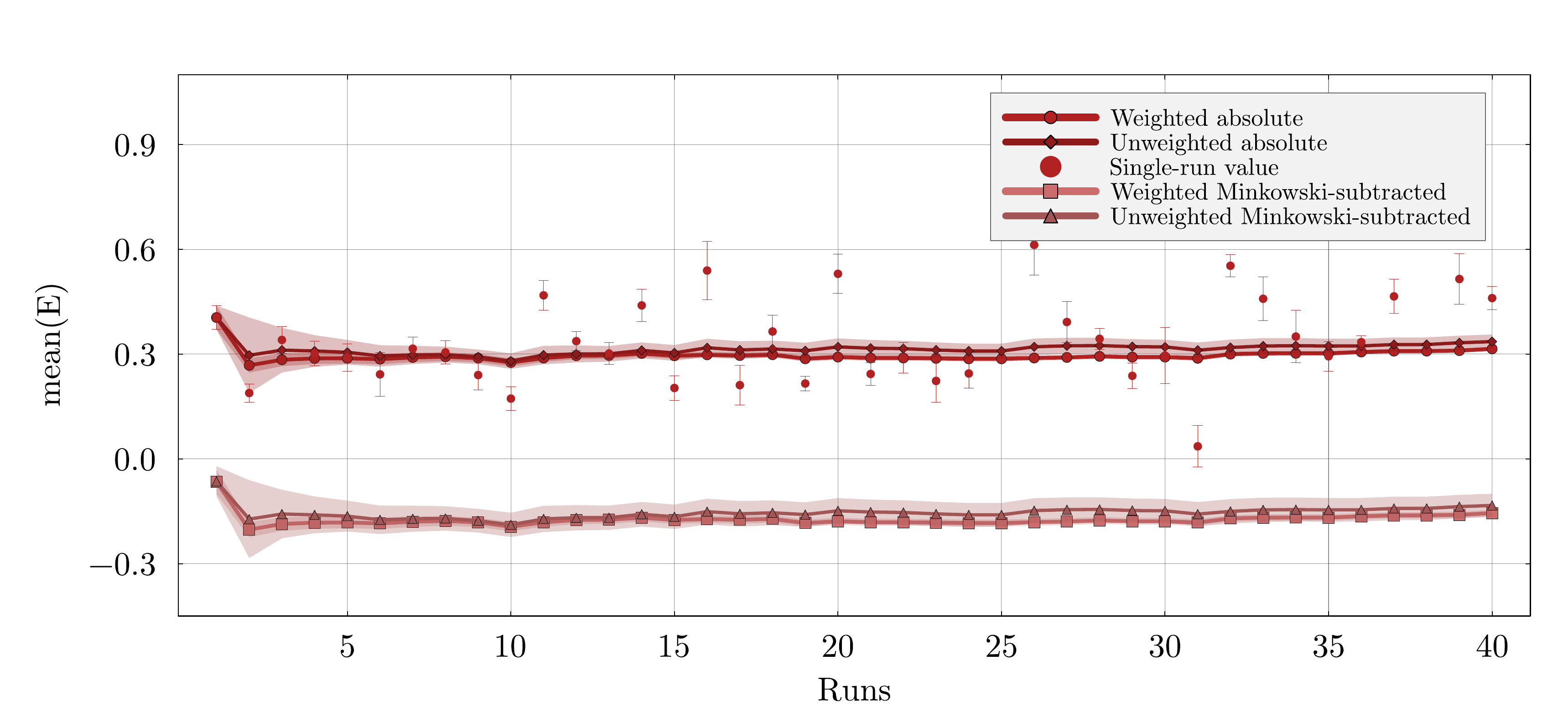}
    \end{center}
    \caption{\label{fig:bhout_jumps_fuzzy} Cumulative expansion in the exterior of Schwarzschild spacetime as a function of included runs of a squared box with $M=3$, for $8$-rung ladders and $N\geq 10$. The lighter red curve on top (ball points) and its band show the cumulative weighted mean of the expansion proxy, while the darker red curve on top (diamond points) and its band show the corresponding cumulative unweighted mean, as it is labeled. The points show the single-run values of the expansion estimate, $\operatorname{mean}(E)$. The two subtraction curves show the corresponding weighted and unweighted results after subtracting the final Minkowski baseline.}
\end{figure}

The results are presented in Figs.~\ref{fig:bhin_jumps_fuzzy} and~\ref{fig:bhout_jumps_fuzzy}. 
In these plots we show both the measured value of the expansion and the value obtained after subtracting the Minkowski baseline. We also display the individual contribution of each sprinkling, together with its corresponding uncertainty.
As in the Minkowski case, the weighted and unweighted averages remain close to one another, with the unweighted result again lying slightly below the weighted one. In the exterior they agree within the error bands, while in the interior the agreement is slightly weaker, but the two curves still remain close.
These plots make clear the effect already anticipated in the Minkowski analysis: in the Schwarzschild case, a significantly larger number of runs fail to contribute, because they do not satisfy the minimum requirement $N_{\min}=10$. This effect is much stronger in the interior than in the exterior, which may be due to our choice of boundary shape of the sprinkling.

For the Schwarzschild interior, after subtracting the Minkowski baseline, we obtain
\begin{equation}
    \overline{\operatorname{mean}(E)}_{\text{Schwarzschild,}\,r<r_S} = -0.33\pm 0.01 \,.
    \label{eq:fuzzy_ladders_schwarzschild_interior}
\end{equation}
Although the Minkowski value obtained with fuzzy ladders lies further away from its theoretical expectation, namely zero, than in the rigid ladder case, the baseline-subtracted value in the Schwarzschild interior becomes clearly negative. 
In this sense, the fuzzy ladder prescription gives a cleaner identification of the expected negative expansion in the Schwarzschild interior. 
For the Schwarzschild exterior, we obtain
\begin{equation} 
    \overline{\operatorname{mean}(E)}_{\text{Schwarzschild,}\,r>r_S} = -0.16 \pm 0.01\,.
    \label{eq:fuzzy_ladders_schwarzschild_exterior}
\end{equation}
In this case, however, the result is clearly worse. 
As shown in Fig.~\ref{fig:fuzzy_initial_distance}, independently of the initial choice of the minimum spatial distance used to search other rungs to compare with a reference one, we do not find a regime analogous to the one found for rigid ladders, where the Schwarzschild exterior value lies above the Minkowski baseline, as expected given the continuum expectation. 
Consequently, the baseline-subtracted exterior value is slightly negative. 
This is not the expected behavior, since the expansion in the Schwarzschild exterior should be positive, even if only slightly, rather than zero or negative. Thus, we find that fuzzy ladders encode one aspect of the discrete expansion correctly, namely that it should be smaller for the interior than for the exterior Schwarzschild region. However, the fuzziness results in large systematic uncertainties which mean that other aspects of the expected behavior cannot be reproduced. 

\bibliographystyle{apsrev4-2}
\bibliography{refs}

\begin{thebibliography}{71}%
\makeatletter
\providecommand \@ifxundefined [1]{%
 \@ifx{#1\undefined}
}%
\providecommand \@ifnum [1]{%
 \ifnum #1\expandafter \@firstoftwo
 \else \expandafter \@secondoftwo
 \fi
}%
\providecommand \@ifx [1]{%
 \ifx #1\expandafter \@firstoftwo
 \else \expandafter \@secondoftwo
 \fi
}%
\providecommand \natexlab [1]{#1}%
\providecommand \enquote  [1]{``#1''}%
\providecommand \bibnamefont  [1]{#1}%
\providecommand \bibfnamefont [1]{#1}%
\providecommand \citenamefont [1]{#1}%
\providecommand \href@noop [0]{\@secondoftwo}%
\providecommand \href [0]{\begingroup \@sanitize@url \@href}%
\providecommand \@href[1]{\@@startlink{#1}\@@href}%
\providecommand \@@href[1]{\endgroup#1\@@endlink}%
\providecommand \@sanitize@url [0]{\catcode `\\12\catcode `\$12\catcode
  `\&12\catcode `\#12\catcode `\^12\catcode `\_12\catcode `\%12\relax}%
\providecommand \@@startlink[1]{}%
\providecommand \@@endlink[0]{}%
\providecommand \url  [0]{\begingroup\@sanitize@url \@url }%
\providecommand \@url [1]{\endgroup\@href {#1}{\urlprefix }}%
\providecommand \urlprefix  [0]{URL }%
\providecommand \Eprint [0]{\href }%
\providecommand \doibase [0]{https://doi.org/}%
\providecommand \selectlanguage [0]{\@gobble}%
\providecommand \bibinfo  [0]{\@secondoftwo}%
\providecommand \bibfield  [0]{\@secondoftwo}%
\providecommand \translation [1]{[#1]}%
\providecommand \BibitemOpen [0]{}%
\providecommand \bibitemStop [0]{}%
\providecommand \bibitemNoStop [0]{.\EOS\space}%
\providecommand \EOS [0]{\spacefactor3000\relax}%
\providecommand \BibitemShut  [1]{\csname bibitem#1\endcsname}%
\let\auto@bib@innerbib\@empty
\bibitem [{\citenamefont {Penrose}(1969)}]{Penrose:1969pc}%
  \BibitemOpen
  \bibfield  {author} {\bibinfo {author} {\bibfnamefont {R.}~\bibnamefont
  {Penrose}},\ }\href {https://doi.org/10.1023/A:1016578408204} {\bibfield
  {journal} {\bibinfo  {journal} {Riv. Nuovo Cim.}\ }\textbf {\bibinfo {volume}
  {1}},\ \bibinfo {pages} {252} (\bibinfo {year} {1969})}\BibitemShut {NoStop}%
\bibitem [{\citenamefont {Penrose}(1979)}]{Penrose:1979azm}%
  \BibitemOpen
  \bibfield  {author} {\bibinfo {author} {\bibfnamefont {R.}~\bibnamefont
  {Penrose}}\ }(\bibinfo {year} {1979})\ pp.\ \bibinfo {pages}
  {581--638}\BibitemShut {NoStop}%
\bibitem [{\citenamefont {Dafermos}\ and\ \citenamefont
  {Luk}(2025)}]{Dafermos:2017dbw}%
  \BibitemOpen
  \bibfield  {author} {\bibinfo {author} {\bibfnamefont {M.}~\bibnamefont
  {Dafermos}}\ and\ \bibinfo {author} {\bibfnamefont {J.}~\bibnamefont {Luk}},\
  }\href {https://doi.org/10.4007/annals.2025.202.2.1} {\bibfield  {journal}
  {\bibinfo  {journal} {Ann. Math. (2)}\ }\textbf {\bibinfo {volume} {202}},\
  \bibinfo {pages} {309} (\bibinfo {year} {2025})},\ \Eprint
  {https://arxiv.org/abs/1710.01722} {arXiv:1710.01722 [gr-qc]} \BibitemShut
  {NoStop}%
\bibitem [{\citenamefont {Sbierski}(2026)}]{Sbierski:2024fko}%
  \BibitemOpen
  \bibfield  {author} {\bibinfo {author} {\bibfnamefont {J.}~\bibnamefont
  {Sbierski}},\ }\href {https://doi.org/10.1007/s00222-025-01387-0} {\bibfield
  {journal} {\bibinfo  {journal} {Invent. Math.}\ }\textbf {\bibinfo {volume}
  {243}},\ \bibinfo {pages} {961} (\bibinfo {year} {2026})},\ \Eprint
  {https://arxiv.org/abs/2409.18838} {arXiv:2409.18838 [gr-qc]} \BibitemShut
  {NoStop}%
\bibitem [{\citenamefont {Van~de Moortel}(2025)}]{VandeMoortel:2025ngd}%
  \BibitemOpen
  \bibfield  {author} {\bibinfo {author} {\bibfnamefont {M.}~\bibnamefont
  {Van~de Moortel}},\ }\href@noop {} {\  (\bibinfo {year} {2025})},\ \Eprint
  {https://arxiv.org/abs/2501.13180} {arXiv:2501.13180 [gr-qc]} \BibitemShut
  {NoStop}%
\bibitem [{\citenamefont {Hawking}(1976)}]{Hawking:1976ra}%
  \BibitemOpen
  \bibfield  {author} {\bibinfo {author} {\bibfnamefont {S.~W.}\ \bibnamefont
  {Hawking}},\ }\href {https://doi.org/10.1103/PhysRevD.14.2460} {\bibfield
  {journal} {\bibinfo  {journal} {Phys. Rev. D}\ }\textbf {\bibinfo {volume}
  {14}},\ \bibinfo {pages} {2460} (\bibinfo {year} {1976})}\BibitemShut
  {NoStop}%
\bibitem [{\citenamefont {Palti}(2019)}]{Palti:2019pca}%
  \BibitemOpen
  \bibfield  {author} {\bibinfo {author} {\bibfnamefont {E.}~\bibnamefont
  {Palti}},\ }\href {https://doi.org/10.1002/prop.201900037} {\bibfield
  {journal} {\bibinfo  {journal} {Fortsch. Phys.}\ }\textbf {\bibinfo {volume}
  {67}},\ \bibinfo {pages} {1900037} (\bibinfo {year} {2019})},\ \Eprint
  {https://arxiv.org/abs/1903.06239} {arXiv:1903.06239 [hep-th]} \BibitemShut
  {NoStop}%
\bibitem [{\citenamefont {van Beest}\ \emph {et~al.}(2022)\citenamefont {van
  Beest}, \citenamefont {Calder{\'o}n-Infante}, \citenamefont {Mirfendereski},\
  and\ \citenamefont {Valenzuela}}]{vanBeest:2021lhn}%
  \BibitemOpen
  \bibfield  {author} {\bibinfo {author} {\bibfnamefont {M.}~\bibnamefont {van
  Beest}}, \bibinfo {author} {\bibfnamefont {J.}~\bibnamefont
  {Calder{\'o}n-Infante}}, \bibinfo {author} {\bibfnamefont {D.}~\bibnamefont
  {Mirfendereski}},\ and\ \bibinfo {author} {\bibfnamefont {I.}~\bibnamefont
  {Valenzuela}},\ }\href {https://doi.org/10.1016/j.physrep.2022.09.002}
  {\bibfield  {journal} {\bibinfo  {journal} {Phys. Rept.}\ }\textbf {\bibinfo
  {volume} {989}},\ \bibinfo {pages} {1} (\bibinfo {year} {2022})},\ \Eprint
  {https://arxiv.org/abs/2102.01111} {arXiv:2102.01111 [hep-th]} \BibitemShut
  {NoStop}%
\bibitem [{\citenamefont {Dou}\ and\ \citenamefont
  {Sorkin}(2003)}]{Dou:2003af}%
  \BibitemOpen
  \bibfield  {author} {\bibinfo {author} {\bibfnamefont {D.}~\bibnamefont
  {Dou}}\ and\ \bibinfo {author} {\bibfnamefont {R.~D.}\ \bibnamefont
  {Sorkin}},\ }\href {https://doi.org/10.1023/A:1023781022519} {\bibfield
  {journal} {\bibinfo  {journal} {Found. Phys.}\ }\textbf {\bibinfo {volume}
  {33}},\ \bibinfo {pages} {279} (\bibinfo {year} {2003})},\ \Eprint
  {https://arxiv.org/abs/gr-qc/0302009} {arXiv:gr-qc/0302009} \BibitemShut
  {NoStop}%
\bibitem [{\citenamefont {Barton}\ \emph {et~al.}(2019)\citenamefont {Barton},
  \citenamefont {Counsell}, \citenamefont {Dowker}, \citenamefont {Gould},
  \citenamefont {Jubb},\ and\ \citenamefont {Taylor}}]{Barton:2019okw}%
  \BibitemOpen
  \bibfield  {author} {\bibinfo {author} {\bibfnamefont {C.}~\bibnamefont
  {Barton}}, \bibinfo {author} {\bibfnamefont {A.}~\bibnamefont {Counsell}},
  \bibinfo {author} {\bibfnamefont {F.}~\bibnamefont {Dowker}}, \bibinfo
  {author} {\bibfnamefont {D.~S.~W.}\ \bibnamefont {Gould}}, \bibinfo {author}
  {\bibfnamefont {I.}~\bibnamefont {Jubb}},\ and\ \bibinfo {author}
  {\bibfnamefont {G.}~\bibnamefont {Taylor}},\ }\href
  {https://doi.org/10.1103/PhysRevD.100.126008} {\bibfield  {journal} {\bibinfo
   {journal} {Phys. Rev. D}\ }\textbf {\bibinfo {volume} {100}},\ \bibinfo
  {pages} {126008} (\bibinfo {year} {2019})},\ \Eprint
  {https://arxiv.org/abs/1909.08620} {arXiv:1909.08620 [gr-qc]} \BibitemShut
  {NoStop}%
\bibitem [{\citenamefont {Broderick}\ \emph {et~al.}(2009)\citenamefont
  {Broderick}, \citenamefont {Loeb},\ and\ \citenamefont
  {Narayan}}]{Broderick:2009ph}%
  \BibitemOpen
  \bibfield  {author} {\bibinfo {author} {\bibfnamefont {A.~E.}\ \bibnamefont
  {Broderick}}, \bibinfo {author} {\bibfnamefont {A.}~\bibnamefont {Loeb}},\
  and\ \bibinfo {author} {\bibfnamefont {R.}~\bibnamefont {Narayan}},\ }\href
  {https://doi.org/10.1088/0004-637X/701/2/1357} {\bibfield  {journal}
  {\bibinfo  {journal} {Astrophys. J.}\ }\textbf {\bibinfo {volume} {701}},\
  \bibinfo {pages} {1357} (\bibinfo {year} {2009})},\ \Eprint
  {https://arxiv.org/abs/0903.1105} {arXiv:0903.1105 [astro-ph.HE]}
  \BibitemShut {NoStop}%
\bibitem [{\citenamefont {Cardoso}\ \emph {et~al.}(2016)\citenamefont
  {Cardoso}, \citenamefont {Franzin},\ and\ \citenamefont
  {Pani}}]{Cardoso:2016rao}%
  \BibitemOpen
  \bibfield  {author} {\bibinfo {author} {\bibfnamefont {V.}~\bibnamefont
  {Cardoso}}, \bibinfo {author} {\bibfnamefont {E.}~\bibnamefont {Franzin}},\
  and\ \bibinfo {author} {\bibfnamefont {P.}~\bibnamefont {Pani}},\ }\href
  {https://doi.org/10.1103/PhysRevLett.116.171101} {\bibfield  {journal}
  {\bibinfo  {journal} {Phys. Rev. Lett.}\ }\textbf {\bibinfo {volume} {116}},\
  \bibinfo {pages} {171101} (\bibinfo {year} {2016})},\ \bibinfo {note}
  {[Erratum: Phys.Rev.Lett. 117, 089902 (2016)]},\ \Eprint
  {https://arxiv.org/abs/1602.07309} {arXiv:1602.07309 [gr-qc]} \BibitemShut
  {NoStop}%
\bibitem [{\citenamefont {Bombelli}\ \emph {et~al.}(1987)\citenamefont
  {Bombelli}, \citenamefont {Lee}, \citenamefont {Meyer},\ and\ \citenamefont
  {Sorkin}}]{Bombelli:1987aa}%
  \BibitemOpen
  \bibfield  {author} {\bibinfo {author} {\bibfnamefont {L.}~\bibnamefont
  {Bombelli}}, \bibinfo {author} {\bibfnamefont {J.}~\bibnamefont {Lee}},
  \bibinfo {author} {\bibfnamefont {D.}~\bibnamefont {Meyer}},\ and\ \bibinfo
  {author} {\bibfnamefont {R.}~\bibnamefont {Sorkin}},\ }\href
  {https://doi.org/10.1103/PhysRevLett.59.521} {\bibfield  {journal} {\bibinfo
  {journal} {Phys. Rev. Lett.}\ }\textbf {\bibinfo {volume} {59}},\ \bibinfo
  {pages} {521} (\bibinfo {year} {1987})}\BibitemShut {NoStop}%
\bibitem [{\citenamefont {Bombelli}\ \emph {et~al.}(2009)\citenamefont
  {Bombelli}, \citenamefont {Henson},\ and\ \citenamefont
  {Sorkin}}]{Bombelli:2006nm}%
  \BibitemOpen
  \bibfield  {author} {\bibinfo {author} {\bibfnamefont {L.}~\bibnamefont
  {Bombelli}}, \bibinfo {author} {\bibfnamefont {J.}~\bibnamefont {Henson}},\
  and\ \bibinfo {author} {\bibfnamefont {R.~D.}\ \bibnamefont {Sorkin}},\
  }\href {https://doi.org/10.1142/S0217732309031958} {\bibfield  {journal}
  {\bibinfo  {journal} {Mod. Phys. Lett. A}\ }\textbf {\bibinfo {volume}
  {24}},\ \bibinfo {pages} {2579} (\bibinfo {year} {2009})},\ \Eprint
  {https://arxiv.org/abs/gr-qc/0605006} {arXiv:gr-qc/0605006} \BibitemShut
  {NoStop}%
\bibitem [{\citenamefont {Bhattacharya}\ \emph {et~al.}(2023)\citenamefont
  {Bhattacharya}, \citenamefont {Mathur},\ and\ \citenamefont
  {Surya}}]{Bhattacharya:2023xnj}%
  \BibitemOpen
  \bibfield  {author} {\bibinfo {author} {\bibfnamefont {A.}~\bibnamefont
  {Bhattacharya}}, \bibinfo {author} {\bibfnamefont {A.}~\bibnamefont
  {Mathur}},\ and\ \bibinfo {author} {\bibfnamefont {S.}~\bibnamefont
  {Surya}},\ }\href {https://doi.org/10.1007/s10714-023-03074-y} {\bibfield
  {journal} {\bibinfo  {journal} {Gen. Rel. Grav.}\ }\textbf {\bibinfo {volume}
  {55}},\ \bibinfo {pages} {32} (\bibinfo {year} {2023})},\ \Eprint
  {https://arxiv.org/abs/2301.06480} {arXiv:2301.06480 [gr-qc]} \BibitemShut
  {NoStop}%
\bibitem [{\citenamefont {Hawking}\ \emph {et~al.}(1976)\citenamefont
  {Hawking}, \citenamefont {King},\ and\ \citenamefont
  {Mccarthy}}]{Hawking:1976fe}%
  \BibitemOpen
  \bibfield  {author} {\bibinfo {author} {\bibfnamefont {S.~W.}\ \bibnamefont
  {Hawking}}, \bibinfo {author} {\bibfnamefont {A.~R.}\ \bibnamefont {King}},\
  and\ \bibinfo {author} {\bibfnamefont {P.~J.}\ \bibnamefont {Mccarthy}},\
  }\href {https://doi.org/10.1063/1.522874} {\bibfield  {journal} {\bibinfo
  {journal} {J. Math. Phys.}\ }\textbf {\bibinfo {volume} {17}},\ \bibinfo
  {pages} {174} (\bibinfo {year} {1976})}\BibitemShut {NoStop}%
\bibitem [{\citenamefont {Malament}(1977)}]{malament1977class}%
  \BibitemOpen
  \bibfield  {author} {\bibinfo {author} {\bibfnamefont {D.~B.}\ \bibnamefont
  {Malament}},\ }\href@noop {} {\bibfield  {journal} {\bibinfo  {journal}
  {Journal of mathematical physics}\ }\textbf {\bibinfo {volume} {18}},\
  \bibinfo {pages} {1399} (\bibinfo {year} {1977})}\BibitemShut {NoStop}%
\bibitem [{\citenamefont {Dowker}(2005)}]{Dowker:2005tz}%
  \BibitemOpen
  \bibfield  {author} {\bibinfo {author} {\bibfnamefont {F.}~\bibnamefont
  {Dowker}},\ }\bibinfo {title} {{Causal sets and the deep structure of
  spacetime}},\ in\ \href {https://doi.org/10.1142/9789812700988_0016} {\emph
  {\bibinfo {booktitle} {{100 Years Of Relativity}: {space-time structure:
  Einstein and beyond}}}},\ \bibinfo {editor} {edited by\ \bibinfo {editor}
  {\bibfnamefont {A.}~\bibnamefont {Ashtekar}}}\ (\bibinfo {year} {2005})\ pp.\
  \bibinfo {pages} {445--464},\ \Eprint {https://arxiv.org/abs/gr-qc/0508109}
  {arXiv:gr-qc/0508109} \BibitemShut {NoStop}%
\bibitem [{\citenamefont {Surya}(2019)}]{Surya:2019ndm}%
  \BibitemOpen
  \bibfield  {author} {\bibinfo {author} {\bibfnamefont {S.}~\bibnamefont
  {Surya}},\ }\href {https://doi.org/10.1007/s41114-019-0023-1} {\bibfield
  {journal} {\bibinfo  {journal} {Living Rev. Rel.}\ }\textbf {\bibinfo
  {volume} {22}},\ \bibinfo {pages} {5} (\bibinfo {year} {2019})},\ \Eprint
  {https://arxiv.org/abs/1903.11544} {arXiv:1903.11544 [gr-qc]} \BibitemShut
  {NoStop}%
\bibitem [{\citenamefont {Surya}(2025)}]{Surya:2025knk}%
  \BibitemOpen
  \bibfield  {author} {\bibinfo {author} {\bibfnamefont {S.}~\bibnamefont
  {Surya}},\ }\href {https://doi.org/10.1007/978-3-031-84420-1} {\bibfield
  {journal} {\bibinfo  {journal} {Lect. Notes Phys.}\ }\textbf {\bibinfo
  {volume} {1036}},\ \bibinfo {pages} {pp.} (\bibinfo {year}
  {2025})}\BibitemShut {NoStop}%
\bibitem [{\citenamefont {Carlip}\ \emph {et~al.}(2023)\citenamefont {Carlip},
  \citenamefont {Carlip},\ and\ \citenamefont {Surya}}]{Carlip:2022nsv}%
  \BibitemOpen
  \bibfield  {author} {\bibinfo {author} {\bibfnamefont {P.}~\bibnamefont
  {Carlip}}, \bibinfo {author} {\bibfnamefont {S.}~\bibnamefont {Carlip}},\
  and\ \bibinfo {author} {\bibfnamefont {S.}~\bibnamefont {Surya}},\ }\href
  {https://doi.org/10.1088/1361-6382/acc50c} {\bibfield  {journal} {\bibinfo
  {journal} {Class. Quant. Grav.}\ }\textbf {\bibinfo {volume} {40}},\ \bibinfo
  {pages} {095004} (\bibinfo {year} {2023})},\ \Eprint
  {https://arxiv.org/abs/2209.00327} {arXiv:2209.00327 [gr-qc]} \BibitemShut
  {NoStop}%
\bibitem [{\citenamefont {Carlip}\ \emph {et~al.}(2024)\citenamefont {Carlip},
  \citenamefont {Carlip},\ and\ \citenamefont {Surya}}]{Carlip:2023zki}%
  \BibitemOpen
  \bibfield  {author} {\bibinfo {author} {\bibfnamefont {P.}~\bibnamefont
  {Carlip}}, \bibinfo {author} {\bibfnamefont {S.}~\bibnamefont {Carlip}},\
  and\ \bibinfo {author} {\bibfnamefont {S.}~\bibnamefont {Surya}},\ }\href
  {https://doi.org/10.1088/1361-6382/ad506e} {\bibfield  {journal} {\bibinfo
  {journal} {Class. Quant. Grav.}\ }\textbf {\bibinfo {volume} {41}},\ \bibinfo
  {pages} {145005} (\bibinfo {year} {2024})},\ \Eprint
  {https://arxiv.org/abs/2311.18238} {arXiv:2311.18238 [gr-qc]} \BibitemShut
  {NoStop}%
\bibitem [{\citenamefont {Surya}(2012)}]{Surya:2011du}%
  \BibitemOpen
  \bibfield  {author} {\bibinfo {author} {\bibfnamefont {S.}~\bibnamefont
  {Surya}},\ }\href {https://doi.org/10.1088/0264-9381/29/13/132001} {\bibfield
   {journal} {\bibinfo  {journal} {Class. Quant. Grav.}\ }\textbf {\bibinfo
  {volume} {29}},\ \bibinfo {pages} {132001} (\bibinfo {year} {2012})},\
  \Eprint {https://arxiv.org/abs/1110.6244} {arXiv:1110.6244 [gr-qc]}
  \BibitemShut {NoStop}%
\bibitem [{\citenamefont {Glaser}\ \emph {et~al.}(2018)\citenamefont {Glaser},
  \citenamefont {O'Connor},\ and\ \citenamefont {Surya}}]{Glaser:2017sbe}%
  \BibitemOpen
  \bibfield  {author} {\bibinfo {author} {\bibfnamefont {L.}~\bibnamefont
  {Glaser}}, \bibinfo {author} {\bibfnamefont {D.}~\bibnamefont {O'Connor}},\
  and\ \bibinfo {author} {\bibfnamefont {S.}~\bibnamefont {Surya}},\ }\href
  {https://doi.org/10.1088/1361-6382/aa9540} {\bibfield  {journal} {\bibinfo
  {journal} {Class. Quant. Grav.}\ }\textbf {\bibinfo {volume} {35}},\ \bibinfo
  {pages} {045006} (\bibinfo {year} {2018})},\ \Eprint
  {https://arxiv.org/abs/1706.06432} {arXiv:1706.06432 [gr-qc]} \BibitemShut
  {NoStop}%
\bibitem [{\citenamefont {Eichhorn}(2018)}]{Eichhorn:2017bwe}%
  \BibitemOpen
  \bibfield  {author} {\bibinfo {author} {\bibfnamefont {A.}~\bibnamefont
  {Eichhorn}},\ }\href {https://doi.org/10.1088/1361-6382/aaa0a3} {\bibfield
  {journal} {\bibinfo  {journal} {Class. Quant. Grav.}\ }\textbf {\bibinfo
  {volume} {35}},\ \bibinfo {pages} {044001} (\bibinfo {year} {2018})},\
  \Eprint {https://arxiv.org/abs/1709.10419} {arXiv:1709.10419 [gr-qc]}
  \BibitemShut {NoStop}%
\bibitem [{\citenamefont {Cunningham}\ and\ \citenamefont
  {Surya}(2020)}]{Cunningham:2019rob}%
  \BibitemOpen
  \bibfield  {author} {\bibinfo {author} {\bibfnamefont {W.~J.}\ \bibnamefont
  {Cunningham}}\ and\ \bibinfo {author} {\bibfnamefont {S.}~\bibnamefont
  {Surya}},\ }\href {https://doi.org/10.1088/1361-6382/ab60b7} {\bibfield
  {journal} {\bibinfo  {journal} {Class. Quant. Grav.}\ }\textbf {\bibinfo
  {volume} {37}},\ \bibinfo {pages} {054002} (\bibinfo {year} {2020})},\
  \Eprint {https://arxiv.org/abs/1908.11647} {arXiv:1908.11647 [gr-qc]}
  \BibitemShut {NoStop}%
\bibitem [{\citenamefont {Eichhorn}(2019)}]{Eichhorn:2019xav}%
  \BibitemOpen
  \bibfield  {author} {\bibinfo {author} {\bibfnamefont {A.}~\bibnamefont
  {Eichhorn}},\ }\href {https://doi.org/10.1088/1742-6596/1275/1/012010}
  {\bibfield  {journal} {\bibinfo  {journal} {J. Phys. Conf. Ser.}\ }\textbf
  {\bibinfo {volume} {1275}},\ \bibinfo {pages} {012010} (\bibinfo {year}
  {2019})},\ \Eprint {https://arxiv.org/abs/1902.00391} {arXiv:1902.00391
  [gr-qc]} \BibitemShut {NoStop}%
\bibitem [{\citenamefont {de~Boer}\ \emph {et~al.}(2022)\citenamefont {de~Boer}
  \emph {et~al.}}]{deBoer:2022zka}%
  \BibitemOpen
  \bibfield  {author} {\bibinfo {author} {\bibfnamefont {J.}~\bibnamefont
  {de~Boer}} \emph {et~al.},\ }\href@noop {} {\  (\bibinfo {year} {2022})},\
  \Eprint {https://arxiv.org/abs/2207.10618} {arXiv:2207.10618 [hep-th]}
  \BibitemShut {NoStop}%
\bibitem [{\citenamefont {Rideout}\ and\ \citenamefont
  {Sorkin}(2000)}]{Rideout:1999ub}%
  \BibitemOpen
  \bibfield  {author} {\bibinfo {author} {\bibfnamefont {D.~P.}\ \bibnamefont
  {Rideout}}\ and\ \bibinfo {author} {\bibfnamefont {R.~D.}\ \bibnamefont
  {Sorkin}},\ }\href {https://doi.org/10.1103/PhysRevD.61.024002} {\bibfield
  {journal} {\bibinfo  {journal} {Phys. Rev. D}\ }\textbf {\bibinfo {volume}
  {61}},\ \bibinfo {pages} {024002} (\bibinfo {year} {2000})},\ \Eprint
  {https://arxiv.org/abs/gr-qc/9904062} {arXiv:gr-qc/9904062} \BibitemShut
  {NoStop}%
\bibitem [{\citenamefont {Zalel}(2024)}]{Zalel:2023uwy}%
  \BibitemOpen
  \bibfield  {author} {\bibinfo {author} {\bibfnamefont {S.}~\bibnamefont
  {Zalel}},\ }\bibinfo {title} {{Covariant Growth Dynamics}}\ (\bibinfo {year}
  {2024})\ \Eprint {https://arxiv.org/abs/2302.10582} {arXiv:2302.10582
  [gr-qc]} \BibitemShut {NoStop}%
\bibitem [{\citenamefont {Myrheim}(1978)}]{Myrheim:1978ce}%
  \BibitemOpen
  \bibfield  {author} {\bibinfo {author} {\bibfnamefont {J.}~\bibnamefont
  {Myrheim}},\ }\href@noop {} {\  (\bibinfo {year} {1978})}\BibitemShut
  {NoStop}%
\bibitem [{\citenamefont {Brightwell}\ and\ \citenamefont
  {Gregory}(1991)}]{Brightwell:1990ha}%
  \BibitemOpen
  \bibfield  {author} {\bibinfo {author} {\bibfnamefont {G.}~\bibnamefont
  {Brightwell}}\ and\ \bibinfo {author} {\bibfnamefont {R.}~\bibnamefont
  {Gregory}},\ }\href {https://doi.org/10.1103/PhysRevLett.66.260} {\bibfield
  {journal} {\bibinfo  {journal} {Phys. Rev. Lett.}\ }\textbf {\bibinfo
  {volume} {66}},\ \bibinfo {pages} {260} (\bibinfo {year} {1991})}\BibitemShut
  {NoStop}%
\bibitem [{\citenamefont {Reid}(2003)}]{Reid:2002sj}%
  \BibitemOpen
  \bibfield  {author} {\bibinfo {author} {\bibfnamefont {D.~D.}\ \bibnamefont
  {Reid}},\ }\href {https://doi.org/10.1103/PhysRevD.67.024034} {\bibfield
  {journal} {\bibinfo  {journal} {Phys. Rev. D}\ }\textbf {\bibinfo {volume}
  {67}},\ \bibinfo {pages} {024034} (\bibinfo {year} {2003})},\ \Eprint
  {https://arxiv.org/abs/gr-qc/0207103} {arXiv:gr-qc/0207103} \BibitemShut
  {NoStop}%
\bibitem [{\citenamefont {Major}\ \emph {et~al.}(2007)\citenamefont {Major},
  \citenamefont {Rideout},\ and\ \citenamefont {Surya}}]{Major:2006hv}%
  \BibitemOpen
  \bibfield  {author} {\bibinfo {author} {\bibfnamefont {S.}~\bibnamefont
  {Major}}, \bibinfo {author} {\bibfnamefont {D.}~\bibnamefont {Rideout}},\
  and\ \bibinfo {author} {\bibfnamefont {S.}~\bibnamefont {Surya}},\ }\href
  {https://doi.org/10.1063/1.2435599} {\bibfield  {journal} {\bibinfo
  {journal} {J. Math. Phys.}\ }\textbf {\bibinfo {volume} {48}},\ \bibinfo
  {pages} {032501} (\bibinfo {year} {2007})},\ \Eprint
  {https://arxiv.org/abs/gr-qc/0604124} {arXiv:gr-qc/0604124} \BibitemShut
  {NoStop}%
\bibitem [{\citenamefont {Sorkin}(2007)}]{Sorkin:2007qi}%
  \BibitemOpen
  \bibfield  {author} {\bibinfo {author} {\bibfnamefont {R.~D.}\ \bibnamefont
  {Sorkin}},\ }\href@noop {} {\ ,\ \bibinfo {pages} {26} (\bibinfo {year}
  {2007})},\ \Eprint {https://arxiv.org/abs/gr-qc/0703099}
  {arXiv:gr-qc/0703099} \BibitemShut {NoStop}%
\bibitem [{\citenamefont {Rideout}\ and\ \citenamefont
  {Wallden}(2009)}]{Rideout:2008rk}%
  \BibitemOpen
  \bibfield  {author} {\bibinfo {author} {\bibfnamefont {D.}~\bibnamefont
  {Rideout}}\ and\ \bibinfo {author} {\bibfnamefont {P.}~\bibnamefont
  {Wallden}},\ }\href {https://doi.org/10.1088/0264-9381/26/15/155013}
  {\bibfield  {journal} {\bibinfo  {journal} {Class. Quant. Grav.}\ }\textbf
  {\bibinfo {volume} {26}},\ \bibinfo {pages} {155013} (\bibinfo {year}
  {2009})},\ \Eprint {https://arxiv.org/abs/0810.1768} {arXiv:0810.1768
  [gr-qc]} \BibitemShut {NoStop}%
\bibitem [{\citenamefont {Major}\ \emph {et~al.}(2009)\citenamefont {Major},
  \citenamefont {Rideout},\ and\ \citenamefont {Surya}}]{Major:2009cw}%
  \BibitemOpen
  \bibfield  {author} {\bibinfo {author} {\bibfnamefont {S.}~\bibnamefont
  {Major}}, \bibinfo {author} {\bibfnamefont {D.}~\bibnamefont {Rideout}},\
  and\ \bibinfo {author} {\bibfnamefont {S.}~\bibnamefont {Surya}},\ }\href
  {https://doi.org/10.1088/0264-9381/26/17/175008} {\bibfield  {journal}
  {\bibinfo  {journal} {Class. Quant. Grav.}\ }\textbf {\bibinfo {volume}
  {26}},\ \bibinfo {pages} {175008} (\bibinfo {year} {2009})},\ \Eprint
  {https://arxiv.org/abs/0902.0434} {arXiv:0902.0434 [gr-qc]} \BibitemShut
  {NoStop}%
\bibitem [{\citenamefont {Benincasa}\ and\ \citenamefont
  {Dowker}(2010)}]{Benincasa:2010ac}%
  \BibitemOpen
  \bibfield  {author} {\bibinfo {author} {\bibfnamefont {D.~M.~T.}\
  \bibnamefont {Benincasa}}\ and\ \bibinfo {author} {\bibfnamefont
  {F.}~\bibnamefont {Dowker}},\ }\href
  {https://doi.org/10.1103/PhysRevLett.104.181301} {\bibfield  {journal}
  {\bibinfo  {journal} {Phys. Rev. Lett.}\ }\textbf {\bibinfo {volume} {104}},\
  \bibinfo {pages} {181301} (\bibinfo {year} {2010})},\ \Eprint
  {https://arxiv.org/abs/1001.2725} {arXiv:1001.2725 [gr-qc]} \BibitemShut
  {NoStop}%
\bibitem [{\citenamefont {Aslanbeigi}\ \emph {et~al.}(2014)\citenamefont
  {Aslanbeigi}, \citenamefont {Saravani},\ and\ \citenamefont
  {Sorkin}}]{Aslanbeigi:2014zva}%
  \BibitemOpen
  \bibfield  {author} {\bibinfo {author} {\bibfnamefont {S.}~\bibnamefont
  {Aslanbeigi}}, \bibinfo {author} {\bibfnamefont {M.}~\bibnamefont
  {Saravani}},\ and\ \bibinfo {author} {\bibfnamefont {R.~D.}\ \bibnamefont
  {Sorkin}},\ }\href {https://doi.org/10.1007/JHEP06(2014)024} {\bibfield
  {journal} {\bibinfo  {journal} {JHEP}\ }\textbf {\bibinfo {volume} {06}},\
  \bibinfo {pages} {024}},\ \Eprint {https://arxiv.org/abs/1403.1622}
  {arXiv:1403.1622 [hep-th]} \BibitemShut {NoStop}%
\bibitem [{\citenamefont {Eichhorn}\ \emph {et~al.}(2019)\citenamefont
  {Eichhorn}, \citenamefont {Surya},\ and\ \citenamefont
  {Versteegen}}]{Eichhorn:2018doy}%
  \BibitemOpen
  \bibfield  {author} {\bibinfo {author} {\bibfnamefont {A.}~\bibnamefont
  {Eichhorn}}, \bibinfo {author} {\bibfnamefont {S.}~\bibnamefont {Surya}},\
  and\ \bibinfo {author} {\bibfnamefont {F.}~\bibnamefont {Versteegen}},\
  }\href {https://doi.org/10.1088/1361-6382/ab114b} {\bibfield  {journal}
  {\bibinfo  {journal} {Class. Quant. Grav.}\ }\textbf {\bibinfo {volume}
  {36}},\ \bibinfo {pages} {105005} (\bibinfo {year} {2019})},\ \Eprint
  {https://arxiv.org/abs/1809.06192} {arXiv:1809.06192 [gr-qc]} \BibitemShut
  {NoStop}%
\bibitem [{\citenamefont {de~Brito}\ \emph {et~al.}(2023)\citenamefont
  {de~Brito}, \citenamefont {Eichhorn},\ and\ \citenamefont
  {Pfeiffer}}]{deBrito:2023axj}%
  \BibitemOpen
  \bibfield  {author} {\bibinfo {author} {\bibfnamefont {G.~P.}\ \bibnamefont
  {de~Brito}}, \bibinfo {author} {\bibfnamefont {A.}~\bibnamefont {Eichhorn}},\
  and\ \bibinfo {author} {\bibfnamefont {C.}~\bibnamefont {Pfeiffer}},\ }\href
  {https://doi.org/10.1140/epjp/s13360-023-04202-y} {\bibfield  {journal}
  {\bibinfo  {journal} {Eur. Phys. J. Plus}\ }\textbf {\bibinfo {volume}
  {138}},\ \bibinfo {pages} {592} (\bibinfo {year} {2023})},\ \Eprint
  {https://arxiv.org/abs/2301.13525} {arXiv:2301.13525 [gr-qc]} \BibitemShut
  {NoStop}%
\bibitem [{\citenamefont {Surya}(2026)}]{Surya:2025mvt}%
  \BibitemOpen
  \bibfield  {author} {\bibinfo {author} {\bibfnamefont {S.}~\bibnamefont
  {Surya}},\ }\href {https://doi.org/10.1103/txbf-hvz3} {\bibfield  {journal}
  {\bibinfo  {journal} {Phys. Rev. D}\ }\textbf {\bibinfo {volume} {113}},\
  \bibinfo {pages} {024034} (\bibinfo {year} {2026})},\ \Eprint
  {https://arxiv.org/abs/2510.19403} {arXiv:2510.19403 [gr-qc]} \BibitemShut
  {NoStop}%
\bibitem [{\citenamefont {Johnston}(2009)}]{Johnston:2009fr}%
  \BibitemOpen
  \bibfield  {author} {\bibinfo {author} {\bibfnamefont {S.}~\bibnamefont
  {Johnston}},\ }\href {https://doi.org/10.1103/PhysRevLett.103.180401}
  {\bibfield  {journal} {\bibinfo  {journal} {Phys. Rev. Lett.}\ }\textbf
  {\bibinfo {volume} {103}},\ \bibinfo {pages} {180401} (\bibinfo {year}
  {2009})},\ \Eprint {https://arxiv.org/abs/0909.0944} {arXiv:0909.0944
  [hep-th]} \BibitemShut {NoStop}%
\bibitem [{\citenamefont {Sorkin}(2011)}]{Sorkin:2011pn}%
  \BibitemOpen
  \bibfield  {author} {\bibinfo {author} {\bibfnamefont {R.~D.}\ \bibnamefont
  {Sorkin}},\ }\href {https://doi.org/10.1088/1742-6596/306/1/012017}
  {\bibfield  {journal} {\bibinfo  {journal} {J. Phys. Conf. Ser.}\ }\textbf
  {\bibinfo {volume} {306}},\ \bibinfo {pages} {012017} (\bibinfo {year}
  {2011})},\ \Eprint {https://arxiv.org/abs/1107.0698} {arXiv:1107.0698
  [gr-qc]} \BibitemShut {NoStop}%
\bibitem [{\citenamefont {Albertini}\ \emph {et~al.}(2024)\citenamefont
  {Albertini}, \citenamefont {Dowker}, \citenamefont {Nasiri},\ and\
  \citenamefont {Zalel}}]{Albertini:2024srq}%
  \BibitemOpen
  \bibfield  {author} {\bibinfo {author} {\bibfnamefont {E.}~\bibnamefont
  {Albertini}}, \bibinfo {author} {\bibfnamefont {F.}~\bibnamefont {Dowker}},
  \bibinfo {author} {\bibfnamefont {A.}~\bibnamefont {Nasiri}},\ and\ \bibinfo
  {author} {\bibfnamefont {S.}~\bibnamefont {Zalel}},\ }\href
  {https://doi.org/10.1103/PhysRevD.109.106014} {\bibfield  {journal} {\bibinfo
   {journal} {Phys. Rev. D}\ }\textbf {\bibinfo {volume} {109}},\ \bibinfo
  {pages} {106014} (\bibinfo {year} {2024})},\ \Eprint
  {https://arxiv.org/abs/2402.08555} {arXiv:2402.08555 [hep-th]} \BibitemShut
  {NoStop}%
\bibitem [{\citenamefont {Dou}(2024)}]{Dou:2023puz}%
  \BibitemOpen
  \bibfield  {author} {\bibinfo {author} {\bibfnamefont {D.}~\bibnamefont
  {Dou}},\ }\bibinfo {title} {{On Horizon Molecules and Entropy in Causal
  Sets}}\ (\bibinfo {year} {2024})\ \Eprint {https://arxiv.org/abs/2307.04150}
  {arXiv:2307.04150 [gr-qc]} \BibitemShut {NoStop}%
\bibitem [{\citenamefont {Yazdi}(2024)}]{Yazdi:2022hhv}%
  \BibitemOpen
  \bibfield  {author} {\bibinfo {author} {\bibfnamefont {Y.~K.}\ \bibnamefont
  {Yazdi}},\ }\bibinfo {title} {{Entanglement Entropy and Causal Set Theory}}\
  (\bibinfo {year} {2024})\ \Eprint {https://arxiv.org/abs/2212.13586}
  {arXiv:2212.13586 [hep-th]} \BibitemShut {NoStop}%
\bibitem [{\citenamefont {Jubb}(2024)}]{Jubb:2023mlv}%
  \BibitemOpen
  \bibfield  {author} {\bibinfo {author} {\bibfnamefont {I.}~\bibnamefont
  {Jubb}},\ }\bibinfo {title} {{Interacting Quantum Scalar Field Theory on a
  Causal Set}}\ (\bibinfo {year} {2024})\ \Eprint
  {https://arxiv.org/abs/2306.12484} {arXiv:2306.12484 [hep-th]} \BibitemShut
  {NoStop}%
\bibitem [{\citenamefont {Glaser}(2024)}]{Glaser:2023pcl}%
  \BibitemOpen
  \bibfield  {author} {\bibinfo {author} {\bibfnamefont {L.}~\bibnamefont
  {Glaser}},\ }\bibinfo {title} {{Computer Simulations of Causal Sets}}\
  (\bibinfo {year} {2024})\ \Eprint {https://arxiv.org/abs/2306.09904}
  {arXiv:2306.09904 [gr-qc]} \BibitemShut {NoStop}%
\bibitem [{\citenamefont {X.}(2024)}]{X:2023ewv}%
  \BibitemOpen
  \bibfield  {author} {\bibinfo {author} {\bibfnamefont {N.}~\bibnamefont
  {X.}},\ }\bibinfo {title} {{Quantum Field Theory on Causal Sets}}\ (\bibinfo
  {year} {2024})\ \Eprint {https://arxiv.org/abs/2306.04800} {arXiv:2306.04800
  [hep-th]} \BibitemShut {NoStop}%
\bibitem [{\citenamefont {Ashmead}\ and\ \citenamefont
  {Reid}(2024)}]{Ashmead:2024pmh}%
  \BibitemOpen
  \bibfield  {author} {\bibinfo {author} {\bibfnamefont {F.}~\bibnamefont
  {Ashmead}}\ and\ \bibinfo {author} {\bibfnamefont {D.~D.}\ \bibnamefont
  {Reid}},\ }\bibinfo {title} {{Estimating the Manifold Dimension of Causal
  Sets}}\ (\bibinfo {year} {2024})\BibitemShut {NoStop}%
\bibitem [{\citenamefont {Ahmed}\ and\ \citenamefont
  {Shafi}(2024)}]{Ahmed:2024gzc}%
  \BibitemOpen
  \bibfield  {author} {\bibinfo {author} {\bibfnamefont {M.}~\bibnamefont
  {Ahmed}}\ and\ \bibinfo {author} {\bibfnamefont {H.}~\bibnamefont {Shafi}},\
  }\bibinfo {title} {{Causal Set Cosmology}}\ (\bibinfo {year}
  {2024})\BibitemShut {NoStop}%
\bibitem [{\citenamefont {Mathur}(2024)}]{Mathur:2024gmk}%
  \BibitemOpen
  \bibfield  {author} {\bibinfo {author} {\bibfnamefont {A.}~\bibnamefont
  {Mathur}},\ }\bibinfo {title} {{Toward the Emergence of Continuum Spacetime
  in Causal Set Theory}}\ (\bibinfo {year} {2024})\BibitemShut {NoStop}%
\bibitem [{\citenamefont {He}\ and\ \citenamefont {Rideout}(2009)}]{He:2008ku}%
  \BibitemOpen
  \bibfield  {author} {\bibinfo {author} {\bibfnamefont {S.}~\bibnamefont
  {He}}\ and\ \bibinfo {author} {\bibfnamefont {D.}~\bibnamefont {Rideout}},\
  }\href {https://doi.org/10.1088/0264-9381/26/12/125015} {\bibfield  {journal}
  {\bibinfo  {journal} {Class. Quant. Grav.}\ }\textbf {\bibinfo {volume}
  {26}},\ \bibinfo {pages} {125015} (\bibinfo {year} {2009})},\ \Eprint
  {https://arxiv.org/abs/0811.4235} {arXiv:0811.4235 [gr-qc]} \BibitemShut
  {NoStop}%
\bibitem [{\citenamefont {Hom{\v{s}}ak}\ and\ \citenamefont
  {Veroni}(2024)}]{Homsak:2024tce}%
  \BibitemOpen
  \bibfield  {author} {\bibinfo {author} {\bibfnamefont {V.}~\bibnamefont
  {Hom{\v{s}}ak}}\ and\ \bibinfo {author} {\bibfnamefont {S.}~\bibnamefont
  {Veroni}},\ }\href {https://doi.org/10.1103/PhysRevD.110.026015} {\bibfield
  {journal} {\bibinfo  {journal} {Phys. Rev. D}\ }\textbf {\bibinfo {volume}
  {110}},\ \bibinfo {pages} {026015} (\bibinfo {year} {2024})},\ \Eprint
  {https://arxiv.org/abs/2404.11670} {arXiv:2404.11670 [gr-qc]} \BibitemShut
  {NoStop}%
\bibitem [{\citenamefont {Olmo}\ and\ \citenamefont
  {Rubiera-Garcia}(2022)}]{Olmo:2022cui}%
  \BibitemOpen
  \bibfield  {author} {\bibinfo {author} {\bibfnamefont {G.~J.}\ \bibnamefont
  {Olmo}}\ and\ \bibinfo {author} {\bibfnamefont {D.}~\bibnamefont
  {Rubiera-Garcia}},\ }\href@noop {} {\  (\bibinfo {year} {2022})},\ \Eprint
  {https://arxiv.org/abs/2209.05061} {arXiv:2209.05061 [gr-qc]} \BibitemShut
  {NoStop}%
\bibitem [{\citenamefont {Eichhorn}\ and\ \citenamefont
  {Held}(2022)}]{Eichhorn:2022bgu}%
  \BibitemOpen
  \bibfield  {author} {\bibinfo {author} {\bibfnamefont {A.}~\bibnamefont
  {Eichhorn}}\ and\ \bibinfo {author} {\bibfnamefont {A.}~\bibnamefont
  {Held}},\ }\href@noop {} {\  (\bibinfo {year} {2022})},\ \Eprint
  {https://arxiv.org/abs/2212.09495} {arXiv:2212.09495 [gr-qc]} \BibitemShut
  {NoStop}%
\bibitem [{\citenamefont {Ashtekar}\ \emph {et~al.}(2023)\citenamefont
  {Ashtekar}, \citenamefont {Olmedo},\ and\ \citenamefont
  {Singh}}]{Ashtekar:2023cod}%
  \BibitemOpen
  \bibfield  {author} {\bibinfo {author} {\bibfnamefont {A.}~\bibnamefont
  {Ashtekar}}, \bibinfo {author} {\bibfnamefont {J.}~\bibnamefont {Olmedo}},\
  and\ \bibinfo {author} {\bibfnamefont {P.}~\bibnamefont {Singh}},\
  }\href@noop {} {\  (\bibinfo {year} {2023})},\ \Eprint
  {https://arxiv.org/abs/2301.01309} {arXiv:2301.01309 [gr-qc]} \BibitemShut
  {NoStop}%
\bibitem [{\citenamefont {Carballo-Rubio}\ \emph {et~al.}(2023)\citenamefont
  {Carballo-Rubio}, \citenamefont {Di~Filippo}, \citenamefont {Liberati},\ and\
  \citenamefont {Visser}}]{Carballo-Rubio:2023mvr}%
  \BibitemOpen
  \bibfield  {author} {\bibinfo {author} {\bibfnamefont {R.}~\bibnamefont
  {Carballo-Rubio}}, \bibinfo {author} {\bibfnamefont {F.}~\bibnamefont
  {Di~Filippo}}, \bibinfo {author} {\bibfnamefont {S.}~\bibnamefont
  {Liberati}},\ and\ \bibinfo {author} {\bibfnamefont {M.}~\bibnamefont
  {Visser}},\ }\href@noop {} {\  (\bibinfo {year} {2023})},\ \Eprint
  {https://arxiv.org/abs/2302.00028} {arXiv:2302.00028 [gr-qc]} \BibitemShut
  {NoStop}%
\bibitem [{\citenamefont {Carballo-Rubio}\ \emph {et~al.}(2025)\citenamefont
  {Carballo-Rubio} \emph {et~al.}}]{Carballo-Rubio:2025fnc}%
  \BibitemOpen
  \bibfield  {author} {\bibinfo {author} {\bibfnamefont {R.}~\bibnamefont
  {Carballo-Rubio}} \emph {et~al.},\ }\href
  {https://doi.org/10.1088/1475-7516/2025/05/003} {\bibfield  {journal}
  {\bibinfo  {journal} {JCAP}\ }\textbf {\bibinfo {volume} {05}},\ \bibinfo
  {pages} {003}},\ \Eprint {https://arxiv.org/abs/2501.05505} {arXiv:2501.05505
  [gr-qc]} \BibitemShut {NoStop}%
\bibitem [{\citenamefont {Bueno}\ \emph {et~al.}(2026)\citenamefont {Bueno},
  \citenamefont {Cano}, \citenamefont {Hennigar},\ and\ \citenamefont
  {Murcia}}]{Bueno:2025zaj}%
  \BibitemOpen
  \bibfield  {author} {\bibinfo {author} {\bibfnamefont {P.}~\bibnamefont
  {Bueno}}, \bibinfo {author} {\bibfnamefont {P.~A.}\ \bibnamefont {Cano}},
  \bibinfo {author} {\bibfnamefont {R.~A.}\ \bibnamefont {Hennigar}},\ and\
  \bibinfo {author} {\bibfnamefont {{\'A}.~J.}\ \bibnamefont {Murcia}},\ }\href
  {https://doi.org/10.1103/8f3j-zcxh} {\bibfield  {journal} {\bibinfo
  {journal} {Phys. Rev. D}\ }\textbf {\bibinfo {volume} {113}},\ \bibinfo
  {pages} {024019} (\bibinfo {year} {2026})},\ \Eprint
  {https://arxiv.org/abs/2509.19016} {arXiv:2509.19016 [gr-qc]} \BibitemShut
  {NoStop}%
\bibitem [{\citenamefont {Eichhorn}\ and\ \citenamefont
  {Fernandes}(2025)}]{Eichhorn:2025pgy}%
  \BibitemOpen
  \bibfield  {author} {\bibinfo {author} {\bibfnamefont {A.}~\bibnamefont
  {Eichhorn}}\ and\ \bibinfo {author} {\bibfnamefont {P.~G.~S.}\ \bibnamefont
  {Fernandes}},\ }\href@noop {} {\  (\bibinfo {year} {2025})},\ \Eprint
  {https://arxiv.org/abs/2508.00686} {arXiv:2508.00686 [gr-qc]} \BibitemShut
  {NoStop}%
\bibitem [{\citenamefont {Carballo-Rubio}(2026)}]{Carballo-Rubio:2025ntd}%
  \BibitemOpen
  \bibfield  {author} {\bibinfo {author} {\bibfnamefont {R.}~\bibnamefont
  {Carballo-Rubio}},\ }\href {https://doi.org/10.1038/s41467-026-69035-6}
  {\bibfield  {journal} {\bibinfo  {journal} {Nature Commun.}\ }\textbf
  {\bibinfo {volume} {17}},\ \bibinfo {pages} {1399} (\bibinfo {year}
  {2026})},\ \Eprint {https://arxiv.org/abs/2507.15920} {arXiv:2507.15920
  [gr-qc]} \BibitemShut {NoStop}%
\bibitem [{\citenamefont {Borissova}\ and\ \citenamefont
  {Carballo-Rubio}(2026)}]{Borissova:2026wmn}%
  \BibitemOpen
  \bibfield  {author} {\bibinfo {author} {\bibfnamefont {J.}~\bibnamefont
  {Borissova}}\ and\ \bibinfo {author} {\bibfnamefont {R.}~\bibnamefont
  {Carballo-Rubio}},\ }\href@noop {} {\  (\bibinfo {year} {2026})},\ \Eprint
  {https://arxiv.org/abs/2602.16773} {arXiv:2602.16773 [gr-qc]} \BibitemShut
  {NoStop}%
\bibitem [{\citenamefont {Hayward}(2006)}]{Hayward:2005gi}%
  \BibitemOpen
  \bibfield  {author} {\bibinfo {author} {\bibfnamefont {S.~A.}\ \bibnamefont
  {Hayward}},\ }\href {https://doi.org/10.1103/PhysRevLett.96.031103}
  {\bibfield  {journal} {\bibinfo  {journal} {Phys. Rev. Lett.}\ }\textbf
  {\bibinfo {volume} {96}},\ \bibinfo {pages} {031103} (\bibinfo {year}
  {2006})},\ \Eprint {https://arxiv.org/abs/gr-qc/0506126}
  {arXiv:gr-qc/0506126} \BibitemShut {NoStop}%
\bibitem [{\citenamefont {Zhou}\ and\ \citenamefont
  {Modesto}(2023)}]{Zhou:2022yio}%
  \BibitemOpen
  \bibfield  {author} {\bibinfo {author} {\bibfnamefont {T.}~\bibnamefont
  {Zhou}}\ and\ \bibinfo {author} {\bibfnamefont {L.}~\bibnamefont {Modesto}},\
  }\href {https://doi.org/10.1103/PhysRevD.107.044016} {\bibfield  {journal}
  {\bibinfo  {journal} {Phys. Rev. D}\ }\textbf {\bibinfo {volume} {107}},\
  \bibinfo {pages} {044016} (\bibinfo {year} {2023})},\ \Eprint
  {https://arxiv.org/abs/2208.02557} {arXiv:2208.02557 [gr-qc]} \BibitemShut
  {NoStop}%
\bibitem [{\citenamefont {Ashtekar}\ and\ \citenamefont
  {Krishnan}(2004)}]{Ashtekar:2004cn}%
  \BibitemOpen
  \bibfield  {author} {\bibinfo {author} {\bibfnamefont {A.}~\bibnamefont
  {Ashtekar}}\ and\ \bibinfo {author} {\bibfnamefont {B.}~\bibnamefont
  {Krishnan}},\ }\href {https://doi.org/10.12942/lrr-2004-10} {\bibfield
  {journal} {\bibinfo  {journal} {Living Rev. Rel.}\ }\textbf {\bibinfo
  {volume} {7}},\ \bibinfo {pages} {10} (\bibinfo {year} {2004})},\ \Eprint
  {https://arxiv.org/abs/gr-qc/0407042} {arXiv:gr-qc/0407042} \BibitemShut
  {NoStop}%
\bibitem [{\citenamefont {Booth}(2005)}]{Booth:2005qc}%
  \BibitemOpen
  \bibfield  {author} {\bibinfo {author} {\bibfnamefont {I.}~\bibnamefont
  {Booth}},\ }\href {https://doi.org/10.1139/p05-063} {\bibfield  {journal}
  {\bibinfo  {journal} {Can. J. Phys.}\ }\textbf {\bibinfo {volume} {83}},\
  \bibinfo {pages} {1073} (\bibinfo {year} {2005})},\ \Eprint
  {https://arxiv.org/abs/gr-qc/0508107} {arXiv:gr-qc/0508107} \BibitemShut
  {NoStop}%
\bibitem [{\citenamefont {Poisson}(2009)}]{Poisson:2009pwt}%
  \BibitemOpen
  \bibfield  {author} {\bibinfo {author} {\bibfnamefont {E.}~\bibnamefont
  {Poisson}},\ }\href {https://doi.org/10.1017/CBO9780511606601} {\emph
  {\bibinfo {title} {{A Relativist's Toolkit: The Mathematics of Black-Hole
  Mechanics}}}}\ (\bibinfo  {publisher} {Cambridge University Press},\ \bibinfo
  {year} {2009})\BibitemShut {NoStop}%
\bibitem [{\citenamefont {Philpott}\ \emph {et~al.}(2009)\citenamefont
  {Philpott}, \citenamefont {Dowker},\ and\ \citenamefont
  {Sorkin}}]{Philpott:2008vd}%
  \BibitemOpen
  \bibfield  {author} {\bibinfo {author} {\bibfnamefont {L.}~\bibnamefont
  {Philpott}}, \bibinfo {author} {\bibfnamefont {F.}~\bibnamefont {Dowker}},\
  and\ \bibinfo {author} {\bibfnamefont {R.~D.}\ \bibnamefont {Sorkin}},\
  }\href {https://doi.org/10.1103/PhysRevD.79.124047} {\bibfield  {journal}
  {\bibinfo  {journal} {Phys. Rev. D}\ }\textbf {\bibinfo {volume} {79}},\
  \bibinfo {pages} {124047} (\bibinfo {year} {2009})},\ \Eprint
  {https://arxiv.org/abs/0810.5591} {arXiv:0810.5591 [gr-qc]} \BibitemShut
  {NoStop}%
\bibitem [{\citenamefont {Surya}\ \emph {et~al.}(2021)\citenamefont {Surya},
  \citenamefont {X},\ and\ \citenamefont {Yazdi}}]{Surya:2020gjj}%
  \BibitemOpen
  \bibfield  {author} {\bibinfo {author} {\bibfnamefont {S.}~\bibnamefont
  {Surya}}, \bibinfo {author} {\bibfnamefont {N.}~\bibnamefont {X}},\ and\
  \bibinfo {author} {\bibfnamefont {Y.~K.}\ \bibnamefont {Yazdi}},\ }\href
  {https://doi.org/10.1088/1361-6382/abf279} {\bibfield  {journal} {\bibinfo
  {journal} {Class. Quant. Grav.}\ }\textbf {\bibinfo {volume} {38}},\ \bibinfo
  {pages} {115001} (\bibinfo {year} {2021})},\ \Eprint
  {https://arxiv.org/abs/2008.07697} {arXiv:2008.07697 [gr-qc]} \BibitemShut
  {NoStop}%
\end{thebibliography}%

\end{document}